\documentclass[12pt]{article}
\usepackage[utf8]{inputenc}
\usepackage{newtxtext} 
\usepackage[scaled=.95]{cabin} 
\usepackage[varqu,varl]{inconsolata} 
\usepackage{amsmath}
\usepackage{amsfonts,url,epsfig,breakurl}
\usepackage{geometry}
\usepackage{sectsty}
\usepackage[normalem]{ulem}
\usepackage{url}
\usepackage{todonotes}

\usepackage{verbatim}

\sectionfont{\sffamily\bfseries\upshape\large}
\subsectionfont{\sffamily\bfseries\upshape\normalsize}
\subsubsectionfont{\sffamily\mdseries\upshape\normalsize}
\makeatletter
\renewcommand\@seccntformat[1]{\csname the#1\endcsname.\quad}
\makeatother\renewcommand{\bibitem}{\vskip2pt\par\hangindent\parindent\hskip-\parindent}

\makeatletter
\def\@maketitle{%
  \begin{center}%
  \let \footnote \thanks
    {\large \@title \par}%
    {\normalsize
      \begin{tabular}[t]{c}%
        \@author
      \end{tabular}\par}%
    {\small \@date}%
  \end{center}%
}
\makeatother

\title{\bf Bayesian workflow\footnote{We thank Berna Devezer, Danielle Navarro, Matthew West, and Ben Bales for helpful suggestions and the National Science Foundation, Institute of Education Sciences, Office of Naval Research, National Institutes of Health, Sloan Foundation, Schmidt Futures, the Canadian Research Chairs program, and the Natural Sciences and Engineering Research Council of Canada for financial support. This work was supported by ELIXIR CZ research infrastructure project (MEYS Grant No.\ LM2015047) including access to computing and storage facilities. Much of Sections \ref{golf} and \ref{sec:orbits} are taken from Gelman (2019) and Margossian and Gelman (2020), respectively.}\vspace{.1in}}

\author{Andrew Gelman\footnote{Department of Statistics, Columbia University, New York.}  \and Aki Vehtari\footnote{Department of Computer Science, Aalto University, Espoo, Finland.}\and Daniel Simpson\footnote{Department of Statistical Sciences, University of Toronto.} \and Charles C. Margossian$^{\dagger}$ \and Bob Carpenter\footnote{Center for Computational Mathematics, Flatiron Institute, New York.} \and Yuling Yao$^{\dagger}$  \and Lauren Kennedy\footnote{Monash University, Melbourne, Australia.} \and Jonah Gabry$^{\dagger}$ \and Paul-Christian Bürkner\footnote{Cluster of Excellence SimTech, University of Stuttgart, Germany.} \and Martin Modrák\footnote{Institute of Microbiology of the Czech Academy of Sciences.} 
\vspace{.1in}}

\date{2 Nov 2020}

\begin{document}\sloppy
\maketitle

\begin{abstract}
  \noindent
The Bayesian approach to data analysis provides a powerful way to
handle uncertainty in all observations, model parameters, and model structure
using probability theory. Probabilistic programming languages make it
easier to specify and fit Bayesian models, but this still leaves us with many options regarding constructing, evaluating, and using these models, along with many remaining challenges in computation. Using Bayesian inference to solve real-world problems requires not only statistical skills, subject matter knowledge, and programming, but also awareness of the decisions made in the process of data analysis. All of these aspects can be understood as part of a tangled workflow of applied Bayesian statistics. Beyond inference, the workflow also includes iterative model building,
model checking, validation and troubleshooting of computational problems, model
understanding, and model comparison.  We review all these aspects of workflow in the context of several examples,
keeping in mind that in practice we will be fitting many models for
any given problem, even if only a subset of them will ultimately be relevant for
our conclusions. 
  

\end{abstract}

\tableofcontents

\section{Introduction}

\subsection{From Bayesian inference to Bayesian workflow}

If mathematical statistics is to be the theory of applied statistics, then any serious discussion of Bayesian methods needs to be clear about how they are used in practice. In particular, we need to clearly separate concepts of Bayesian inference from Bayesian data analysis and, critically, from full Bayesian workflow (the object of our attention).

\begin{figure}
  \centerline{\includegraphics[width=\textwidth]{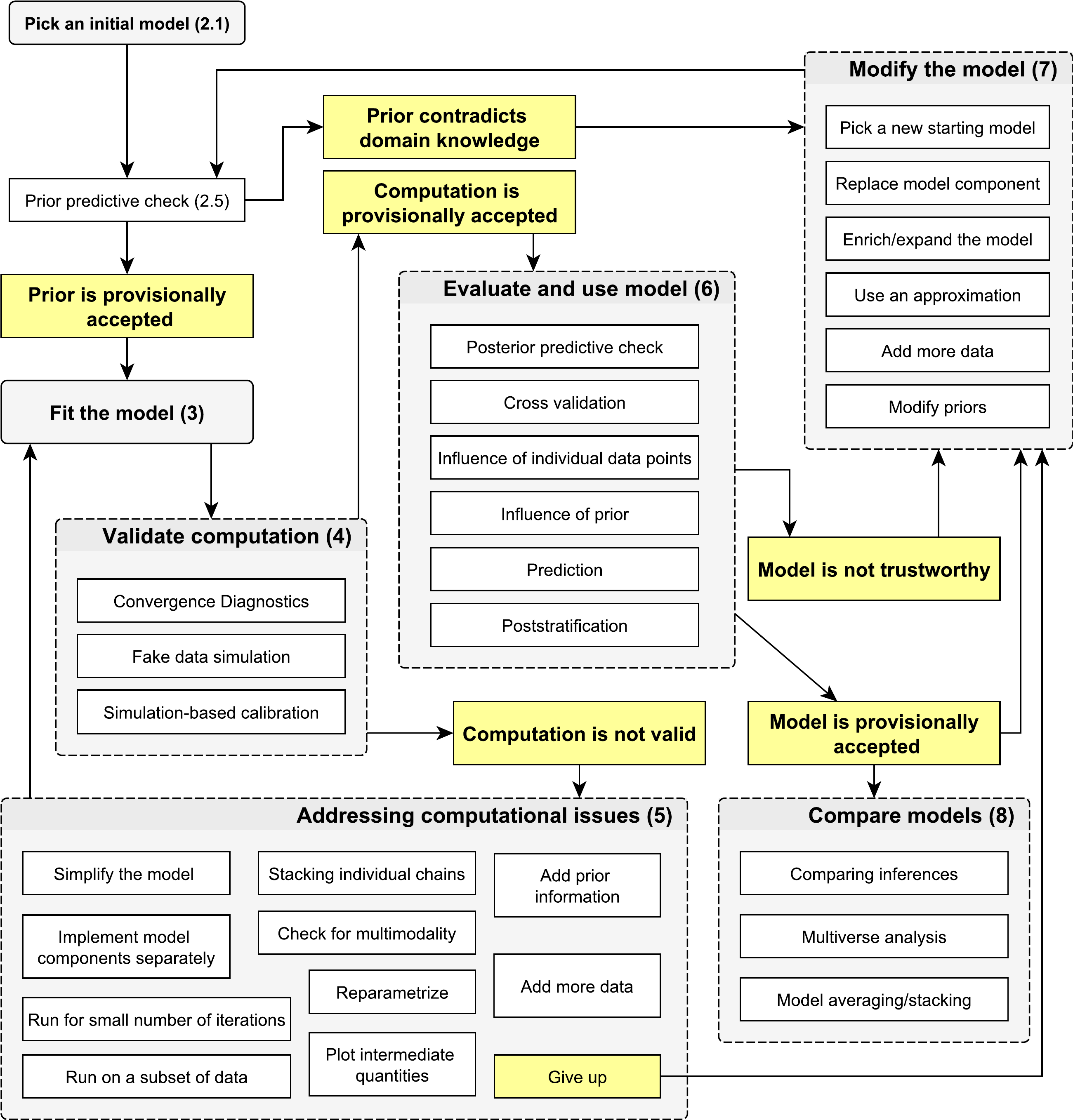}}
\caption{\em Overview of the steps we currently consider in Bayesian workflow. Numbers in brackets refer to sections of this paper where the steps are discussed. The chart aims to show possible steps and paths an individual analysis \emph{may} go through, with the understanding that any particular analysis will most likely not involve all of these steps. One of our goals in studying workflow is to understand how these ideas fit together so they can be applied more systematically.}
\label{flowchart}
\end{figure}

{\em Bayesian inference} is just the formulation and computation of conditional probability or probability densities, $p(\theta|y)\propto p(\theta)p(y|\theta)$.  
{\em Bayesian workflow} includes the three steps of model building, inference, and model checking/improvement, along with the comparison of different models, not just for the purpose of model choice or model averaging but more importantly to better understand these models. That is, for example, why some models have trouble predicting certain aspects of the data, or why uncertainty estimates of important parameters can vary across models.  Even when we have a model we like, it will be useful to compare its inferences to those from simpler and more complicated models as a way to understand what the model is doing. Figure \ref{flowchart} provides an outline. An extended Bayesian workflow would also include pre-data design of data collection and measurement and after-inference decision making, but we focus here on modeling the data.

In a typical Bayesian workflow we end up fitting a series of models, some of which are in retrospect poor choices (for reasons including poor fit to data; lack of connection to relevant substantive theory or practical goals; priors that are too weak, too strong, or otherwise inappropriate; or simply from programming errors), some of which are useful but flawed (for example, a regression that adjusts for some confounders but excludes others, or a parametric form that captures some but not all of a functional relationship), and some of which are ultimately worth reporting.  The hopelessly wrong models and the seriously flawed models are, in practice, unavoidable steps along the way toward fitting the useful models.  Recognizing this can change how we set up and apply statistical methods.

\subsection{Why do we need a Bayesian workflow?}
We need a Bayesian workflow, rather than mere Bayesian inference, for several reasons.
\begin{itemize}
\item Computation can be a challenge, and we often need to work through various steps including fitting simpler or alternative models, approximate computation that is less accurate but faster, and exploration of the fitting process, in order to get to inferences that we trust.
\item In difficult problems we typically do not know ahead of time what model we want to fit, and even in those rare cases that an acceptable model has been chosen ahead of time, we will generally want to expand it as we gather more data or want to ask more detailed questions of the data we have. 
\item Even if our data were static, and we knew what model to fit, and we had no problems fitting it, we still would want to understand the fitted model and its relation to the data, and that understanding can often best be achieved by comparing inferences from a series of related models. 
\item Sometimes different models yield different conclusions, without one of them being clearly favourable. In such cases, presenting multiple models is helpful to illustrate the uncertainty in model choice.
\end{itemize}

\subsection{``Workflow'' and its relation to statistical theory and practice}


``Workflow'' has different meanings in different contexts.  For the purposes of this paper it should suffice that workflow is more general than an example but less precisely specified than a method. 
We have been influenced by the ideas about workflow in computing that are in the air, including statistical developments such as the tidyverse which are not particularly Bayesian but have a similar feel of experiential learning (Wickham and Groelmund, 2017).  Many of the recent developments in machine learning have a similar plug-and-play feel:  they are easy to use, easy to experiment with, and users have the healthy sense that fitting a model is a way of learning something from the data without representing a commitment to some probability model or set of statistical assumptions.

\begin{figure}
  \centerline{
  \fbox{
    \begin{minipage}{0.7\textwidth}
\vspace{.2\baselineskip}
Example $\cdots\,$ Case study $\cdots\,$ Workflow $\cdots\,$  Method $\cdots\,$ Theory
\vspace{.1\baselineskip}
  \end{minipage}
}
}
\caption{\em Meta-workflow of statistical methodology, representing the way in which new ideas first appear in examples, then get formalized into case studies, codified as workflows, are given general implementation as algorithms, and become the subject of theories.}\label{workflow}
\end{figure}

Figure \ref{workflow} shows our perspective on the development of statistical methodology as a process of increasing codification, from example to case study to workflow to method to theory.  Not all methods will reach these final levels of mathematical abstraction, but looking at the history of statistics we have seen new methods being developed in the context of particular examples, stylized into case studies, set up as templates or workflows for new problems, and, when possible, formalized, coded, and studied theoretically.


One way to understand Figure \ref{workflow} is through important ideas in the history of statistics that have moved from left to right along that path.
There have been many ideas that started out as hacks or statistics-adjacent tools and eventually were formalized as methods and brought into the core of statistics.  Multilevel modeling is a formalization of what has been called empirical Bayes estimation of prior distributions, expanding the model so as to fold inference about priors into a fully Bayesian framework.  Exploratory data analysis can be understood as a form of predictive model checking (Gelman, 2003).  Regularization methods such as lasso (Tibshirani, 1996) and horseshoe  (Piironen et al., 2020) have replaced ad hoc variable selection tools in regression.  Nonparametric models such as Gaussian processes (O'Hagan, 1978, Rasumussen and Williams, 2006) can be thought of as Bayesian replacements for procedures such as kernel smoothing.  In each of these cases and many others, a framework of statistical methodology has been expanded to include existing methods, along the way making the methods more modular and potentially useful.

The term ``workflow'' has been gradually coming into use in statistics and data science; see for example Liu et al.\ (2005), Lins et al.\ (2008), Long (2009), and Turner and Lambert (2015).  Related ideas of workflow are in the air in software development and other fields of informatics; recent discussions for practitioners include Wilson et al.\ (2014, 2017). Applied statistics (not just Bayesian statistics) has become increasingly computational and algorithmic, and this has placed workflow at the center of statistical practice (see, for example, Grolemund and Wickham, 2017, Bryan, 2017, and Yu and Kumbier, 2020), as well as in application areas (for example, Lee et al., 2019, discuss modeling workflow in psychology research).  ``Bayesian workflow'' has been expressed as a general concept by Savage (2016), Gabry et al.\ (2019), and Betancourt (2020a).  Several of the individual components of Bayesian workflow were discussed by Gelman (2011) but not in a coherent way.  In addition there has been development of Bayesian workflow for particular problems, as by Shi and Stevens (2008) and Chiu et al.\ (2017).


In this paper we go through several aspects of Bayesian workflow with the hope that these can ultimately make their way into routine practice and automatic software.  We set up much of our workflow in the probabilistic programming language Stan (Carpenter et al., 2017, Stan Development Team, 2020), but similar ideas apply in other computing environments.\footnote{Wikipedia currently lists more than 50 probabilistic programming frameworks: \url{en.wikipedia.org/wiki/Probabilistic_programming}.}

\subsection{Organizing the many aspects of Bayesian workflow}
Textbook presentations of statistical workflow are often linear, with different paths corresponding to different problem situations.  For example, a clinical trial in medicine conventionally begins with a sample size calculation and analysis plan, followed by data collection, cleaning, and statistical analysis, and concluding with the reporting of $p$-values and confidence intervals.  An observational study in economics might begin with exploratory data analysis, which then informs choices of transformations of variables, followed by a set of regression analyses and then an array of alternative specifications and robustness studies.

The statistical workflow discussed in this article is more tangled than the usual data analysis workflows presented in textbooks and research articles.  The additional complexity comes in several places and there are many sub-workflows inside the higher level workflow:
\begin{enumerate}
\item Computation to fit a complex model can itself be difficult, requiring a certain amount of experimentation to solve the problem of computing, approximating, or simulating from the desired posterior distribution, along with checking that the computational algorithm did what it was intended to do.
\item With complex problems we typically have an idea of a general model that is more complex than we can easily computationally fit (for example including features such as correlations, hierarchical structure, and parameters varying over time), and so we start with a model that we know is missing some important features, in the hope it will be computationally easier, with the understanding that we will gradually add in features.
\item Relatedly, we often consider problems where the data are not fixed, either because data collection is ongoing or because we have the ability to draw in related datasets, for example new surveys in a public opinion analysis or data from other experiments in a drug trial.  Adding new data often requires model extensions to allow parameters to vary or to extend functional forms, as for example a linear model might fit well at first but then break down with data are added under new conditions.
\item Beyond all the challenges of fitting and expansion, models can often be best understood by comparing to inferences under alternative models.  Hence our workflow includes tools for understanding and comparing multiple models fit to the same data.
\end{enumerate}

Statistics is all about uncertainty.  In addition to the usual uncertainties in the data and model parameters, we are often uncertain whether we are fitting our models correctly, uncertain about how best to set up and expand our models, and uncertain in their interpretation.  Once we go beyond simple preassigned designs and analysis, our workflow can be disorderly, with our focus moving back and forth between data exploration, substantive theory, computing, and interpretation of results.  Thus, any attempt to organize the steps of workflow will oversimplify and many sub-workflows are complex enough to justify their own articles or book chapters.

We discuss many aspects of workflow, but practical considerations---especially available time, computational resources and the severity of penalty for being wrong---can compel a practitioner to take shortcuts. Such shortcuts can make interpretation of results more difficult, but we must be aware that they will be taken, and not fitting a model at all could be worse than fitting it using an approximate computation (where approximate can be defined as not giving exact summaries of the posterior distribution even in the limit of infinite compute time). Our aim in describing statistical workflow is thus also to explicitly understand various shortcuts as approximations to the full workflow, letting practitioners to make more informed choices about where to invest their limited time and energy.

\subsection{Aim and structure of this article}

There is all sorts of tacit knowledge in applied statistics that does not always make it into published papers and textbooks.  The present article is intended to put some of these ideas out in the open, both to improve applied Bayesian analyses and to suggest directions for future development of theory, methods, and software.

Our target audience is (a) practitioners of applied Bayesian statistics, especially users of probabilistic programming languages such as Stan, and (b) developers of methods and software intended for these users.  We are also targeting researchers of Bayesian theory and methods, as we believe that many of these aspects of workflow have been under-studied.

In the rest of the paper we go more slowly through individual aspects of Bayesian workflow as outlined in Figure~\ref{flowchart}, starting with steps to be done before a model is fit (Section \ref{before_fitting}), through fitting, debugging and evaluating models (Sections \ref{fitting}--\ref{evaluating}), and then modifying models (Section \ref{building}) and understanding and comparing a series of models (Section \ref{comparing}).

Sections \ref{golf} and \ref{sec:orbits} then go through these steps in two examples, one in which we add features step by step to a model of golf putting, and one in which we go through a series of investigations to resolve difficulties in fitting a simple model of planetary motion.  The first of these examples shows how new data can motivate model improvements, and also illustrates some of the unexpected challenges that arise when expanding a model.  The second example demonstrates the way in which challenges in computation can point to modeling difficulties.  These two small examples do not illustrate all the aspects of Bayesian workflow, but they should at least suggest that there could be a benefit to systematizing the many aspects of Bayesian model development.
We conclude in Section \ref{discussion} with some general discussion and our responses to potential criticism of the workflow.

\section{Before fitting a model}\label{before_fitting}

\subsection{Choosing an initial model}\label{sec:initial_model}

The starting point of almost all analyses is to adapt what has been done before, using a model from a textbook or case study or published paper that has been applied to a similar problem (a strongly related concept in software engineering is software design pattern). Using a model taken from some previous analysis and altering it can be seen as a shortcut to effective data analysis, and by looking at the results from the model template we know in which direction of the model space there are likely to be useful elaborations or simplifications. 
Templates can save time in model building and computing, and we should also take into account the cognitive load for the person who needs to understand the results. Shortcuts  are important for humans as well as computers, and shortcuts help explain why the typical workflow is iterative (see more in Section \ref{sec:justification_of_iterative}).  Similarly, if we were to try to program a computer to perform data analysis automatically, it would have to work through some algorithm to construct models, and the building blocks of such an algorithm would represent templates of a sort.  Despite the negative connotations of ``cookbook analysis,'' we think templates can be useful as starting points and comparison points to more elaborate analyses.  Conversely, we should recognize that theories are not static, and the process of development of scientific theories is not the same as that of statistical models (Navarro, 2020). 

Sometimes our workflow starts with a simple model with the aim to add features later (modeling varying parameters, including measurement errors, correlations, and so forth). Other times we start with a big model and aim to strip it down in next steps, trying to find something that is simple and understandable that still captures key features of the data. Sometimes we even consider multiple completely different approaches to modeling the same data and thus have multiple starting points to choose from.

\subsection{Modular construction}
\label{sec:modular}

A Bayesian model is built from modules which can often be viewed as placeholders to be replaced as necessary.  For example, we model data with a normal distribution and then replace this with a longer-tailed or mixture distribution; we model a latent regression function as linear and replace it with nonlinear splines or Gaussian processes; we can treat a set of observations as exact and then add a measurement-error model; we can start with a weak prior and then make it stronger when we find the posterior inference includes unrealistic parameter values.  Thinking of components as placeholders takes some of the pressure off the model-building process, because you can always go back and generalize or add information as necessary. 

The idea of modular construction goes against a long-term tradition in the statistical literature where whole models were given names and a new name was given every time a slight change to an existing model was proposed. Naming model modules rather than whole models makes it easier to see connections between seemingly different models and adapt them to the specific requirements of the given analysis project. 

\subsection{Scaling and transforming the parameters}\label{scaling}

We like our parameters to be interpretable for both practical and ethical reasons. This leads to wanting them on natural scales and modeling them as independent, if possible, or with an interpretable dependence structure, as this facilitates the use of informative priors (Gelman, 2004).  It can also help to separate out the scale so that the unknown parameters are scale-free.  For example, in a problem in pharmacology (Weber et al., 2018)  we had a parameter that we expected would take on values of approximately $50$ on the scale of measurement; following the principle of scaling we might set up a model on $\log(\theta/50)$, so that $0$ corresponds to an interpretable value ($50$ on the original scale) and a difference of $0.1$, for example, on the log scale corresponds to increasing or decreasing by approximately $10\%$.  This sort of transformation is not just for ease of interpretation; it also sets up the parameters in a way that readies them for effective hierarchical modeling.  As we build larger models, for example by incorporating data from additional groups of patients or additional drugs, it will make sense to allow parameters to vary by group (as we discuss in Section \ref{generalization}), and partial pooling can be more effective on scale-free parameters.  For example, a model in toxicology required the volume of the liver for each person in the study. Rather than fitting a hierarchical model to these volumes, we expressed each as the volume of the person multiplied by the proportion of volume that the liver; we would expect these scale-free factors to vary less across people and so the fitted model can do more partial pooling compared to the result from modeling absolute volumes. The scaling transformation is a decomposition that facilitates effective hierarchical modeling.

In many cases we can put parameters roughly on unit scale by using logarithmic or logit transformations or by standardizing, subtracting a center and dividing by a scale.
If the center and scale are themselves computed from the data, as we do for default priors in regression coefficients in rstanarm (Gabry et al., 2020a), we can consider this as an approximation to a hierarchical model in which the center and scale are hyperparameters that are estimated from the data. 

More complicated transformations can also serve the purpose of making parameters more interpretable and thus facilitating the use of prior information; Riebler et al.\ (2016) give an example for a class of spatial correlation models, and Simpson et al.\ (2017) consider this idea more generally.

\subsection{Prior predictive checking}
\label{sec:prior_pred_checks}

\begin{figure}
\centerline{
   \includegraphics[width=.8\textwidth]{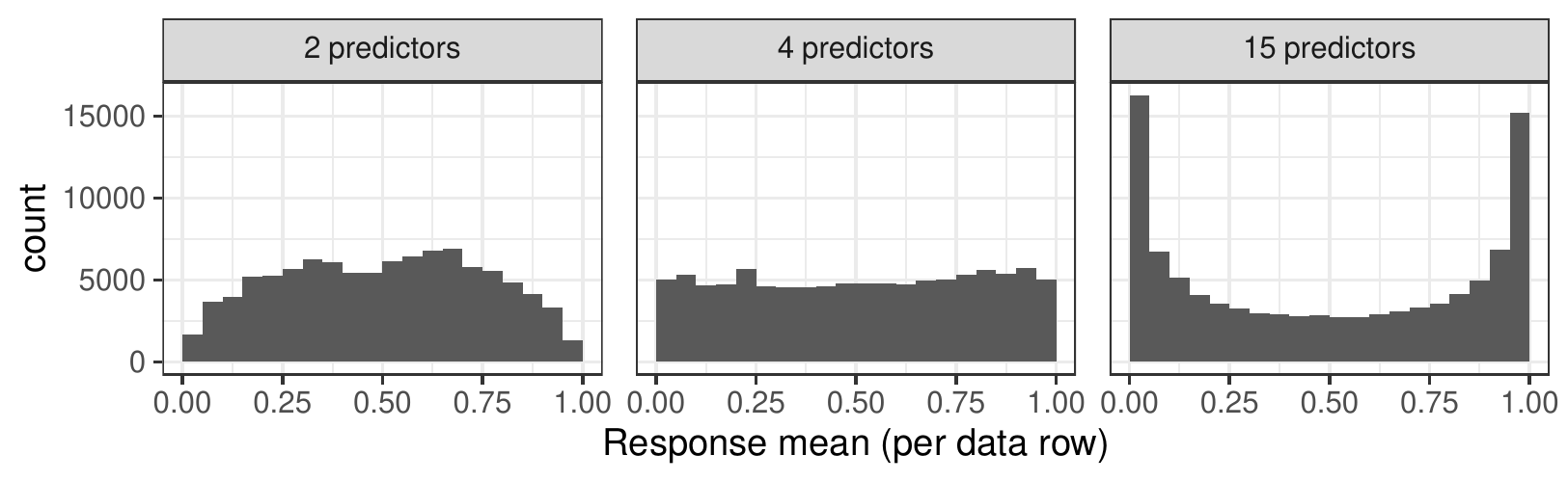}
 }
 \vspace{-.1in}
\caption{\em Demonstration of the use of prior predictive checking to gain understanding of non-obvious features of a model.  The above graphs correspond to logistic regression models with $100$ data points and $2$, $4$, or $15$ binary covariates.  In each case, the regression coefficients were given independent $\mbox{normal}(0,1)$ prior distributions.  For each model, we performed a prior predictive check, $1000$ times simulating the vector of coefficients $\theta$ from the prior, then simulating a dataset $y$ from the logistic model, and then summarizing this by the mean of the simulated data, $\bar{y}$.  Each plot shows the prior predictive distribution of this summary statistic, that is, the $1000$ simulations of $\bar{y}$.  When the number of covariates in the model is small, this prior predictive distribution is spread out, indicating that the model is compatible with a wide range of regimes of data.  But as the number of covariates increases, the posterior predictive distribution becomes concentrated near $\bar{y}=0$ or $1$, indicating that weak priors on the individual coefficients of the model imply a strong prior on this particular predictive quantity.  If we wanted a more moderate prior predictive distribution on $\bar{y}$, the prior on the coefficients would need to be strongly concentrated near zero.}
\label{logistic_sim}
\end{figure}

Prior predictive checks are a useful tool to understand the implications of a prior distribution in the context of a generative model (Box, 1980, Gabry et al., 2019; see also Section~\ref{priors} for details on how to work with prior distributions). In particular, because prior predictive checks make use of simulations from the model rather than observed data, they provide a way to refine the model without using the data multiple times. Figure \ref{logistic_sim} shows a simple prior predictive check for a logistic regression model. The simulation shows that even independent priors on the individual coefficients have different implications as the number of covariates in the model increases. This is a general phenomenon in regression models where as the number of predictors increases, we need stronger priors on model coefficients (or enough data) if we want to push the model away from extreme predictions. 

A useful approach is to consider priors on outcomes and then derive a corresponding joint prior on parameters (see, e.g., Piironen and Vehtari, 2017, and Zhang et al., 2020). More generally, joint priors allow us to control the overall complexity of larger parameter sets, which helps generate more sensible prior predictions that would be hard or impossible to achieve with independent priors.

\begin{figure}
\centerline{
   \includegraphics[width=.8\textwidth]{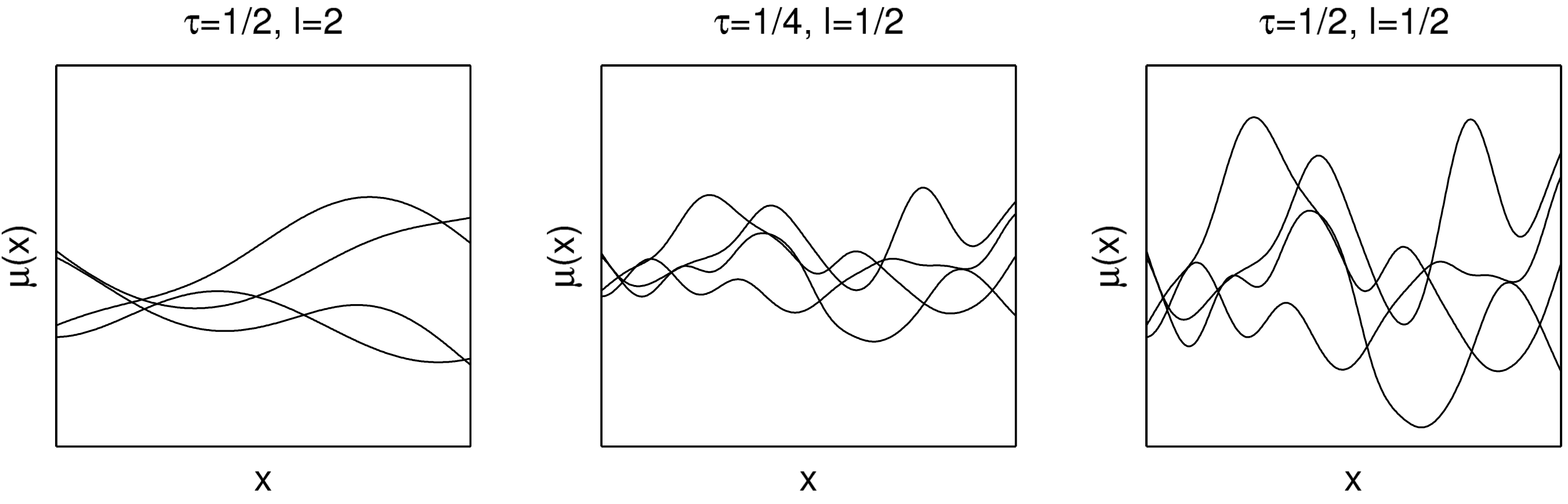}
 }
 \vspace{-.1in}
\caption{\em Prior predictive draws from the Gaussian process model with squared exponential covariance function and different values of the amplitude parameter $\tau$ and the length scale parameter $l$. From Gelman et al.\ (2013).}
\label{gp_sim}
\end{figure}

Figure \ref{gp_sim} shows an example of prior predictive checking for three choices of prior distribution for a Gaussian process model (Rasmussen and Willams, 2006).  This sort of simulation and graphical comparison is useful when working with any model and essential when setting up unfamiliar or complicated models. 

Another benefit of prior predictive simulations is that they can be used to elicit expert prior knowledge on the measurable quantities of interest, which is often easier than soliciting expert opinion on model parameters that are not observable (O'Hagan et al., 2006). 

Finally, even when we skip computational prior predictive checking, it might be useful to think about how the priors we have chosen would affect a hypothetical simulated dataset.

\subsection{Generative and partially generative models}

Fully Bayesian data analysis requires a generative model---that is, a joint probability distribution for all the data and parameters.  The point is subtle:  Bayesian {\em inference} does not actually require the generative model; all it needs from the data is the likelihood, and different generative models can have the same likelihood. But Bayesian {\em data analysis} requires the generative model to be able to perform predictive simulation and model checking (Sections \ref{sec:prior_pred_checks}, \ref{fake}, \ref{sec:sbc}, \ref{sec:post_pred_checks}, and \ref{crossvalidation}), and Bayesian {\em workflow} will consider a series of generative models.

For a simple example, suppose we have data $y\sim\mbox{binomial}(n,\theta)$, where $n$ and $y$ are observed and we wish to make inference about $\theta$.  For the purpose of Bayesian inference it is irrelevant if the data were sampled with fixed $n$ (binomial sampling) or sampled until a specified number of successes occurred (negative binomial sampling):  the two likelihoods are equivalent for the purpose of estimating $\theta$ because they differ only by a multiplicative factor that depends on $y$ and $n$ but not $\theta$.  However, if we want to simulate new data from the predictive model, the two models are different, as the binomial model yields replications with a fixed value of $n$ and the negative binomial model yields replications with a fixed value of $y$.  Prior and posterior predictive checks (Sections \ref{sec:prior_pred_checks} and \ref{sec:post_pred_checks}) will look different under these two different generative models.

This is not to say that the Bayesian approach is necessarily better;  the assumptions of a generative model can increase inferential efficiency but can also go wrong, and this motivates much of our workflow.

It is common in Bayesian analysis to use models that are not \emph{fully} generative. For example, in regression we will typically model an outcome $y$ given predictors $x$ without a generative model for $x$.  Another example is survival data with censoring, where the censoring process is not usually modeled.  When performing predictive checks for such models, we either need to condition on the observed predictors or else extend the model to allow new values of the predictors to be sampled. It is also possible that there is no stochastic generative process for some parts of the model, for example if $x$ has been chosen by a deterministic design of experiment.

Thinking in terms of generative models can help illuminate the limitations of what can be learned from the observations. For example, we might want to model a temporal process with a complicated autocorrelation structure, but if our actual data are spaced far apart in time, we might not be able to distinguish this model from a simpler process with nearly independent errors.

In addition, Bayesian models that use improper priors are not fully generative, in the sense that they do not have a joint distribution for data and parameters and it would not be possible to sample from the prior predictive distribution.  When we do use improper priors, we think of them as being placeholders or steps along the road to a full Bayesian model with a proper joint distribution over parameters and data.

In applied work, complexity often arises from incorporating different sources of data.  For example, we fit a Bayesian model for the 2020 presidential election using state and national polls, partially pooling toward a forecast based on political and economic ``fundamentals'' (Morris, Gelman, and Heidemanns, 2020).  The model includes a stochastic process for latent time trends in state and national opinion.  Fitting the model using Stan yields posterior simulations which are used to compute probabilities for election outcomes. The Bayesian model-based approach is superficially similar to poll aggregations such as described by Katz (2016), which also summarize uncertainty by random simulations.  The difference is that our model could be run forward to generate polling data; it is not just a data analysis procedure but also provides a probabilistic model for public opinion at the national and state levels.

Thinking more generally, we can consider a progression from least to most generative models.  At one extreme are
completely non-generative methods which are defined simply as data summaries, with no model for the data at all.  Next come classical statistical models, characterized by probability distributions $p(y;\theta)$ for data $y$ given parameters $\theta$, but with no probability distribution for $\theta$.  At the next step are the Bayesian models we usually fit, which are generative on $y$ and $\theta$ but include additional unmodeled data $x$ such as sample sizes, design settings, and hyperparameters; we write such models as $p(y,\theta|x)$.  The final step would be a completely generative model $p(y, \theta, x)$ with no ``left out'' data, $x$.

In statistical workflow we can move up and down this ladder, for example starting with an unmodeled data-reduction algorithm and then formulating it as a probability model, or starting with the inference from a probability model, considering it as a data-based estimate, and tweaking it in some way to improve performance.  In Bayesian workflow we can move data in and out of the model, for example taking an unmodeled predictor $x$ and allowing it to have measurement error, so that the model then includes a new level of latent data (Clayton, 1992, Richardson and Gilks, 1993).

\section{Fitting a model}\label{fitting}

Traditionally, Bayesian computation has been performed using a combination of analytic calculation and normal approximation.  Then in the 1990s, it became possible to perform Bayesian inference for a wide range of models using Gibbs and Metropolis algorithms (Robert and Casella, 2011).  The current state of the art algorithms for fitting open-ended Bayesian models include variational inference (Blei and Kucukelbir, 2017), sequential Monte Carlo (Smith, 2013), and Hamiltonian Monte Carlo (HMC; Neal, 2011, Betancourt, 2017a).  Variational inference is a generalization of the expectation-maximization (EM) algorithm and can, in the Bayesian context, be considered as providing a fast but possibly inaccurate approximation to the posterior distribution. Variational inference is the current standard for computationally intensive models such as deep neural networks. Sequential Monte Carlo is a generalization of the Metropolis algorithm that can be applied to any Bayesian computation, and HMC is a different generalization of Metropolis that uses gradient computation to move efficiently through continuous probability spaces.

In the present article we focus on fitting Bayesian models using HMC and its variants, as implemented in Stan and other probabilistic programming languages. While similar principles should apply also to other software and other algorithms, there will be differences in the details.

To safely use an inference algorithm in Bayesian workflow, it is vital that the algorithm provides strong diagnostics to determine when the computation is unreliable. In the present paper we discuss such diagnostics for HMC.

\subsection{Initial values, adaptation, and warmup}

Except in the simplest case, Markov chain simulation algorithms operate in multiple stages.  First there is a warmup phase which is intended to move the simulations from their possibly unrepresentative initial values to something closer to the region of parameter space where the log posterior density is close to its expected value, which is related to the concept of ``typical set'' in information theory (Carpenter, 2017).  
Initial values are not supposed to matter in the asymptotic limit, but they can matter in practice, and a wrong choice can threaten the validity of the results.

Also during warmup there needs to be some procedure to set the algorithm's tuning parameters; this can be done using information gathered from the warmup runs.  Third is the sampling phase, which ideally is run until multiple chains have mixed (Vehtari et al., 2020).

When fitting a model that has been correctly specified, warmup thus has two purposes: (a) to run through a transient phase to reduce the bias due to dependence on the initial values, and (b) to provide information about the target distribution to use in setting tuning parameters.
In model exploration, warmup has a third role, which is to quickly flag computationally problematic models.

\subsection{How long to run an iterative algorithm}

We similarly would like to consider decisions in the operation of iterative algorithms in the context of the larger workflow.  Recommended standard practice is to run at least until $\widehat{R}$, the measure of mixing of chains, is less than 1.01 for all parameters and quantities of interest (Vehtari et al., 2020), and to also monitor the multivariate mixing statistic $R^*$ (Lambert and Vehtari, 2020).  There are times when earlier stopping can make sense in the early stages of modeling.
For example, it might seem like a safe and conservative choice to run MCMC until the effective sample size is in the thousands or Monte Carlo standard error is tiny in comparison to the required precision for parameter interpretation---but if this takes a long time, it limits the number of models that can be fit in the exploration stage. More often than not, our model also has some big issues (especially coding errors) that become apparent after running only a few iterations, so that the remaining computation is wasted. In this respect, running many iterations for a newly-written model is similar to premature optimization in software engineering.  For the final model, the required number of iterations depends on the desired Monte Carlo accuracy for the quantities of interest.

Another choice in computation is how to best make use of available parallelism, beyond the default of running 4 or 8 separate chains on multiple cores. Instead of increasing the number of iterations, effective  variance reduction can also be obtained by increasing the number of parallel chains (see, e.g., Hoffman and Ma, 2020).

\subsection{Approximate algorithms and approximate models}
\label{sec:approximate}

Bayesian inference typically involves intractable integrals, hence the need for approximations.  Markov chain simulation is a form of approximation where the theoretical error  approaches zero as the number of simulations increases.  If our chains have mixed, we can make a good estimate of the Monte Carlo standard error (Vehtari et al., 2020), and for practical purposes we  often treat these computations as exact.

\begin{figure}
  \centerline{\includegraphics[width=.6\textwidth]{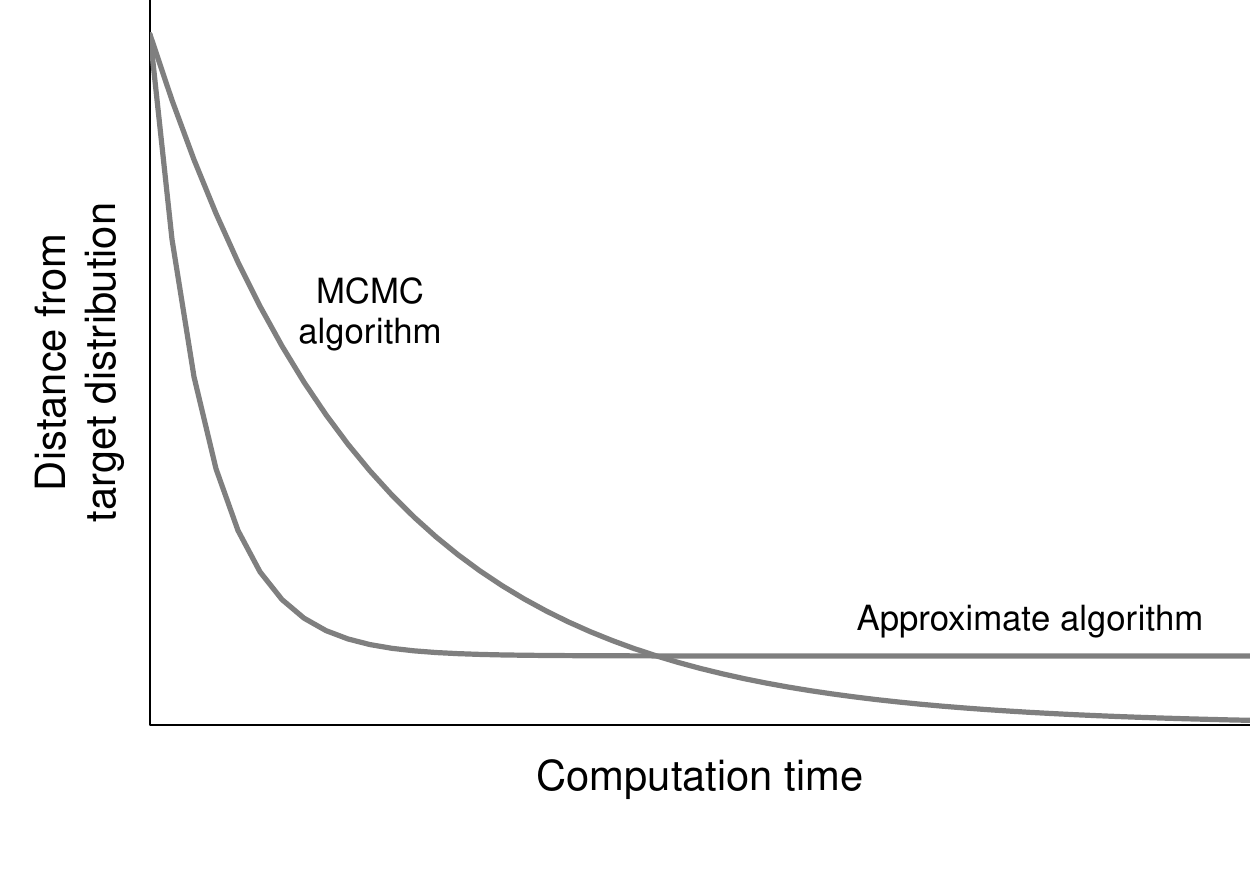}}
  \vspace{-.2in}
\caption{\em Idealized sketch of the tradeoff between approximate algorithms and MCMC in Bayesian computation.  If the goal is to get the best fit to the target distribution, MCMC should ultimately win out.  But if we are currently fitting just one in a series of models, it can make sense to use approximate algorithms so as to be able to move through model space more rapidly.  Which of these algorithms performs better depends on the time budget of the user and where the two curves intersect.}
\label{twocurves}
\end{figure}

Unfortunately, running MCMC to convergence is not always a scalable solution as data and models get large, hence the desire for faster approximations.  Figure \ref{twocurves} shows the resulting tradeoff between speed and accuracy.  This graph is only conceptual; in a real problem, the positions of these lines would be unknown, and indeed in some problems an approximate algorithm can perform worse than MCMC even at short time scales. 

Depending on where we are in the workflow, we have different requirements of our computed posteriors. Near the end of the workflow, where we are examining fine-scale and delicate features, we require accurate exploration of the posterior distribution. This usually requires MCMC. On the other hand, at the beginning of the workflow, we can frequently make our modeling decisions based on large-scale features of the posterior that can be accurately estimated using relatively simple methods such as empirical Bayes, linearization or Laplace approximation, nested approximations like INLA (Rue et al., 2009), or even sometimes data-splitting methods like expectation propagation (Vehtari, Gelman, Siivola, et al., 2020), mode-finding approximations like variational inference (Kucukelbir et al., 2017), or penalized maximum likelihood. The point is to use a suitable tool for the job and to not try to knock down a retaining wall using a sculptor's chisel. 

All of these approximate methods have at least a decade of practical experience, theory, and diagnostics behind them. There is no one-size-fits-all approximate inference algorithm, but when a workflow includes relatively well-understood components such as generalized linear models, multilevel regression, autoregressive time series models, or Gaussian processes, it is often possible to construct an appropriate approximate algorithm. Furthermore, depending on the specific approximation being used, generic diagnostic tools described by Yao et al.\ (2018a) and Talts et al.\ (2020) can be used to verify that a particular approximate algorithm reproduces the features of the posterior that you care about for a specific model.

An alternative view is to understand an approximate algorithm as an exact algorithm for an approximate model. In this sense, a workflow is a sequence of steps in an abstract computational scheme aiming to infer some ultimate, unstated model.  More usefully, we can think of things like empirical Bayes approximations as replacing a model's prior distributions with a particular data-dependent point-mass prior. Similarly a Laplace approximation can be viewed as a data-dependent linearization of the desired model, while a nested Laplace approximation (Rue et al., 2009, Margossian et al., 2020a) uses a linearized conditional posterior in place of the posited conditional posterior.

\subsection{Fit fast, fail fast}
\label{sec:fast}

An important intermediate goal  is to be able to {\em fail fast} when fitting bad models. This can be considered as a shortcut that avoids spending a lot of time for (near) perfect inference for a bad model. There is a large literature on approximate algorithms to fit the desired model fast, but little on algorithms designed to waste as little time as possible on the models that we will ultimately abandon.  We believe it is important to evaluate methods on this criterion, especially because inappropriate and poorly fitting models can often be more difficult to fit.

\begin{figure}
\centerline{\includegraphics[width=.4\textwidth]{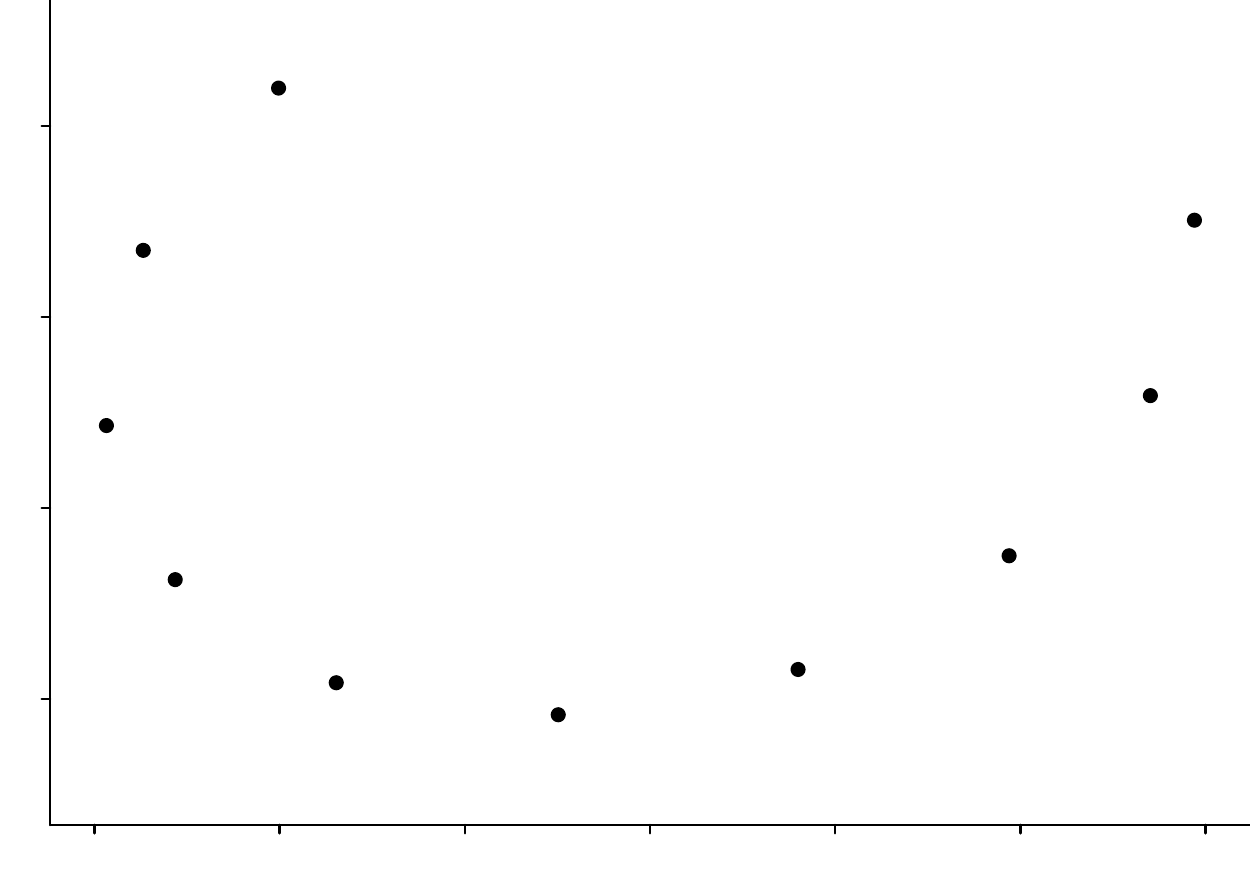}\hspace{.1\textwidth}\includegraphics[width=.4\textwidth]{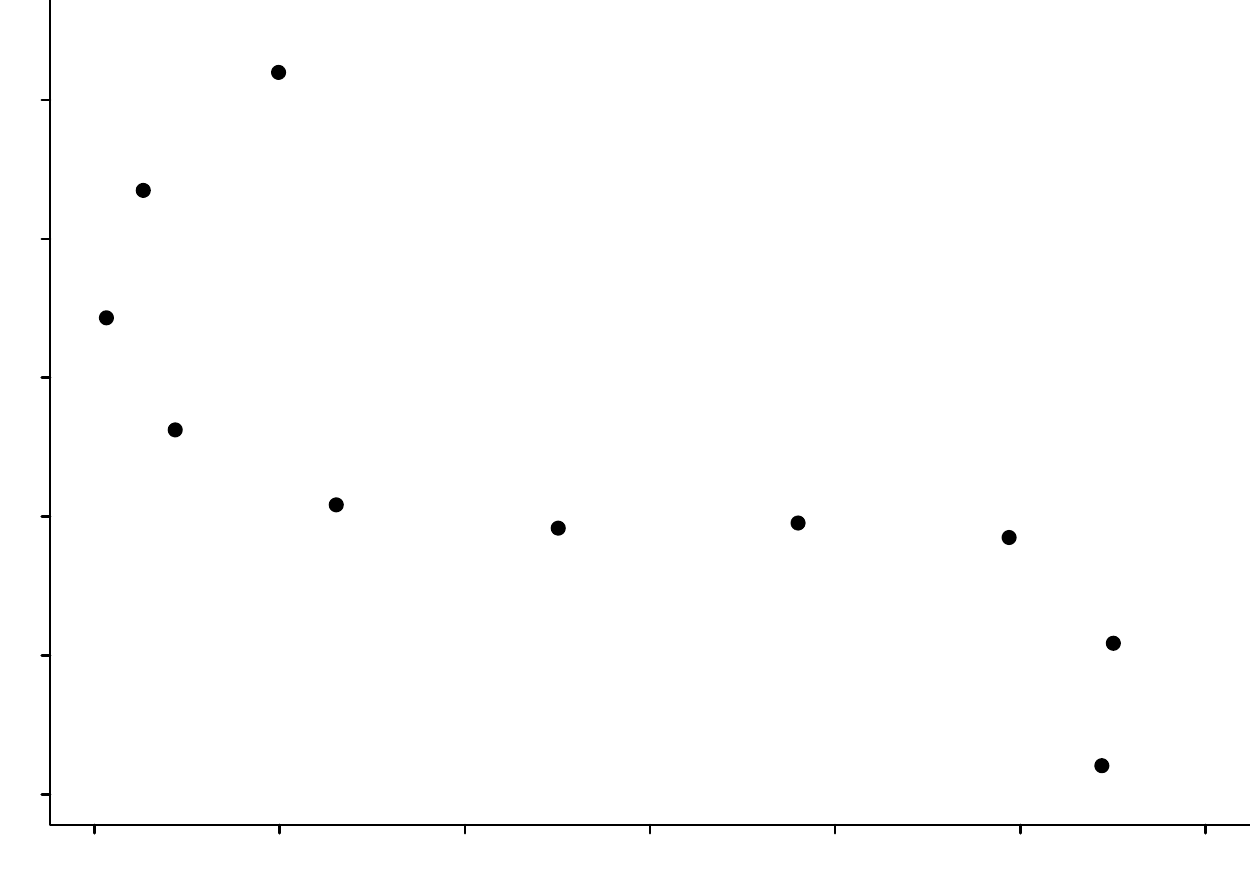}}
\caption{\em Illustration of the need for ``fit fast, fail fast'':  (a) Idealized data representing measurements of planetary orbit which could be fit as an ellipse with measurement error, (b) Measurements of a hypothetical orbit that was perturbed by a Death Star.  In the second example, it would be challenging to fit a single ellipse to the data---but we have no particular interest in an omnibus elliptical fit in any case.  We would like any attempt to fit an ellipse to the second dataset to fail fast.}
\label{orbits}
\end{figure}

For a simple idealized example, suppose you are an astronomer several centuries ago fitting ellipses to a planetary orbit based on 10 data points measured with error.  Figure \ref{orbits}a shows the sort of data that might arise, and just about any algorithm will fit reasonably well.  For example, you could take various sets of five points and fit the exact ellipse to each, and then take the average of these fits.  Or you could fit an ellipse to the first five points, then perturb it slightly to fit the sixth point, then perturb that slightly to fit the seventh, and so forth.  Or you could implement some sort of least squares algorithm.

Now suppose some Death Star comes along and alters the orbit---in this case, we are purposely choosing an unrealistic example to create a gross discrepancy between model and data---so that your 10 data points look like Figure \ref{orbits}b.  In this case, convergence will be much harder to attain.  If you start with the ellipse fit to the first five points, it will be difficult to take any set of small perturbations that will allow the curve to fit the later points in the series.  But, more than that, even if you could obtain a least squares solution, any ellipse would be a terrible fit to the data.  It's just an inappropriate model.  If you fit an ellipse to these data, you should want the fit to fail fast so you can quickly move on to something more reasonable.

This example has, in extreme form, a common pattern of difficult statistical computations, that fitting to different subsets of the data yields much different parameter estimates.

\section{Using constructed data to find and understand problems}\label{computation}

The first step in validating computation is to check that the model actually finishes the fitting process in an acceptable time frame and the convergence diagnostics are reasonable.  In the context of HMC, this is primarily the absence of divergent transitions, $\widehat{R}$ diagnostic near 1, and sufficient effective sample sizes for the central tendency, the tail quantiles, and the energy (Vehtari et al., 2020). However, those diagnostics cannot protect against a probabilistic program that is computed correctly but encodes a different model than the user intended. 

The main tool we have for ensuring that the statistical computation is done reasonably well is to actually fit the model to some data and check that the fit is good. Real data can be awkward for this purpose because modeling issues can collide with computational issues and make it impossible to tell if the problem is the computation or the model itself. To get around this challenge, we first explore models by fitting them to simulated data.

\subsection{Fake-data simulation}\label{fake}

Working in a controlled setting where the true parameters are known can help us understand our data model and priors, what can be learned from an experiment, and the validity of the applied inference methods.
The basic idea is to check whether our procedure recovers the correct parameter values when fitting fake data.
Typically we choose parameter values that seem reasonable a priori and then simulate a fake dataset of the same size, shape, and structure as the original data. We next fit the model to the fake data to check several things.

The first thing we check isn't strictly computational, but rather an aspect of the design of the data. For all parameters, we check to see if 
the observed data are able  to provide additional information beyond the prior. The procedure is to  simulate some fake data from the model with
fixed, known parameters and then see whether our method comes close to reproducing the known truth. We can look at point estimates and also the coverage of posterior intervals.

If our fake-data check fails, in the sense that the inferences are not close to the assumed parameter values or if there seem to be model components that are not gaining any information from the data (Lindley, 1956, Goel and DeGroot, 1981), we recommend breaking down the model. Go simpler and simpler until we get the model to work. Then, from there, we can try to identify the problem, as illustrated in Section \ref{sec:simple_and_complex}.

The second thing that we check is if the true parameters  can be recovered to roughly within the uncertainty  implied by the fitted posterior distribution.  This will not be possible if the data are not informative for a parameter, but it should typically happen otherwise. It is not possible to run a single fake data simulation, compute the associated posterior distribution, and declare that everything works well. We will see in the next section that a more elaborate setup is needed. However, a single simulation run will often reveal blatant errors. For instance, if the code has an error in it and fits the wrong model this will often be clear from a catastrophic failure of parameter recovery. 

The third thing that we can do is use fake data simulations to understand how the behavior of a model can change across different parts of the parameter space. 
In this sense, a statistical model can contain many stories of how the data get generated.
As illustrated in Section \ref{adding_prior_info}, the data are informative about the parameters for a sum of 
declining exponentials when the exponents are well separated, but not so informative 
when the two components are close to each other. This sort of instability contingent on parameter values
is also a common phenomenon in differential equation models, as can been seen in Section~\ref{sec:orbits}.  
For another example, the posterior distribution of a hierarchical model looks much different at the neck than at the mouth of the funnel geometry implied by the hierarchical prior.   Similar issues arise in Gaussian process models, depending on the length scale of the process and the resolution of the data.

All this implies that fake data simulation can be particularly relevant in the zone of the parameter space that is predictive of the data.
This in turn suggests a two-step procedure in which we first fit the model to real data, then draw parameters from the resulting posterior distribution to use in fake-data checking.  The statistical properties of such a procedure are unclear but in practice we have found such checks to be helpful, both for revealing problems 
with the computation or model, and for providing some reassurance when the fake-data-based inferences do 
reproduce the assumed parameter value.

To carry this idea further, we may break our method by coming up with fake data that cause our procedure to give bad answers. This sort of simulation-and-exploration can be the first step in a deeper understanding of an inference method, which can be valuable even for a practitioner who plans to use this method for just one applied problem. It can also be useful for building up a set of 
more complex models to possibly explore later.

Fake-data simulation is a crucial component of our workflow because it is, arguably, the only point where we can directly check that our inference on latent variables is reliable.
When fitting the model to real data, we do not by definition observe the latent variables. 
Hence we can only evaluate how our model fits the observed data.
If our goal is not merely prediction but estimating the latent variables, examining predictions only helps us so much.
This is especially true of overparameterized models, where wildly different parameter values can yield comparable predictions (e.g. Section~\ref{adding_prior_info} and Gelman et al, 1996).
In general, even when the model fit is good, we should only draw conclusions about the estimated latent variables with caution.
Fitting the model to simulated data 
helps us better understand what the model can and cannot learn about the latent process in a controlled setting where we know the ground truth. If a model is able to make good inference on fake-data generated from that very model, this provides no guarantee that its inference on real data will be sensible. But if a model is unable to make good inference on such fake data, then it's hopeless to expect the model to provide reasonable inference on real data. Fake-data simulations provide an upper bound of what can be learned about a latent process.

\subsection{Simulation-based calibration}\label{sec:sbc}

\begin{figure}
    \centering
    \includegraphics[width = 6in]{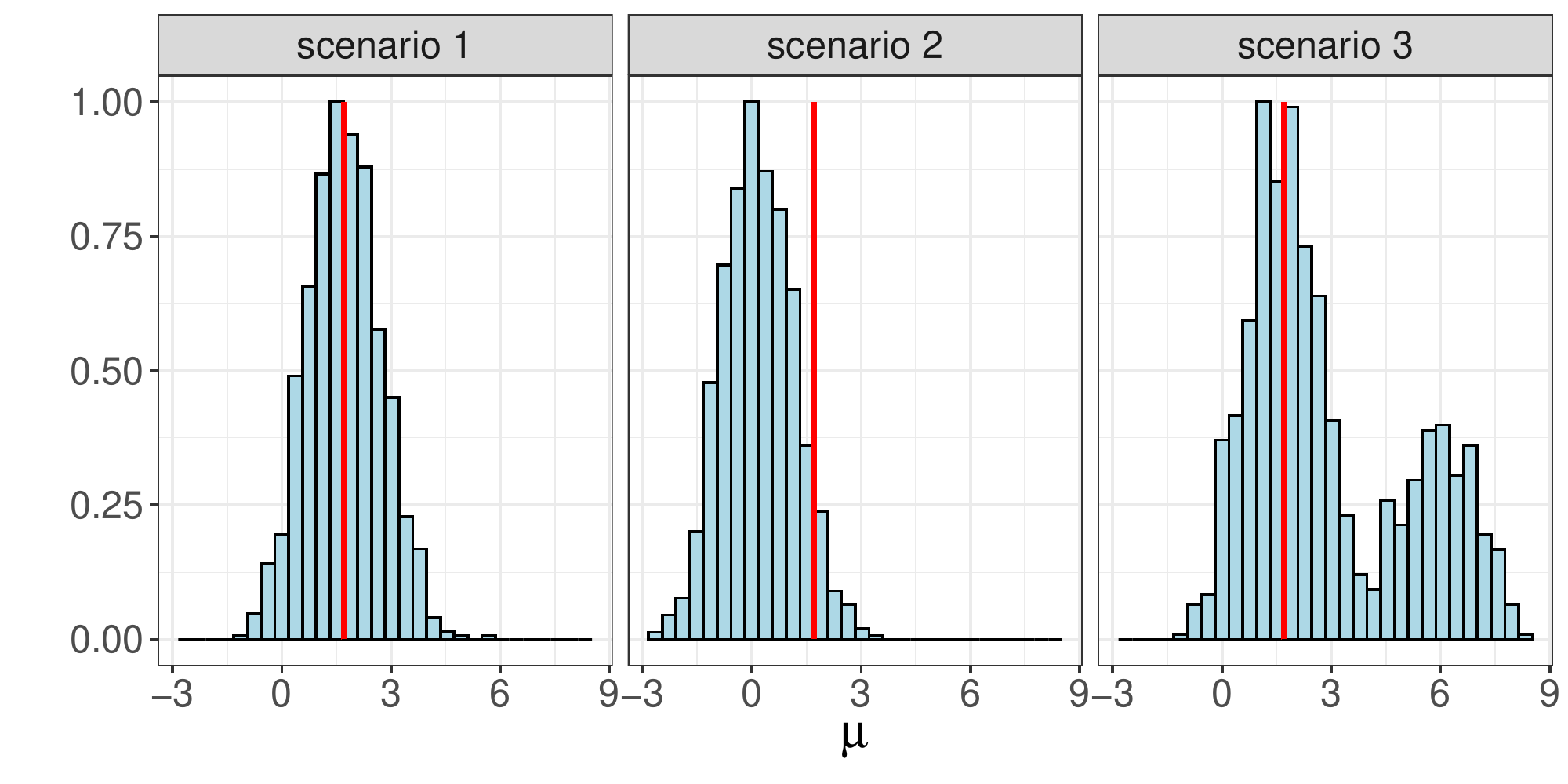}
    \vspace{-.1in}
    \caption{\em Comparison of a posterior distribution to the assumed true parameter value. When fitting the model to simulated data, we can examine whether the posterior sample (blue histogram) comports with the true parameter value (red line). In scenario 1, the posterior is centered at the true value, which suggests the fit is reasonable. In scenario 2, the true parameter is now in the tail of the posterior distribution. It is unclear whether this indicates a fault in our inference. In scenario 3, the posterior is multimodal and it becomes evident that comparing the posterior to a single point cannot validate the inference algorithm. A more comprehensive approach, such as simulation-based calibration, can teach us more.}
    \label{fig:sbc_point}
\end{figure}

There is a formal, and at times practical, issue when comparing the result of Bayesian inference, a posterior distribution, to a single (true) point, as illustrated in Figure~\ref{fig:sbc_point}.

Using a single fake data simulation to test a model will not necessarily ``work,'' even if the computational algorithm is working correctly.  The problem here arises not just because with one simulation anything can happen (there is a 5\% chance that a random draw will be outside a 95\% uncertainty interval) but also because Bayesian inference will in general only be calibrated when averaging over the prior, not for any single parameter value. Furthermore, parameter recovery may fail not because the algorithm fails, but because the observed data are not providing information that could update the uncertainty quantified by the prior for a particular parameter. If the prior and posterior are approximately unimodal and the chosen parameter value is from the center of the prior, we can expect overcoverage of posterior intervals.

A more comprehensive approach than what we present in Section~\ref{fake} is {\em simulation-based calibration} (SBC; Cook et al., 2006, Talts et al., 2020).
In this scheme, the model parameters are drawn from the prior; then data are simulated conditional on these parameter values; then the model is fit to data; and finally the obtained posterior is compared to the simulated parameter values that were used to generate the data.
By repeating this procedure several times,
it is possible to check the coherence of the inference algorithm.
The idea is that by performing Bayesian inference across a range of datasets simulated using parameters drawn from the prior, we should recover the prior.
Simulation-based calibration is useful to evaluate how closely approximate algorithms match the theoretical posterior even in cases when the posterior is not tractable.

While in many ways superior to benchmarking against a truth point, simulation-based calibration requires fitting the model multiple times, which incurs a substantial computational cost, especially if we do not use extensive parallelization. 
In our view, simulation-based calibration and truth-point benchmarking are two ends of a spectrum.
Roughly, a single truth-point benchmark will possibly flag gross problems, but it does not guarantee anything.
As we do more experiments, it is possible to see finer and finer problems in the computation. It is an open 
research question to understand SBC with a small number of draws. We expect that abandoning random draws for 
a more designed exploration of the prior would make the method more efficient, especially in models with 
a relatively small number of parameters.

A serious problem with SBC is that it clashes somewhat with most modelers' tendency to specify their priors 
wider than they believe necessary. The slightly conservative nature of weakly informative priors can cause the 
data sets simulated during SBC to occasionally be extreme. Gabry et al.\ (2019) give an example in which fake
air pollution datasets were simulated where the pollution is denser than a black hole. These extreme data sets
can cause an algorithm that works well on realistic data to fail dramatically. But this isn't really a problem
with the computation so much as a problem with the prior. 

One possible way around this is to ensure that the priors are very tight. However, a pragmatic idea
is to keep the priors and compute reasonable parameter values using the real data. This can be done 
either through rough estimates or by computing the actual posterior. We then suggest widening out the 
estimates slightly and using these as a prior for the SBC. This will ensure that all of the 
simulated data will be as realistic as the model allows.


\subsection{Experimentation using constructed data}

A good way to understand a model is to fit it to data simulated from different scenarios.

For a simple case, suppose we are interested in the statistical properties of linear regression fit to data from alternative distributional forms.  To start, we could specify data $x_i,\ i=1,\dots,n$, draw coefficients $a$ and $b$ and a residual standard deviation $\sigma$ from our prior distribution, simulate data from $y_i\sim\mbox{normal}(a + bx_i, \sigma)$, and fit the model to the simulated data.  Repeat this 1000 times and we can check the coverage of interval estimates:  that's a version of simulation-based calibration.  We could then fit the same model but simulating data using different assumptions, for example drawing independent data points $y_i$ from the $t_4$ distribution rather than the normal.  This will then fail simulation-based calibration---the wrong model is being fit---but the interesting question here is, how bad will these inferences be?  One could, for example, use SBC simulations to examine coverage of posterior 50\% and 95\% intervals for the coefficients.

\begin{figure}
    \centering
    \includegraphics[width = .4\textwidth]{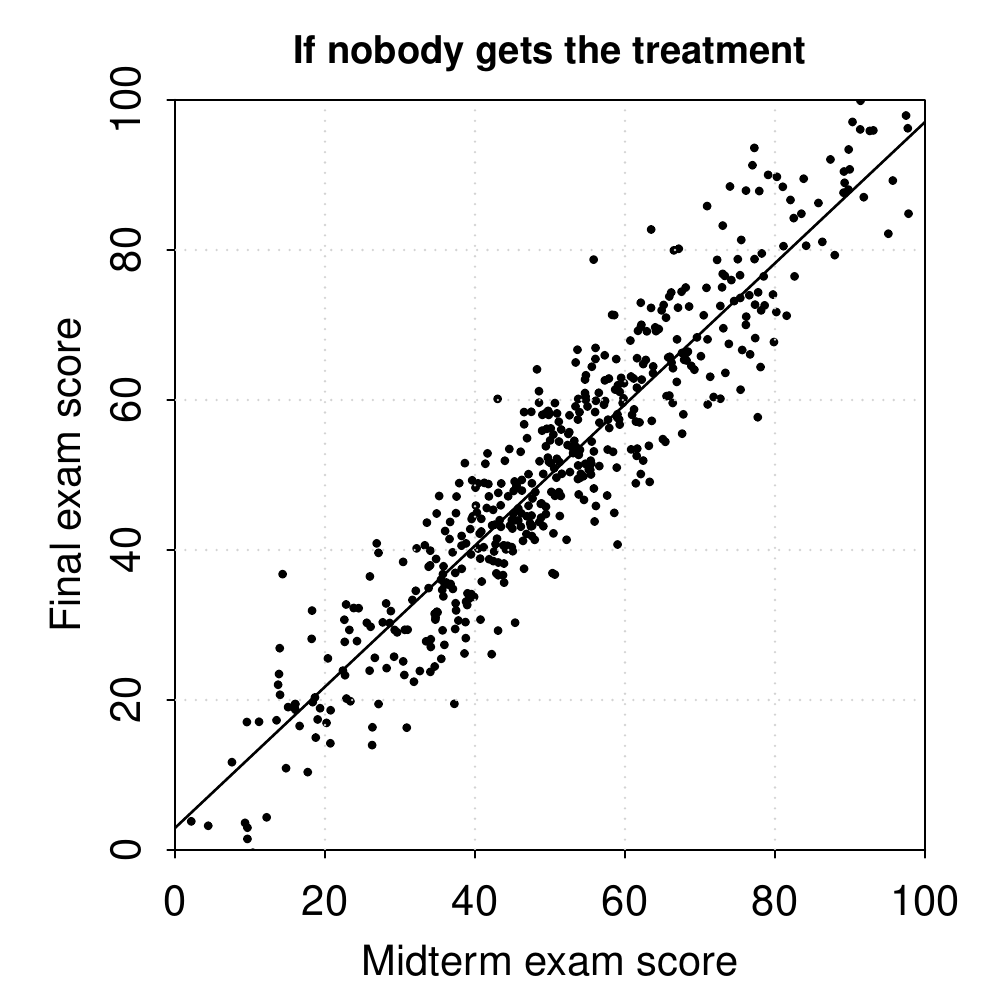}\hspace{.1\textwidth}  \includegraphics[width = .4\textwidth]{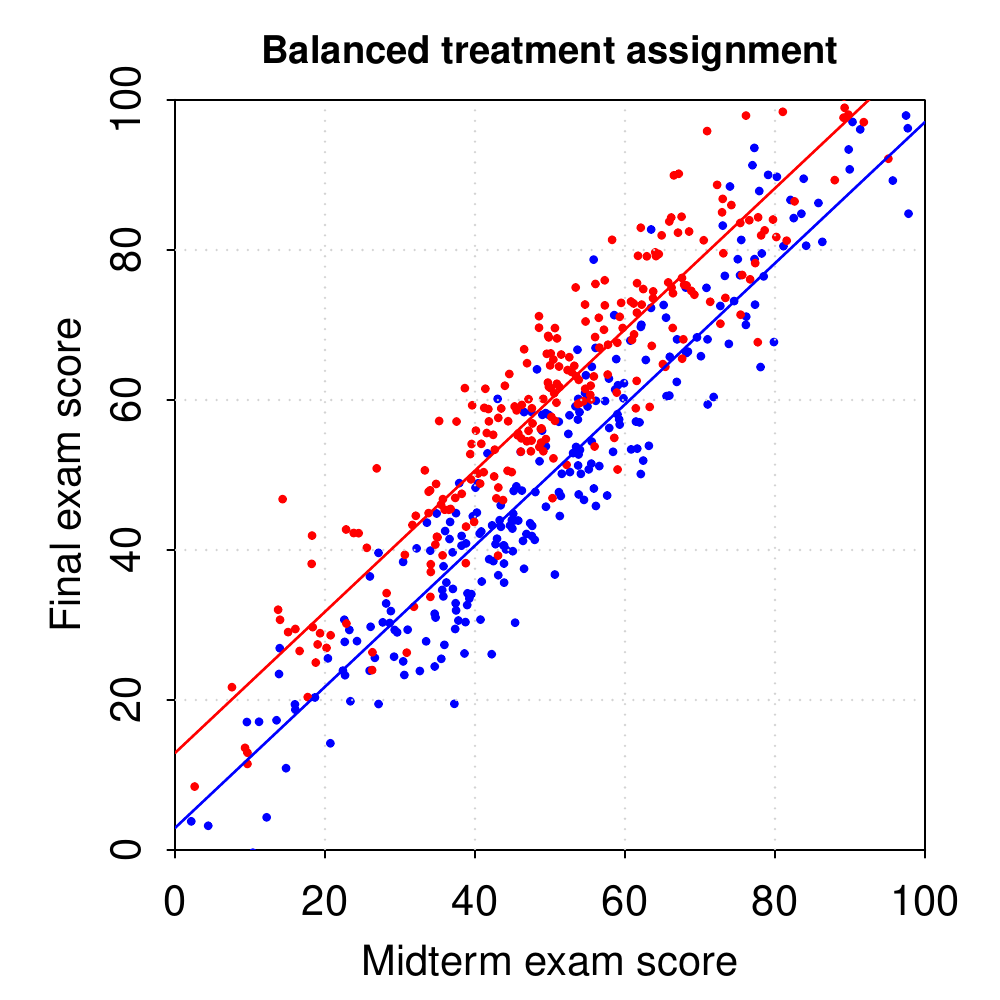}
    \vspace{-.1in}
    \caption{\em Simulated data on 500 students from a hypothetical study of the effect of an educational intervention.}
    \label{students1}
\end{figure}

For a more elaborate example, we perform a series of simulations to understand assumptions and bias correction in observational studies.  We start with a scenario in which 500 students in a class take a midterm and final exam.  We simulate the data by first drawing students' true abilities $\eta_i$ from a $\mbox{normal}(50, 20)$ distribution, then drawing the two exam scores $x_i,y_i$ as independent $\mbox{normal}(\eta_i, 10)$ distributions.  This induces the two scores to have a correlation of $\frac{20^2}{20^2 + 5^2}=0.94$; we designed the simulation with this high value to make patterns apparent in the graphs.  Figure \ref{students1}a displays the result, along with the underlying regression line, $\mbox{E}(y|x)$. We then construct a hypothetical randomized experiment of a treatment performed after the midterm that would add 10 points to any student's final exam score.  We give each student an equal chance of receiving the treatment or control.  Figure \ref{students1}b shows the simulated data and underlying regression lines.

In this example we can estimate the treatment effect by simply taking the difference between the two groups, which for these simulated data yields an estimate of 10.7 with standard error 1.8.  Or we can fit a regression adjusting for midterm score, yielding an estimate of $9.7\pm 0.6$.

\begin{figure}
    \centering
    \includegraphics[width = .5\textwidth]{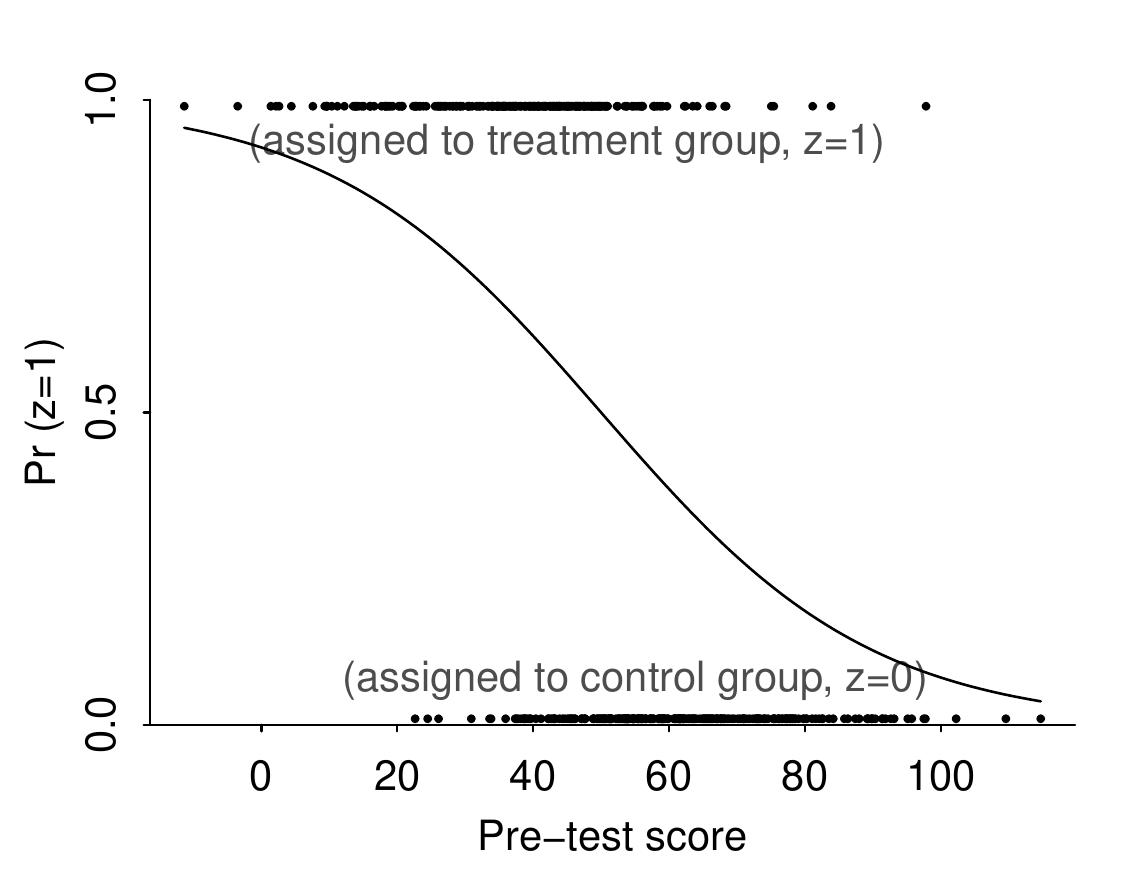}\includegraphics[width = .4\textwidth]{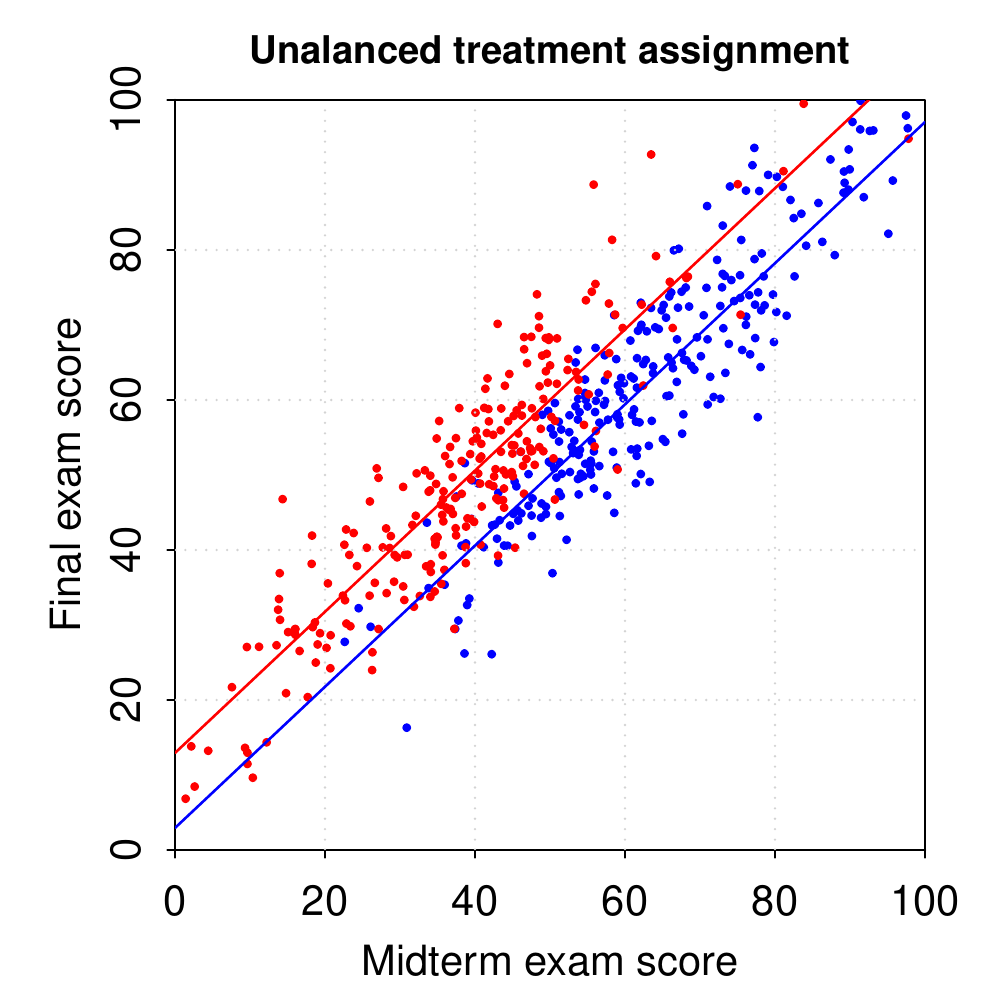}
    \vspace{-.1in}
    \caption{\em Alternative simulation in which the treatment assignment is unbalanced, with less well-performing students being more likely to receive the treatment.}
    \label{students2}
\end{figure}
We next consider an unbalanced assignment mechanism in which the probability of receiving the treatment depends on the midterm score:  $\mbox{Pr}(z=1)=\mbox{logit}^{-1}((x-50)/10)$.  Figure \ref{students2}a shows simulated treatment assignments for the 200 students and Figure \ref{students2}a displays the simulated exam scores.  The underlying regression lines are the same as before, as this simulation changes the distribution of $z$ but not the model for $y|x,z$.  Under this design, the treatment is preferentially given to the less well-performing students, hence a simple comparison of final exam scores will give a poor estimate.  For this particular simulation, the difference in average grades comparing the two groups is $-13.8\pm 1.5$, a terrible inference given that the true effect is, by construction, 10.  In this case, though, the linear regression adjusting for $x$ recovers the treatment effect, yielding an estimate of $9.7\pm 0.8$.

\begin{figure}
    \centering
    \includegraphics[width = .4\textwidth]{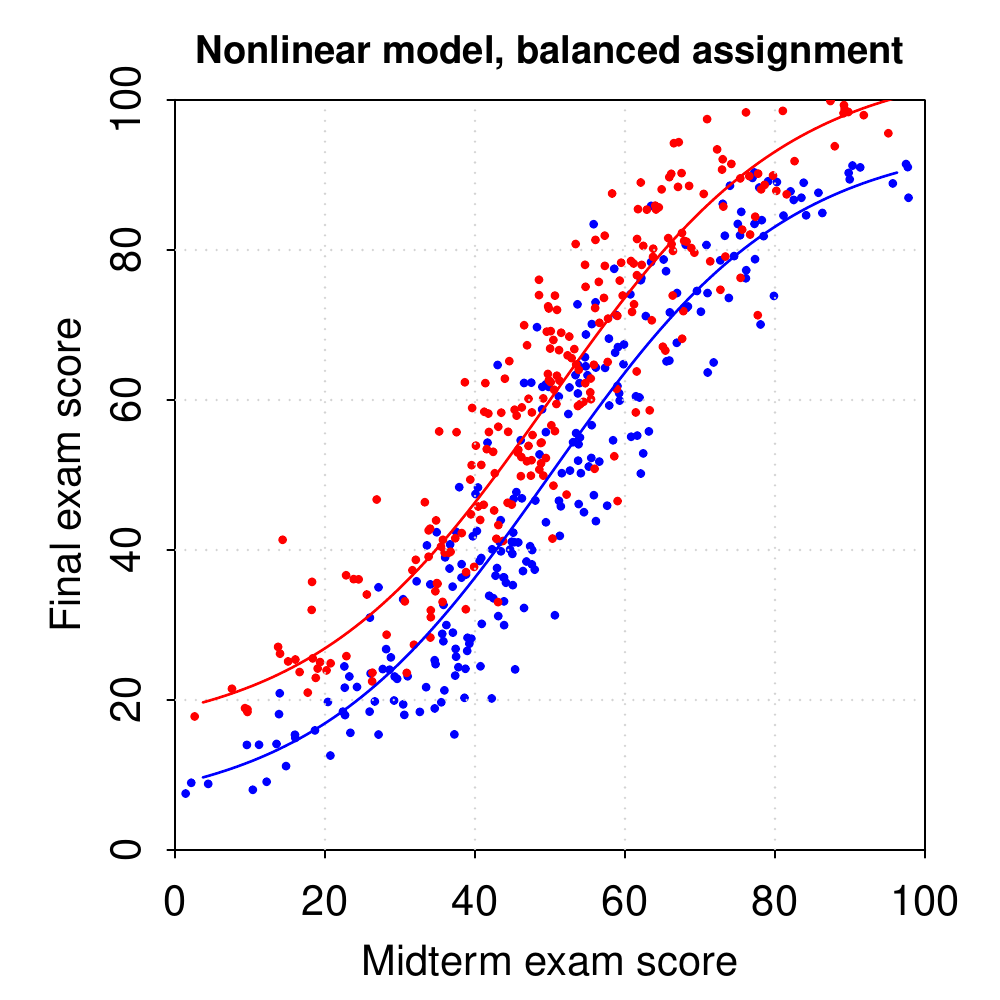}\hspace{.1\textwidth}  \includegraphics[width = .4\textwidth]{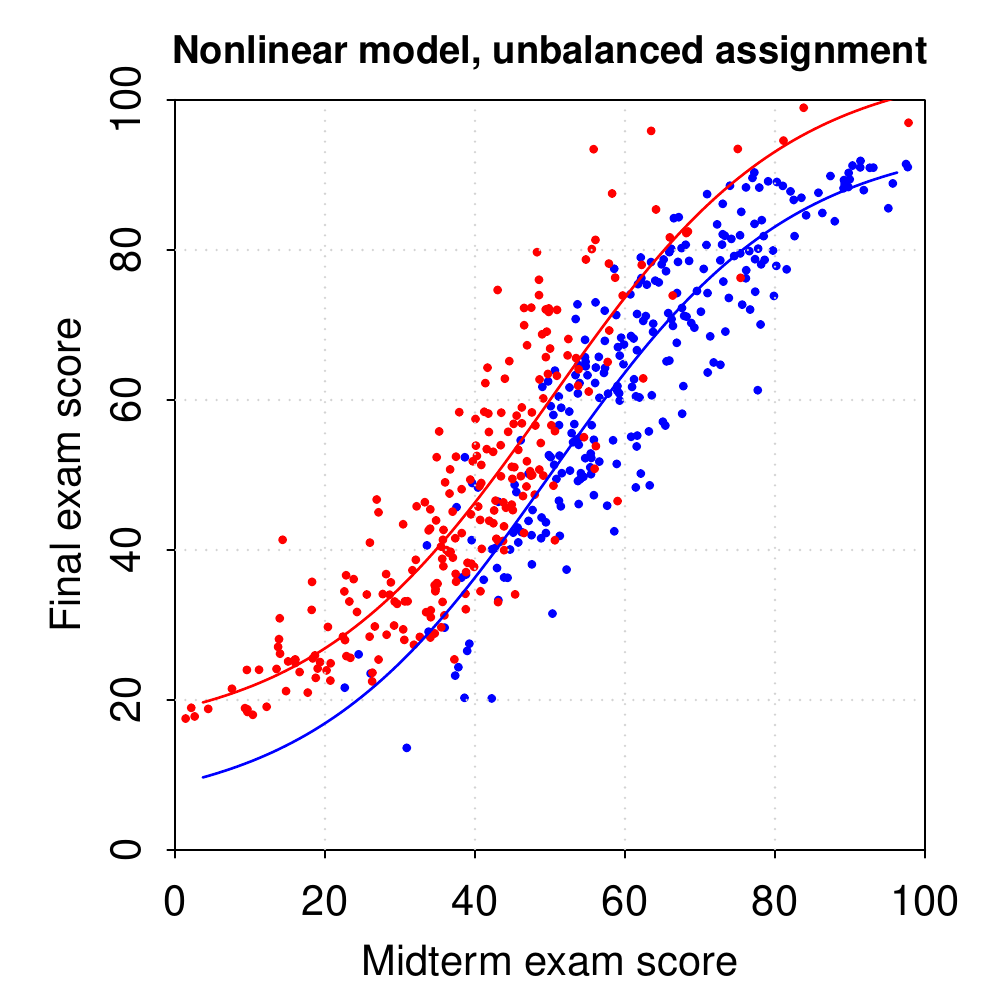}
    \vspace{-.1in}
    \caption{\em Again comparing two different treatment assignments, this time in a setting where the relation between final and midterm exam scores is nonlinear.}    \label{students3}
\end{figure}

But this new estimate is sensitive to the functional form of the adjustment for $x$.  We can see this by simulating data from an alternative model in which the true treatment effect is 10 but the function $\mbox{E}(y|x,z)$ is no longer linear.  In this case we construct such a model by 
drawing the midterm exam score given true ability from $\mbox{normal}(\eta_i, 10)$ as before, but transforming the final final exam score, as displayed in Figure \ref{students3}. We again consider two hypothetical experiments:  Figure \ref{students3}a shows data from the completely randomized assignment, and Figure \ref{students3}b displays the result using the unbalanced treatment assignment rule from Figure \ref{students2}a.  Both graphs show the underlying regression curves as well.

What happens when we now fit a linear regression to estimate the treatment effect?  The estimate from the design in Figure \ref{students3}a is reasonable:  even though the linear model is wrong and thus the resulting estimate is not fully statistically efficient, the balance in the design ensures that on average the specification errors will cancel, and the estimate is $10.5\pm 0.8$.  But the unbalanced design has problems:  even after adjusting for $x$ in the linear regression, the estimate is $7.3\pm 0.9$.

In the context of the present article, the point of this example is to demonstrate how simulation of a statistical system under different conditions can give us insight, not just about computational issues but also about data and inference more generally.  One could go further in this particular example by considering varying treatment effects, selection on unobservables, and other complications, and this is generally true that such theoretical explorations can be considered indefinitely to address whatever concerns might arise.

\section{Addressing computational problems}\label{addressing_computation}

\subsection{The folk theorem of statistical computing}\label{folk}
When you have computational problems, often there’s a problem with your model (Yao, Vehtari, and Gelman, 2020).  Not always---sometimes you will have a model that is legitimately difficult to fit---but many cases of poor convergence correspond to regions of parameter space that are not of substantive interest or even to a nonsensical model. An example of pathologies in irrelevant regions of parameter space is given in Figure \ref{orbits}. Examples of fundamentally problematic models would be bugs in code or using a Gaussian-distributed varying intercept for each individual observation in a Gaussian or logistic regression context, where they cannot be informed by data. Our first instinct when faced with a problematic model should not be to throw more computational resources on the model (e.g., by running the sampler for more iterations or reducing the step size of the HMC algorithm), but to check whether our model contains some substantive pathology.

\subsection{Starting at simple and complex models and meeting in the middle}\label{sec:simple_and_complex}

\begin{figure}
\centerline{\includegraphics[width=.6\textwidth]{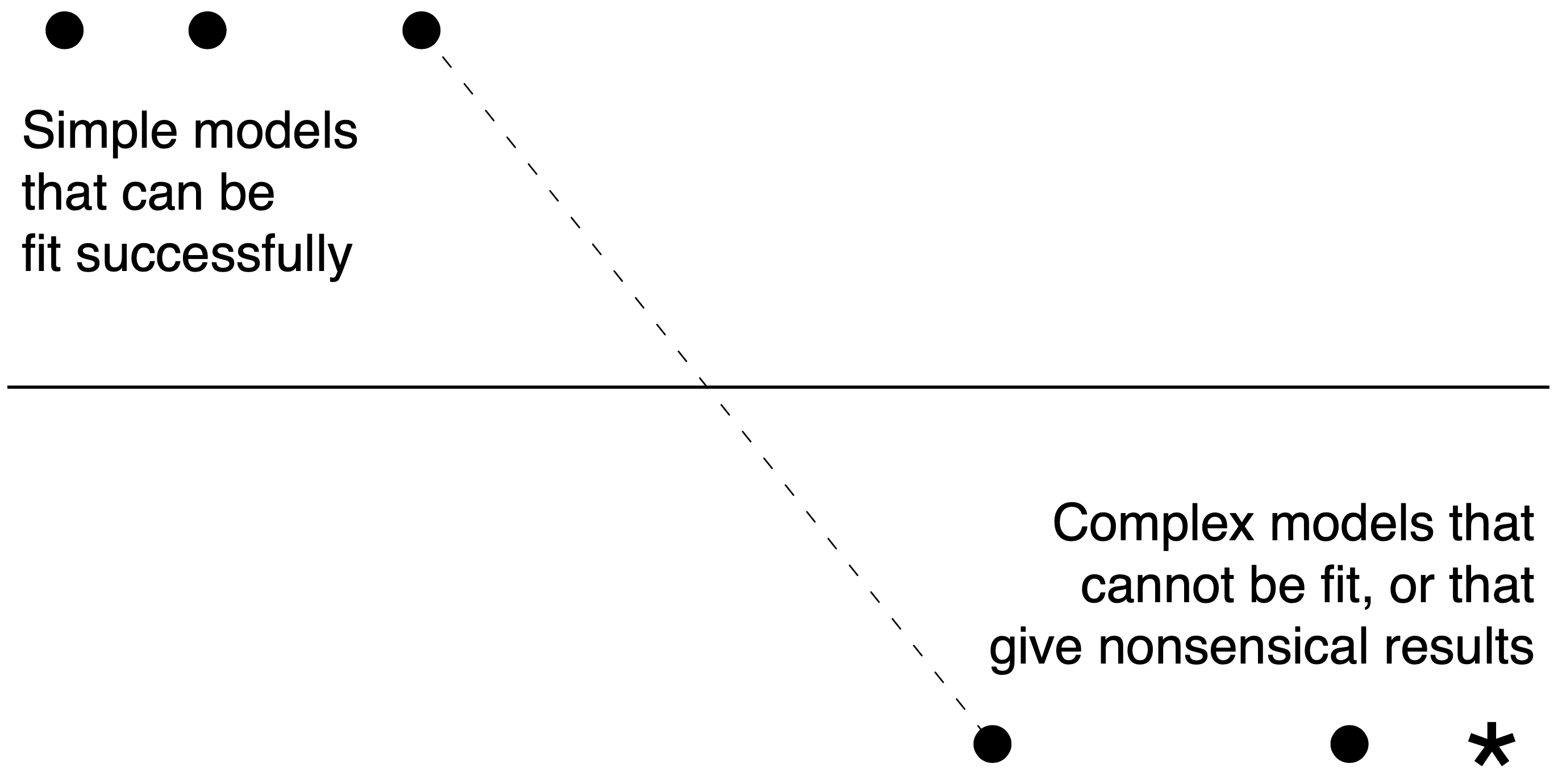}}
\caption{\em Diagram of advice for debugging. The asterisk on the lower right represents the scenario in which problems arise when trying to fit the desired complex model.  The dots on the upper left represent successes at fitting various simple versions, and the dots on the lower right represent failures at fitting various simplifications of the full model.  The dotted line represents the idea that the problems can be identified somewhere between the simple models that fit and the complex models that don't.  From Gelman and Hill (2007).}
\label{debugging}
\end{figure}

Figure \ref{debugging} illustrates a commonly useful approach to debugging.  The starting point is that a model is not performing well, possibly not converging or being able to reproduce the true parameter values in fake-data simulation, or not fitting the data well, or yielding unreasonable inferences.  The path toward diagnosing the problem is to move from two directions:  to gradually simplify the poorly-performing model, stripping it down until you get something that works; and from the other direction starting with a simple and well-understood model and gradually adding features until the problem appears.  Similarly, if the model has multiple components (e.g., a differential equation and a linear predictor for parameters of the equation), it is usually sensible to perform a sort of unit test by first making sure each component can be fit separately, using simulated data.

We may never end up fitting the complex model we had intended to fit at first, either because it was too difficult to fit using currently available computational algorithms, or because existing data and prior information are not informative enough to allow useful inferences from the model, or simply because the process of model exploration lead us in a different direction than we had originally planned.

\subsection{Getting a handle on models that take a long time to fit}

We generally fit our models using HMC, which can run slowly for various reasons, including expensive gradient evaluations as in differential equation models, high dimensionality of the parameter space, or a posterior geometry in which HMC steps that are efficient in one part of the space are too large or too small in other regions of the posterior distribution.  Slow computation is often a sign of other problems, as it indicates a poorly-performing HMC. However the very fact that the fit takes long means the model is harder to debug. 

For example, we recently received a query in the Stan users group regarding a multilevel logistic regression with 35,000 data points, 14 predictors, and 9 batches of varying intercepts, which failed to finish running after several hours in Stan using default settings from rstanarm.

We gave the following suggestions, in no particular order:
\begin{itemize}
    \item Simulate fake data from the model and try fitting the model to the fake data (Section \ref{fake}). Frequently a badly specified model is slow, and working with simulated data allows us not to worry about lack of fit.
\item Since the big model is too slow, you should start with a smaller model and build up from there.  First fit the model with no varying intercepts. Then add one batch of varying intercepts, then the next, and so forth.
\item Run for 200 iterations, rather than the default (at the time of this writing). Eventually you can run for 2000 iterations, but there is no point in doing that while you’re still trying to figure out what’s going on.  If, after 200 iterations, $\widehat{R}$ is large, you can run longer, but there is no need to start with 2000.
\item Put at least moderately informative priors on the regression coefficients and group-level variance parameters (Section \ref{priors}).
\item Consider some interactions of the group-level predictors. It seems strange to have an additive model with 14 terms and no interactions.  This suggestion may seem irrelevant to the user's concern about speed---indeed, adding interactions should only increase run time---but it is a reminder that ultimately the goal is to make predictions or learn something about the underlying process, not merely to get some arbitrary pre-chosen model to converge.
\item Fit the model on a subset of your data.  Again, this is part of the general advice to understand the fitting process and get it to work well, before throwing the model at all the data.
\end{itemize}
The common theme in all these tips is to think of any particular model choice as provisional, and to recognize that data analysis requires many models to be fit in order to gain control over the process of computation and inference for a particular applied problem.

\subsection{Monitoring intermediate quantities}
\label{sec:intermediate_quant}
Another useful approach to diagnosing model issues is to save intermediate quantities in our computations and plot them along with other MCMC output, for example using bayesplot (Gabry et al., 2020a) or ArviZ (Kumar et al., 2019).  These displays are an alternative to inserting print statements inside the code.  In our experience, we typically learn more from a visualization than from a stream of numbers in the console.

One problem that sometimes arises is chains getting stuck in out-of-the-way places in parameter space where the posterior density is very low and where it can be baffling why the MCMC algorithm does not drift back toward the region where the log posterior density is near its expectation and where most of the posterior mass is.
Here it can be helpful to look at predictions from the model given these parameter values to understand what is going wrong, as we illustrate in Section \ref{sec:orbits}.  But the most direct approach is to plot the expected data conditional on the parameter values in these stuck chains, and then to transform the gradient of the parameters to the gradient of the expected data.  This should give some insight as to how the parameters map to expected data in the relevant regions of the posterior distribution.

\subsection{Stacking to reweight poorly mixing chains}

In practice, often our MCMC algorithms mix just fine.  Other times, the simulations quickly move to unreasonable areas of parameter space, indicating the possibility of model misspecification, non-informative or weakly informative observations, or just difficult geometry.

But it is also common to be in an intermediate situation where multiple chains are slow to mix but they are in a generally reasonable range.
In this case we can use stacking to combine the simulations, using cross validation to assign weights to the different chains (Yao, Vehtari, and Gelman, 2020).  This will have the approximate effect of discarding chains that are stuck in out-of-the-way low-probability modes of the target distribution. The result from stacking is not necessarily equivalent, even asymptotically, to fully Bayesian inference, but it serves many of the same goals, and is especially suitable during the model exploration phase, allowing us to move forward and spend more time and energy in other part of Bayesian workflow without getting hung up on exactly fitting one particular model. In addition, non-uniform stacking weights, when used in concert with traceplots and other diagnostic tools, can help us understand where to focus that effort in an iterative way.

\subsection{Posterior distributions with multimodality and other difficult geometry}

We can roughly distinguish four sorts of problems with MCMC related to multimodality and other difficult posterior geometries:
\begin{itemize}
\item Effectively disjoint posterior volumes, where all but one of the modes have near-zero mass.  An example appears in Section \ref{sec:orbits}.  In such problems, the minor modes can be avoided using judicious choices of initial values for the simulation, adding prior information or hard constraints for parameters or they can be pruned by approximately estimating the mass in each mode. 
\item Effectively disjoint posterior volumes of high probability mass that are trivially symmetric, such as label switching in a mixture model.  It is standard practice here to restrict the model in some way to identify the mode of interest; see for example Bafumi et al.\ (2005) and Betancourt (2017b).
\item Effectively disjoint posterior volumes of high probability mass that are different. For example in a model of gene regulation (Modrák, 2018), some data admit two distinct regulatory regimes with opposite signs of the effect, while an effect close to zero has much lower posterior density.  This problem is more challenging.  In some settings, we can use stacking (predictive model averaging) as an approximate solution, recognizing that this is not completely general as it requires defining a predictive quantity of interest. A more fully Bayesian alternative is to divide the model into pieces by introducing a strong mixture prior and then fitting the model separately given each of the components of the prior. Other times the problem can be addressed using strong priors that have the effect of ruling out some of the possible modes.
\item A single posterior volume of high probability mass with an arithmetically unstable tail.  If you initialize near the mass of the distribution, there should not be problems for most inferences.  If there is particular interest in extremely rare events, then the problem should be reparameterized anyway, as there is a limit to what can be learned from the usual default effective sample size of a few hundred to a few thousand.
\end{itemize}

\subsection{Reparameterization}

Generally, an HMC-based sampler will work best if its mass matrix is appropriately tuned and the geometry of the joint posterior distribution is relatively uninteresting, in that it has no sharp corners, cusps, or other irregularities. This is easily satisfied for many classical models, where results like the Bernstein-von Mises theorem suggest that the posterior will become fairly simple when there is enough data. Unfortunately, the moment a model becomes even slightly complex, we can no longer guarantee that we will have enough data to reach this asymptotic utopia (or, for that matter, that a Bernstein-von Mises theorem holds). For these models, the behavior of HMC can be greatly improved by judiciously choosing a parameterization that makes the posterior geometry simpler.

For example, hierarchical models can have difficult funnel pathologies in the limit when group-level variance parameters approach zero (Neal, 2011), but in many such problems these computational difficulties can be resolved using reparameterization, following the principles discussed by Meng and van Dyk (2001); see also Betancourt and Girolami (2015).

\subsection{Marginalization}

Challenging geometries in the posterior distribution are often due to interactions between parameters.
An example is the above-mentioned funnel shape, which we may observe when plotting the joint density of the group-level scale parameter, $\phi$, and the individual-level mean, $\theta$.
By contrast, the marginal density of $\phi$ is well behaved.
Hence we can efficiently draw MCMC samples from the marginal posterior,
$$
  p(\phi | y) = \int_\Theta 
    p(\phi, \theta | y) d \theta.
$$
To draw posterior draws with MCMC, Bayes' rule teaches us we only need the marginal likelihood, $p(y|\phi)$.
It is then possible to recover posterior draws for $\theta$ by doing exact sampling from the conditional distribution $p(\theta|\phi, y)$, at a small computational cost.
This marginalization strategy is notably used for Gaussian processes with a normal likelihood (e.g. Rasmussen and Williams, 2006, Betancourt, 2020b).

In general, the densities $p(y |\phi)$ and $p(\theta |\phi, y)$ are not available to us.
Exploiting the structure of the problem, we can approximate these distributions using a Laplace approximation, notably for latent Gaussian models (e.g., Tierney and Kadane, 1986, Rasmussen and Williams, 2006, Rue et al., 2009, Margossian et al., 2020b).
When coupled with HMC, this marginalization scheme can, depending on the cases, be more effective than reparameterization, if sometimes at the cost of a bias; for a discussion on the topic, see Margossian et al.\ (2020a).

\subsection{Adding prior information}
\label{adding_prior_info}

Often the problems in computation can be fixed by including prior information that is already available but which had not yet been included in the model, for example, because prior elicitation from domain experts has substantial cost (O'Hagan et al., 2006, Sarma and Kay, 2020) and we started with some template model and prior (Section \ref{sec:initial_model}). In some cases, running for more iterations can also help.  But many fitting problems go away when reasonable prior information is added, which is not to say that the primary use of priors is to resolve fitting problems.

We may have assumed (or hoped) that the data would sufficiently informative for all parts of the model, but with careful inspection or as the byproduct of computational diagnostics, we may find out that this is not the case. In classical statistics, models are sometimes classified
as identifiable or non-identifiable, but this can be misleading (even after adding intermediate categories such as partial or weak identifiability), as the amount of information that can be learned from observations depends also on the specific realization of the data that was actually obtained. In addition, ``identification'' is formally defined in statistics as an asymptotic property, but in Bayesian inference we care about inference with finite data, especially given that our models often increase in size and complexity as more data are included into the analysis.  Asymptotic results can supply some insight into finite-sample performance, but we generally prefer to consider the posterior distribution that is in front of us.
Lindley (1956) and Goel and DeGroot (1981) discuss how to measure the information provided by an experiment as how different the posterior is from the prior.  If the data are not informative on some aspects of the model, we may improve the situation by providing more information via priors. Furthermore, we often prefer to use a model with parameters that can be updated by the information in the data instead of a model that may be closer to the truth but where data are not able to provide sufficient information.  In Sections \ref{crossvalidation}--\ref{influence} we discuss tools for assessing the informativeness of individual data points or hyperparameters.

We illustrate with the problem of estimating the sum of declining exponentials with unknown decay rates.  This task is a well-known ill-conditioned problem in numerical analysis and also arises in applications such as pharmacology (Jacquez, 1972).
We assume data 
$$
y_i  = (a_1e^{-b_1x_i} + a_2e^{-b_2x_i})\times e^{\epsilon_i}, \mbox{ for } i=1,\dots,n,
$$
with independent errors $\epsilon_i\sim\mbox{normal}(0,\sigma)$.
The coefficients $a_1$, $a_2$, and the residual standard deviation $\sigma$ are constrained to be positive.  The parameters $b_1$ and $b_2$ are also positive---these are supposed to be declining, not increasing, exponentials---and are also constrained to be ordered, $b_1<b_2$, so as to uniquely define the two  model components.

\begin{figure} \centerline{\includegraphics[width=.45\textwidth]{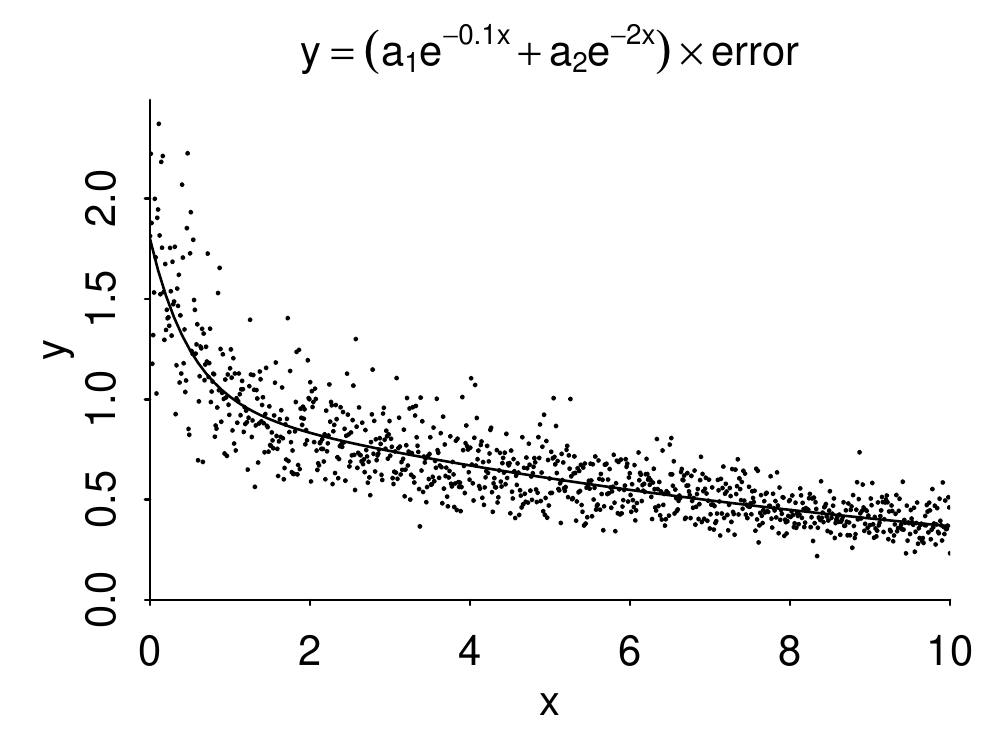}\hspace{.05\textwidth}\includegraphics[width=.45\textwidth]{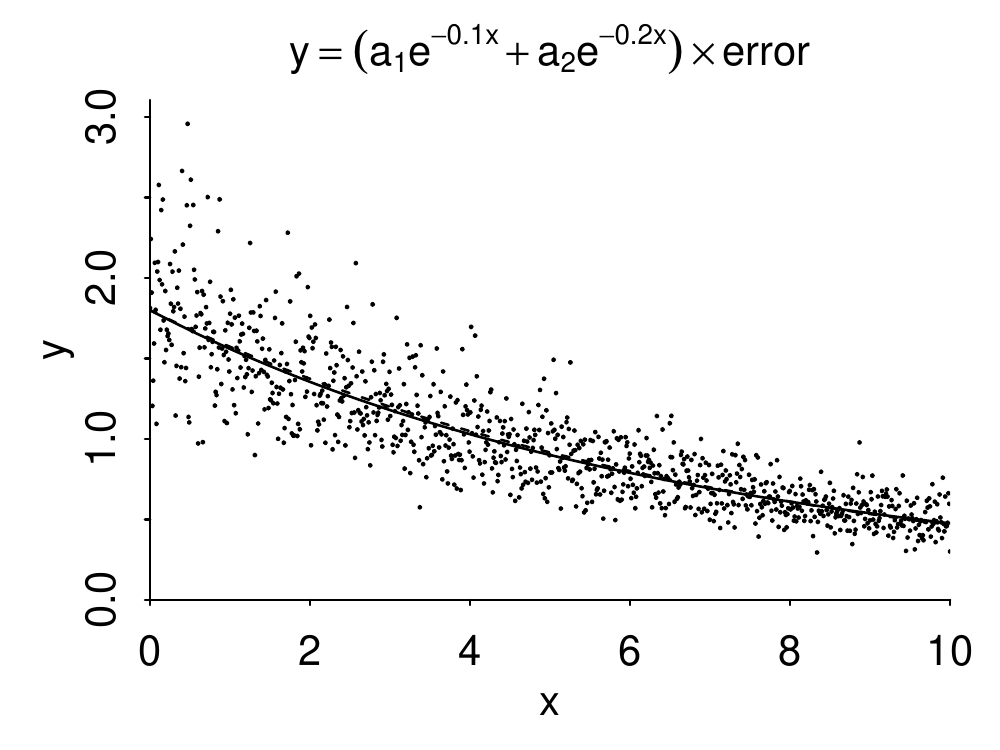}}
  \vspace{-.2in}
\caption{\em Simulated data from the model, $y = (a_1e^{-b_1x} + a_2e^{-b_2x})*\mbox{error}$:  (a) setting $(b_1,b_2)=(0.1, 2.0)$, and (b) setting $(b_1,b_2)=(0.1, 0.2)$.  In addition, the dotted line in the second plot shows the curve, $y=1.8e^{-0.135x}$, which almost exactly coincides with the true curve over the range of these data.  It was easy to fit the model to the data in the left graph and recover the true parameter values.  For the data in the right graph, however, the observations did not provide information to reduce the uncertainty sufficiently, and and the model could only be stably fit using a prior distribution that provided that needed missing information.}
\label{exponentials}
\end{figure}

We start by simulating fake data from a model where the two curves should be cleanly distinguished, setting $b_1=0.1$ and $b_2=2.0$, a factor of 20 apart in scale.  We simulate 1000 data points where the predictor values $x$ are uniformly spaced from 0 to 10, and, somewhat arbitrarily, set $a_1=1.0, a_2=0.8, \sigma=0.2$.  Figure \ref{exponentials}a shows true curve and the simulated data.  We then use Stan to fit the model from the data.  The simulations run smoothly and the posterior inference recovers the five parameters of the model, which is no surprise given that the data have been simulated from the model we are fitting.

But now we make the problem slightly more difficult.  The model is still, by construction, true, but instead
of setting $(b_1,b_2)$ to $(0.1, 2.0)$, we make them $(0.1, 0.2)$, so now only a factor of 2 separates the scales of the two declining exponentials.  The simulated data are shown in Figure \ref{exponentials}b.  But now when we try to fit the model in Stan, we get terrible convergence.  The two declining exponentials have become very hard to tell apart, as we have indicated in the graph by also including the curve, $y=1.8e^{-0.135x}$, which is essentially impossible to distinguish from the true model given these data.

We can make this computation more stable by adding prior information.  For example, default $\mbox{normal}(0,1)$ priors on all the parameters does sufficient regularization, while still being weak if the model has been set up so the parameters are roughly on unit scale, as discussed in Section \ref{scaling}.  In this case we are assuming this has been done.  One could also check the sensitivity of the posterior inferences to the choice of prior, either by comparing to alternatives such as $\mbox{normal}(0,0.5)$ and $\mbox{normal}(0,2)$ or by differentiating the inferences with respect to the prior scale, as discussed in Section \ref{influence}.

\begin{figure}
\centerline{
   \includegraphics[width=\textwidth]{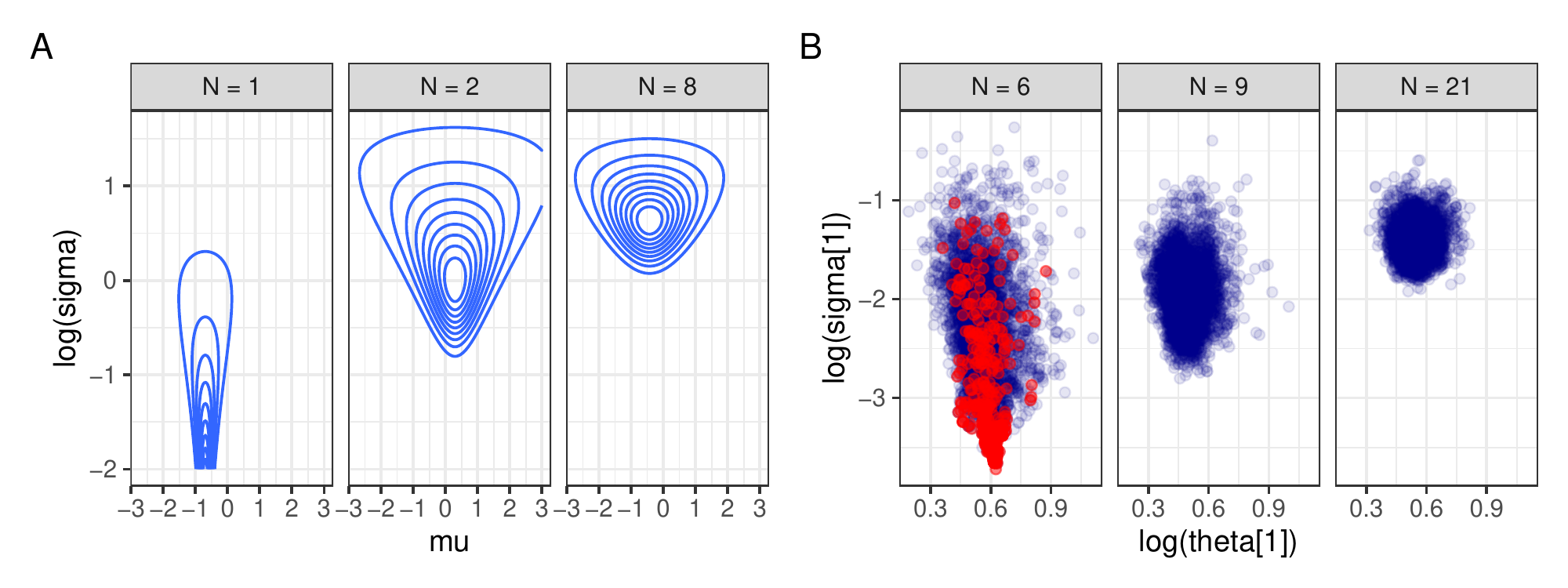}
 }
 \vspace{-.15in}
\caption{\em Example of how adding data changes the geometry of the posterior distribution. (A) Normal likelihood as a function of mean $\mu$ and standard deviation $\sigma$ with increasing number of available observations $N$ drawn from the standard normal distribution. For $N = 1$ the likelihood increases without bounds and  for $N = 2$ the geometry is still funnel-like which can cause computational problems. For $N = 8$ the funnel-like shape is mostly suppressed. (B) Practical example using an Lotka-Volterra model of population dynamics ($8$ parameters total) as described in Carpenter (2018), showing scatter plots of posterior samples of two parameters drawn from fitting the model in Stan. The fit with $6$ data points shows a funnel-like shape. The red points indicate divergent transitions which signal that the sampler encountered difficult geometry and the results are not trustworthy. Stan becomes able to fit the model when $9$ data points are used, despite the slightly uneven geometry. The model is well behaved when fit to $21$ data points.}
\label{fig:underdetermined}
\end{figure}

Using an informative normal distribution for the prior adds to the tail-log-concavity of the posterior density, which leads to a quicker MCMC mixing time. But the informative prior does not represent a tradeoff of model bias vs.\ computational efficiency; rather, in this case the model fitting is improved as the computation burden is eased, an instance of the folk theorem discussed in Section \ref{folk}.  More generally, there can be strong priors with thick tails, and thus the tail behavior is not guaranteed to be more log concave.  On the other hand, the prior can be weak in the context of the likelihood while still guaranteeing a log-concave tail.  This is related to the point that the informativeness of a model depends on what questions are being asked.
 
Consider four steps on a ladder of abstraction:
\begin{enumerate}
\item Poor mixing of MCMC;
\item Difficult geometry as a mathematical explanation for the above;
\item Weakly informative data for some parts of the model as a statistical explanation for the above;
\item Substantive prior information as a solution to the above.
\end{enumerate}
Starting from the beginning of this ladder, we have computational troubleshooting; starting from the end, computational workflow.

As another example, when trying to avoid the funnel pathologies of hierarchical models in the limit when group-level variance parameters approach zero, one could use zero-avoiding priors (for example, lognormal or inverse gamma distributions) to avoid the regions of high curvature of the likelihood; a related idea is discussed for penalized marginal likelihood estimation by Chung et al.\ (2013, 2014). Zero-avoiding priors can make sense when such prior information is available---such as for the length-scale parameter of a Gaussian process (Fuglstad et al., 2019)---but we want to be careful when using such a restriction merely to make the algorithm run faster.  At the very least, if we use a restrictive prior to speed computation, we should make it clear that this is information being added to the model.

More generally, we have found that poor mixing of statistical fitting algorithms can often be fixed by stronger regularization. This does not come free, though:  in order to effectively regularize without blurring the very aspects of the model we want to estimate, we need some subject-matter knowledge---actual prior information. Haphazardly tweaking the model until computational problems disappear is dangerous and can threaten the validity of inferences without there being a good way to diagnose the problem.

\subsection{Adding data}
\label{adding_more_data}

Similarly to adding prior information, one can constrain the model by adding new data sources that are handled within the model.  For example, a calibration experiment can inform the standard deviation of a response. 

In other cases, models that are well behaved for larger datasets can have computational issues in small data regimes; Figure~\ref{fig:underdetermined} shows an example. While the funnel-like shape of the posterior in such cases looks similar to the funnel in hierarchical models, this pathology is much harder to avoid, and we can often  only  acknowledge that the full model is not informed by the data and a simpler model needs to be used. Betancourt (2018) further discusses this issue.

\section{Evaluating and using a fitted model}\label{evaluating}

Once a model has been fit, the workflow of evaluating that fit is more convoluted, because there are many different things that can be checked, and each of these checks can lead in many directions. Statistical models can be fit with multiple goals in mind, and statistical methods are developed for different groups of users. The aspects of a model that needs to be checked will depend on the application.

\subsection{Posterior predictive checking}
\label{sec:post_pred_checks}

Posterior predictive checking is analogous to prior predictive checking (Section \ref{sec:prior_pred_checks}), but the parameter draws used in the simulations come from the posterior distribution rather than the prior. While prior predictive checking is a way to understand a model and the implications of the specified priors, posterior predictive checking also allows one to examine the fit of a model to real data (Box, 1980, Rubin, 1984, Gelman, Meng, and Stern, 1996).

\begin{figure}[ht]
\centerline{
   \includegraphics[width=.8\textwidth]{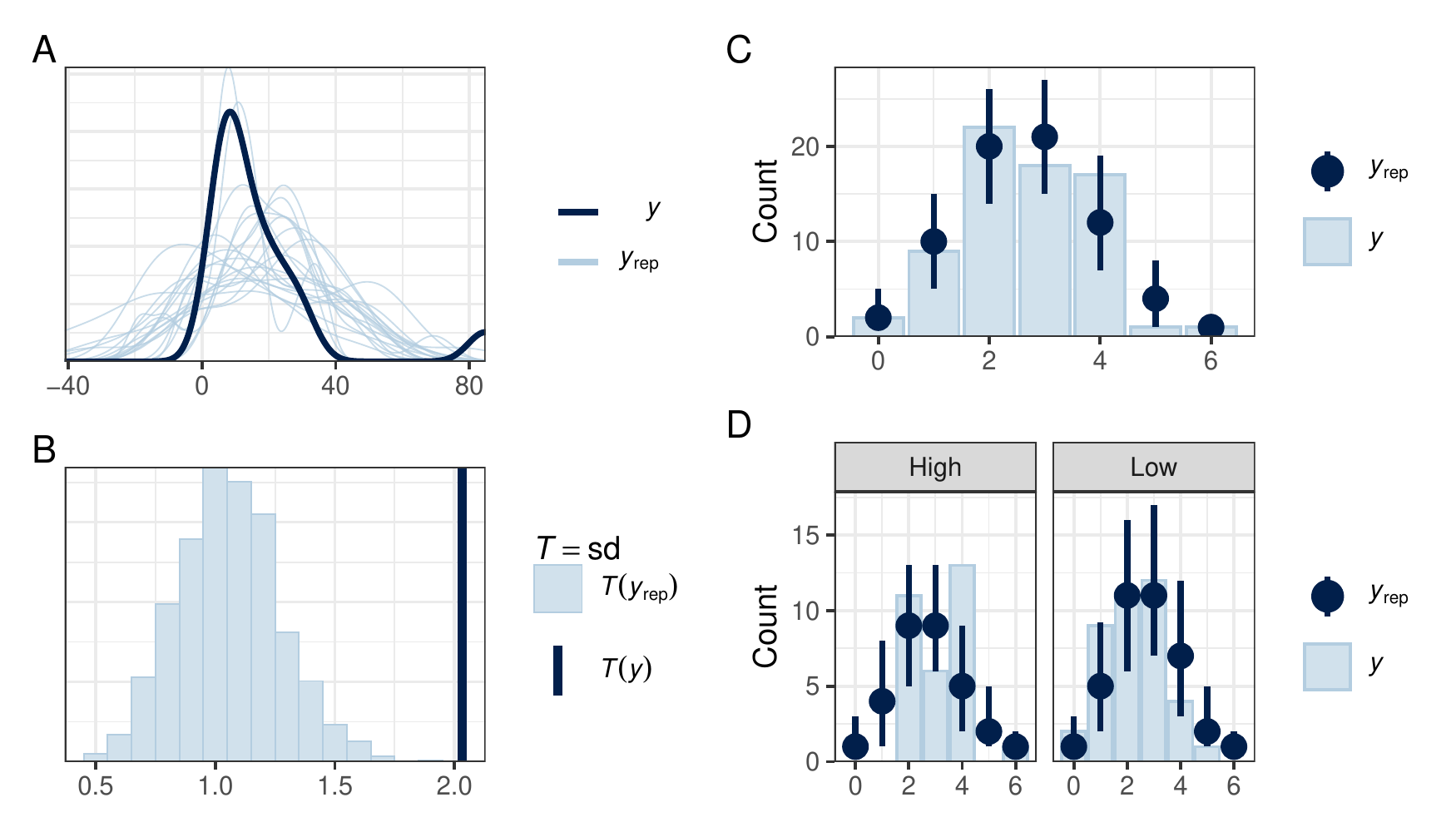}
 }
 \vspace{-.2in}
\caption{\em Examples of diagnosing model misfit with simple posterior predictive checks as implemented in the bayesplot R package. In all plots $y$ is plotted based on the input data and $y_{rep}$ based on the distribution of posterior simulations. (A) A ``density'' check for a normal distribution being fit to data simulated from a lognormal distribution: the tails of the $y_{rep}$ behave very differently than for $y$. (B) A ``statistic'' check for a binomial distribution being fit to data simulated from a beta-binomial distribution. Histogram of standard deviation (sd) of the $y_{rep}$ datasets is compared to sd of $y$. The check shows that the data have larger sd than what the model can handle. (C) A ``bars'' check for discrete data (note the switch of colors between $y$ and $y_{rep}$). This check looks good: the distribution of frequencies of individual counts in $y$ falls well within those in $y_{rep}$. (D) The same model and data but with the check grouped by a covariate that was not included in the model but in fact influenced the response. The High subgroup systematically deviates from the range of $y_{rep}$, indicating that there is an additional source of variability that the model fails to capture.}
\label{ppc_basic}
\end{figure}

When comparing simulated datasets from the posterior predictive distribution to the actual dataset, if the dataset we are analyzing is unrepresentative of the posterior predictive distribution, this indicates a failure of the model to describe an aspect of the data. The most direct checks compare the simulations from the predictive distribution to the full distribution of the data or a summary statistic computed from the data or subgroups of the data, especially for groupings not included in the model (see Figure \ref{ppc_basic}). There is no general way to choose which checks one should perform on a model, but running a few such direct checks is a good safeguard against gross misspecification. There is also no general way to decide when a check that fails requires adjustments to the model. Depending on the goals of the analysis and the costs and benefits specific to the circumstances, we may tolerate that the model fails to capture certain aspects of the data or it may be essential to invest in improving the model. In general, we try to find ``severe tests'' (Mayo, 2018): checks that are likely to fail if the model would give misleading answers to the questions we care most about. 

\begin{figure}[ht]
\centerline{
   \includegraphics[width=.8\textwidth]{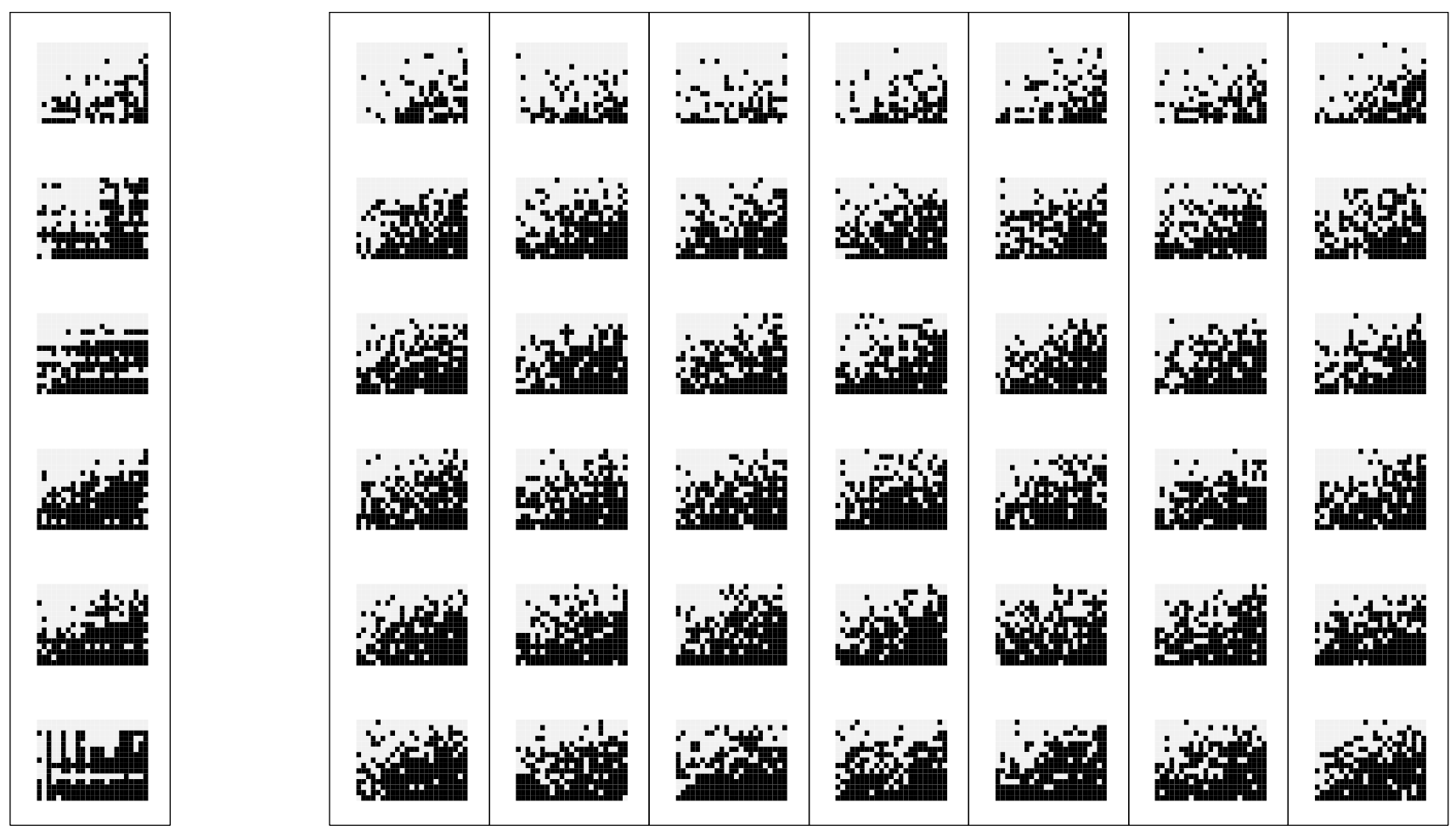}
 }
 \vspace{-.1in}
\caption{\em Example of a posterior predictive check.  The left column shows observed data from a psychology experiment, displayed as a $15\times 23$ array of binary responses from each of 6 participants, ordered by row, column, and person.  The right columns show 7 replicated datasets from the posterior predictive distribution of the fitted model.  Each replicated dataset has been ordered, as this is part of the display.  The check reveals some patterns in the observed that do not appear in the replications, indicating a aspect of lack of fit of model to data.  From Gelman et al.\ (2013).}
\label{ppc_sim}
\end{figure}

Figure \ref{ppc_sim} shows a more involved posterior predictive check from an applied project.  This example demonstrates the way that predictive simulation can be combined with graphical display, and it also gives a sense of the practical challenges of predictive checking, in that it is often necessary to come up with a unique visualization tailored to the specific problem at hand.

\subsection{Cross validation and influence of individual data points and subsets of the data}\label{crossvalidation}

Posterior predictive checking is often sufficient for revealing model misfit, but as it uses data both for model fitting and misfit evaluation, it can be overly optimistic. In cross validation, part of the data is left out, the model is fit to the remaining data, and predictive performance is checked on the left-out data. This  improves predictive checking diagnostics, especially for flexible models (for example, overparameterized models with more parameters than observations).

Three diagnostic approaches using cross validation that we have found useful for further evaluating models are 
\begin{enumerate}
\item calibration checks using the cross validation predictive distribution,
\item identifying which observations or groups of observations are most difficult to predict, 
\item identifying how influential particular observations are, that is, how much information they provide on top of other observations. 
\end{enumerate}

In all three cases, efficient approximations to leave-one-out cross-validation using importance sampling can facilitate practical use by removing the need to re-fit the model when each data point is left out (Vehtari et al., 2017, Paananen et al., 2020).

\begin{figure}
\centerline{
   \includegraphics[width=.42\textwidth]{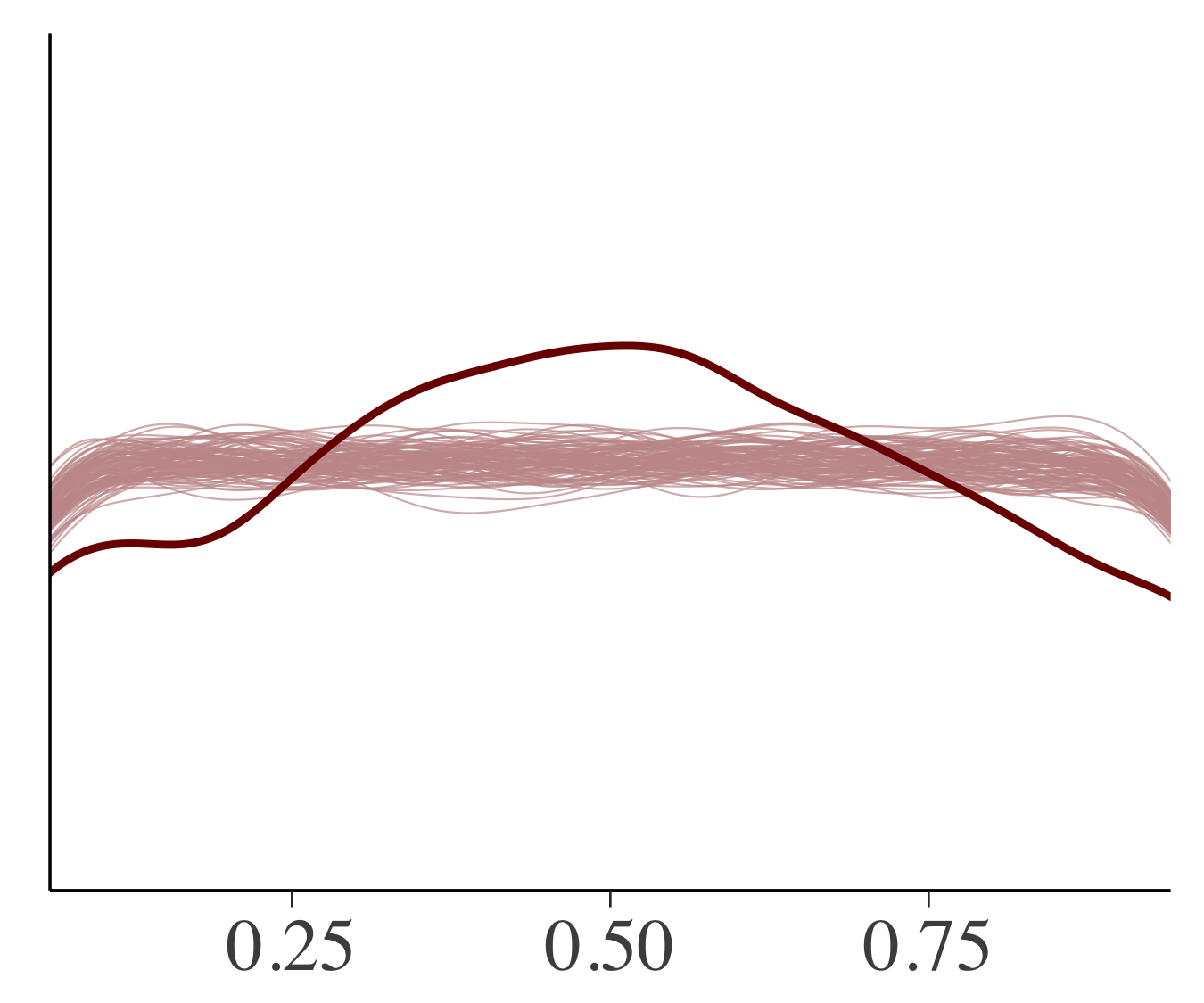}
   \vspace{-.1in}
}
\caption{\em Leave-one-out cross validation probability integral transformation (LOO-PIT) plot evaluating the calibration of predictions  from a fitted model. Under perfect calibration, LOO-PIT values would be uniform. In this case the values are concentrated near the middle, indicating predictive distributions that are too wide. From Gabry et al.\ (2019).}
\label{fig:loo-pit}
\end{figure}

Although perfect calibration of predictive distributions is not the ultimate goal of Bayesian inference, looking at how well calibrated leave-one-out cross validation (LOO-CV) predictive distributions are, can reveal opportunities to improve the model. While posterior predictive checking often compares the marginal distribution of the predictions to the data distribution, leave-one-out cross validation predictive checking looks at the calibration of conditional predictive distributions. Under a good calibration, the conditional cumulative probability distribution of the predictive distributions (also known as probability integral transformations, PIT) given the left-out observations are uniform. Deviations from uniformity can reveal, for example, under or overdispersion of the predictive distributions. Figure \ref{fig:loo-pit} shows an example from Gabry et al.\ (2019) where leave-one-out cross validation probability integral transformation (LOO-PIT) values are too concentrated near the middle, revealing that the predictive distributions are overdispersed compared to the actual conditional observations.

In addition to looking at the calibration of the conditional predictive distributions, we can also look at which observations are hard to predict and see if there is a pattern or explanation for why some are harder to predict than others. This approach can reveal potential problems in the data or data processing, or point to directions for model improvement (Vehtari et al., 2017, Gabry et al., 2019). We illustrate with an analysis of a survey of residents from a small area in Bangladesh that was affected by arsenic in drinking water. Respondents with elevated arsenic levels in their wells were asked if they were interested in switching to water from a neighbor's well, and a series of models were fit to predict this binary response given household information (Vehtari et al., 2017).

\begin{figure}
\centerline{
   \includegraphics[width=.42\textwidth]{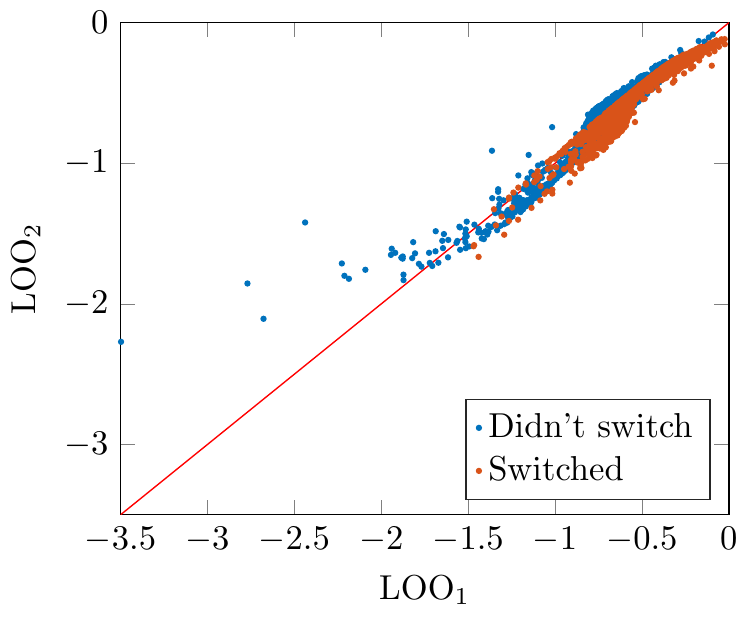}\hspace{1cm}
   \includegraphics[width=.438\textwidth]{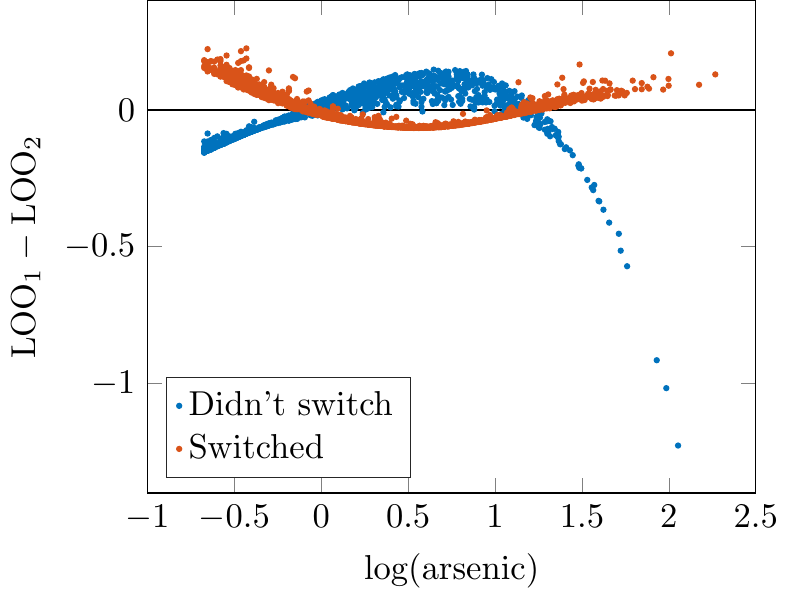}
   \vspace{-.1in}
}
\caption{\em Logistic regression example, comparing two models in terms of their pointwise contributions to leave one out (LOO) cross validation error:  
(a) comparing contributions of LOO directly; (b) plotting the difference in LOO as a function of a key predictor 
(the existing arsenic level).  To aid insight, we have colored the data according to the (binary) output, with red
corresponding to $y=1$ and blue representing $y=0$.  For any given data point, one model will fit better than 
another, but for this example the graphs reveal that the difference in LOO between the models arises from 
the linear model's poor predictions for 10--15 particular data points. From Vehtari et al.\ (2017).}
\label{fig:arsenic2}
\end{figure}

Figure \ref{fig:arsenic2} compares the pointwise log scores for two of the models. The scattered blue dots on 
the left side of Figure \ref{fig:arsenic2}a and on the lower right of Figure \ref{fig:arsenic2}b correspond to
data points for which one of the models fits particularly poorly---that is, large negative contributions to the expected log predictive density.  We can sum all of the pointwise differences to yield an estimated difference in expected log predictive densities
${\rm elpd}_{\rm loo}$ of 16.4 with a standard error of just 4.4, but beyond that we can use this plot to find {\em which} data points create problems for the model, in this case 10--15 non-switchers with very high existing arsenic levels.

In this particular example, we did not follow up on this modeling issue, because even more elaborate models that fit the data better do not change the conclusions and thus would not change any recommended actions in Bangladesh.  Gabry et al.\ (2019) provide an example where LOO-CV indicated problems that motivated efforts to improve the statistical model. 

The above two approaches focus on the predictions, but we can also look at how parameter inferences change when each data point is  left out, which provides a sense of the influence of each observation.  As with cross validation more generally, this approach has limitations if the data are clustered or otherwise structured so that multiple points would need to be removed to have an effect, but it can still be part of general Bayesian workflow, given that it is computationally inexpensive and can be valuable in many applied settings.  Following this cross validation idea, the influence of an individual data point $y_{i}$ can be summarized according to properties of the distribution of the importance weights computed when approximating LOO-CV (see Vehtari et al., 2017 for details on the approximation and the corresponding implementation in the loo R package (Vehtari et al. 2020)). 

An alternative approach to importance weighting is to frame the removal of data points as a gradient in a larger model space.  Suppose we have a simple independent likelihood, $\prod_{i=1}^n p(y_i|\theta)$, and we work with the more general form, $\prod_{i=1}^n p(y_i|\theta)^{\alpha_i}$, which reduces to the likelihood of the original model when $\alpha_i=1$ for all $i$.  Leave-one-out cross validation corresponds to setting $\alpha_i=0$ for one observation at a time.  But another option, discussed by Giordano et al.\ (2018) and implemented by Giordano (2018), is to compute the gradient of the augmented log likelihood as a function of $\alpha$:  this can be interpreted as a sort of differential cross validation or influence function.

Cross validation for multilevel (hierarchical) models requires more thought.  Leave-one-out is still possible, but it does not always match our inferential goals.  For example, when performing multilevel regression for adjusting political surveys, we are often interested in estimating opinion at the state level.  A model can show real improvements at the state level with this being undetectable at the level of cross validation of individual observations (Wang and Gelman, 2016). Millar (2018), Merkle, Furr, and Rabe-Hesketh (2019), and Vehtari (2019) demonstrate different cross validation variants and their approximations in hierarchical models, including leave-one-unit-out and leave-one-group-out.  
In applied problems we have performed a mix, holding out some individual observations and some groups and then evaluating predictions at both levels (Price et al., 1996). 

Unfortunately, approximating such cross validation procedures using importance sampling tends to be much harder than in the leave-one-out case. This is because more observations are left out at a time which implies stronger changes in the posterior distributions from the full to the subsetted model. As a result, we may have to rely on more costly model refits to obtain leave-one-unit-out and leave-one-group-out cross validation results.

\subsection{Influence of prior information}\label{influence}

Complex models can be difficult to understand, hence the need for exploratory model analysis (Unwin et al., 2003, Wickham, 2006) and explainable AI (Chen et al., 2018, Gunning, 2017, Rudin, 2018), which complements methods for evaluating, comparing, and averaging models using a range of interrelated approaches, including cross validation, stacking, boosting, and Bayesian evaluation.  In this section we discuss methods to understand how posterior inferences under a fitted model depend on the data and priors.

A statistical model can be understood in two ways:  generatively and inferentially. From the generative perspective, we want to understand how the parameters map to the data.  We can perform prior predictive simulation to visualize possible data from the model (as in Figure \ref{gp_sim}).  From the inferential perspective, we want to understand the path from inputs (data and prior distributions) to outputs (estimates and uncertainties).

The most direct way to understand the influence of priors is to run sensitivity analysis by refitting the model with multiple priors, this can however be prohibitively expensive if the models take a long time to fit. However, there are some shortcuts.

One approach is to compute the shrinkage between prior and posterior, for example, by comparing prior to posterior standard deviations for each parameter or by comparing prior and posterior quantiles. If the data relative to the prior are informative for a particular parameter, shrinkage for that parameter should be stronger. This type of check
has been extensively developed in the literature; see, for example,
Nott et al.\ (2020).

Another approach is to use importance sampling to approximate the posterior of the new model using the posterior of old model, provided the two posteriors are similar enough for importance sampling to bridge (Vehtari et al., 2019, Paananen et al., 2020). And if they are not, this is also valuable information in itself (see Section \ref{crossvalidation}). 

\begin{figure}
  \centerline{\includegraphics[width=.9\textwidth]{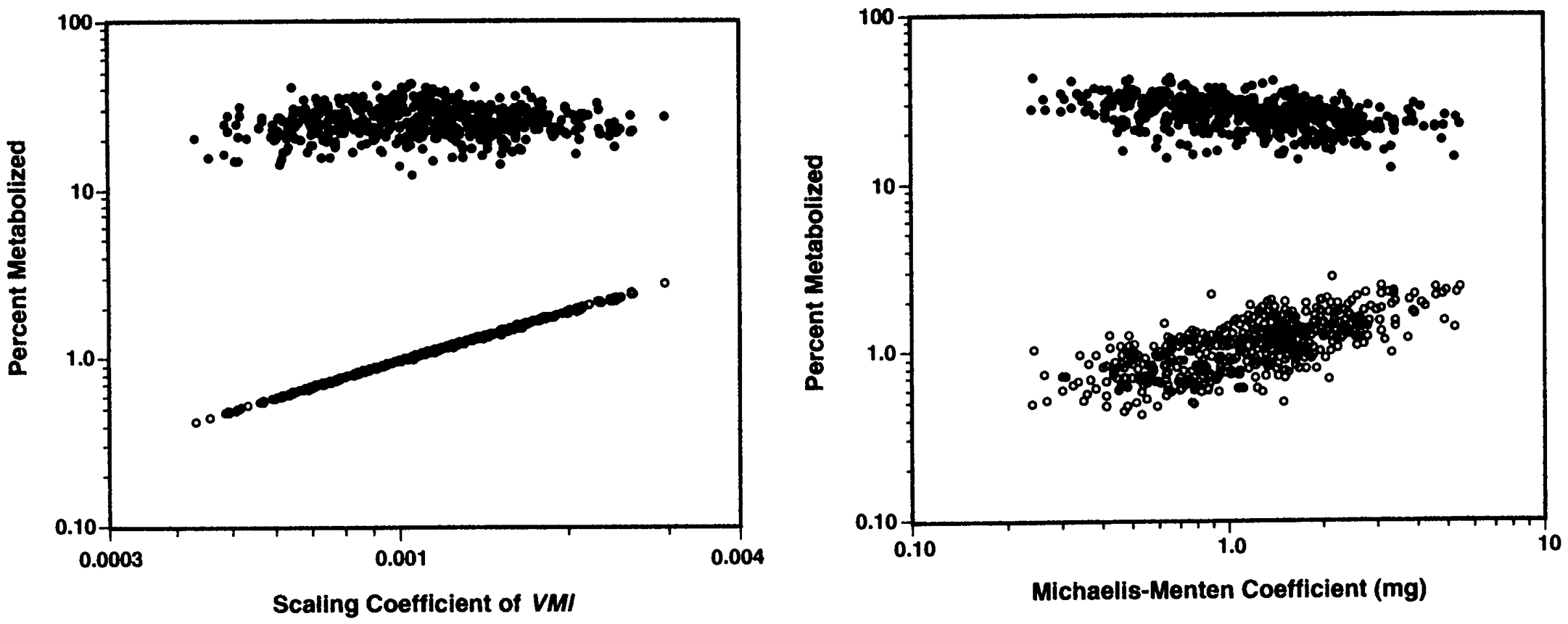}}
  \vspace{-.1in}
  \caption{\em Example of static sensitivity analysis from a statistical model fit to a problem in toxicology.  Each graph shows posterior simulation draws of the percent metabolized under two conditions (hence the clusters of points at the top and bottom of each graph), plotted vs.\ two of the parameters in the model.  The plots reveal little sensitivity to either parameter on inference for percent metabolized under one condition, but strong sensitivity for the other.  This sort of graph can be used to assess sensitivity to alternative prior distributions, without requiring the model to be re-fit. From Gelman, Bois, and Jiang (1996).}
\label{static}
\end{figure}

Yet another approach is to perform {\em static sensitivity analysis}, which is a way to study sensitivity of posterior inferences to perturbations in the prior without requiring that the model be re-fit using alternative prior distributions (Gelman, Bois, and Jiang, 1996; see Figure~\ref{static} for an example).  Each graph in Figure~\ref{static} shows posterior simulations revealing the posterior dependence of a quantity of interest (in this example, the percentage of a toxin that is metabolized by the body) as a function of a parameter in the model.

Consider Figure \ref{static} as four scatterplots, as each of the two graphs is really two plots superimposed, one for a condition of low exposure to the toxin and one for high exposure.
Each of these four plots can be interpreted in two ways.  First, the direct interpretation shows the posterior correlation between the predictive quantity of interest (the percentage metabolized in the body) and a particular parameter (for example, the scaling coefficient of the toxin's metabolism). 
Second, each scatterplot can be read indirectly to reveal sensitivity of the quantity plotted on the $y$-axis to the prior of the parameter plotted on the $x$-axis.  The interpretation goes as follows:  a change in the prior distribution for the parameter plotted on the $x$-axis can be performed by reweighting of the points on the graph according to the ratio of the new prior to that used in the analysis.  With these graphs, the importance weighing can be visualized implicitly:  the impact of changes in the prior distribution can be seen based on the dependence in the scatterplot.

The mapping of prior and data to posterior can also be studied more formally, as discussed in Section \ref{crossvalidation}. 

\subsection{Summarizing inference and propagating uncertainty}

Bayesian inference is well suited for problems with latent variables and other settings with unresolvable uncertainty.  In addition, we often use hierarchical models that include batches of parameters representing variation.  For example, when reporting the results from our election forecasting model, we are interested in uncertainty in the forecast votes and also variation among states.

Unfortunately, the usual ways of displaying Bayesian inference do not fully capture the multiple levels of variation and uncertainty in our inferences.  A table or even a graph of parameter estimates, uncertainties, and standard errors is only showing one-dimensional margins, while graphs of marginal posterior distributions are unwieldy for models with many parameters and also fail to capture the interplay between uncertainty and variation in a hierarchical model.

To start with, we should follow general principles of good statistical practice and graph data and fitted models, both for the ``exploratory data analysis'' purpose of uncovering unexpected patterns in the data and also more directly to understand how the model relates to the data used to fit it.

\begin{figure}
 \centerline{\includegraphics[width=0.6\textwidth]{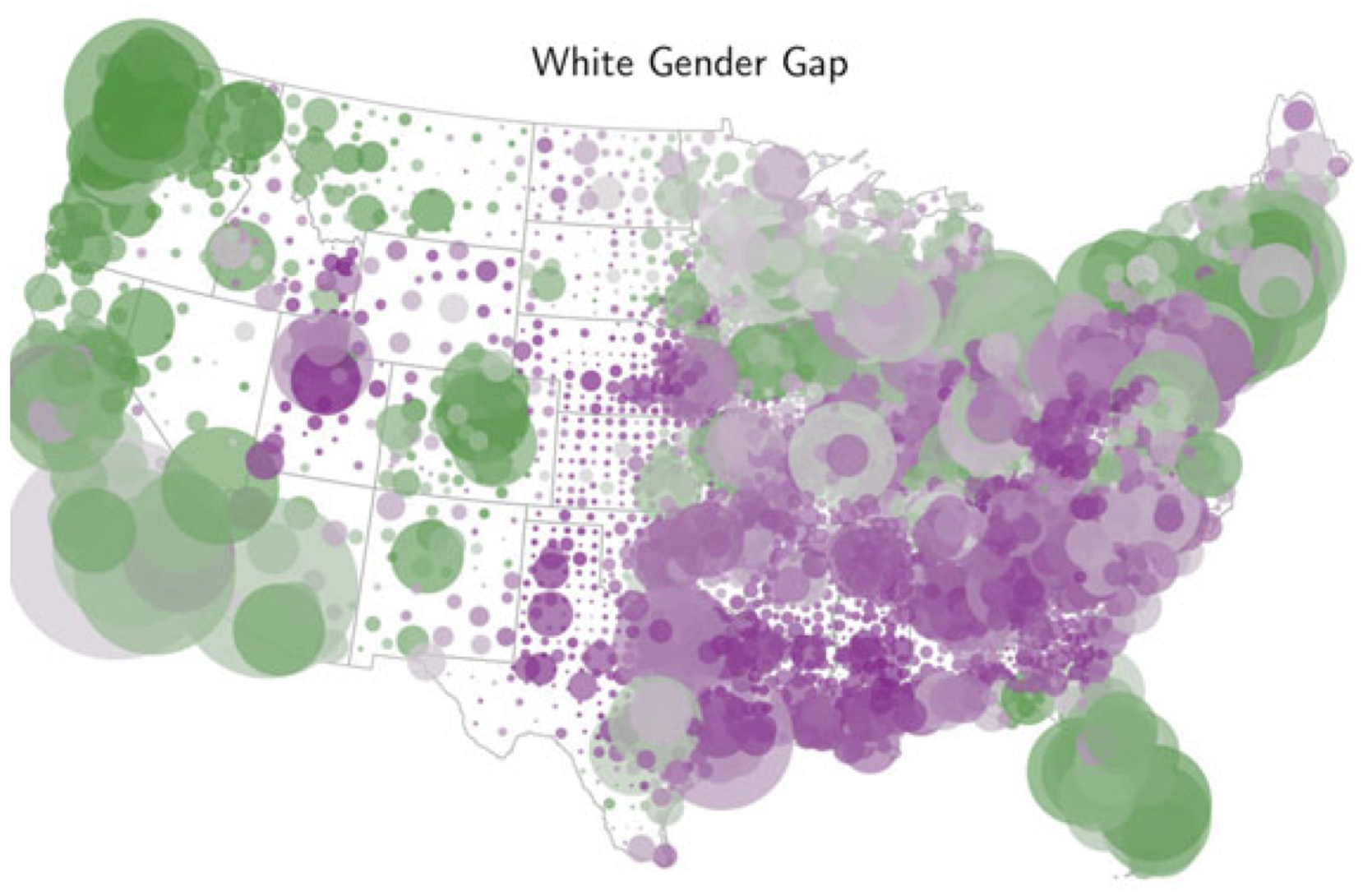}
 \includegraphics[width=0.4\textwidth]{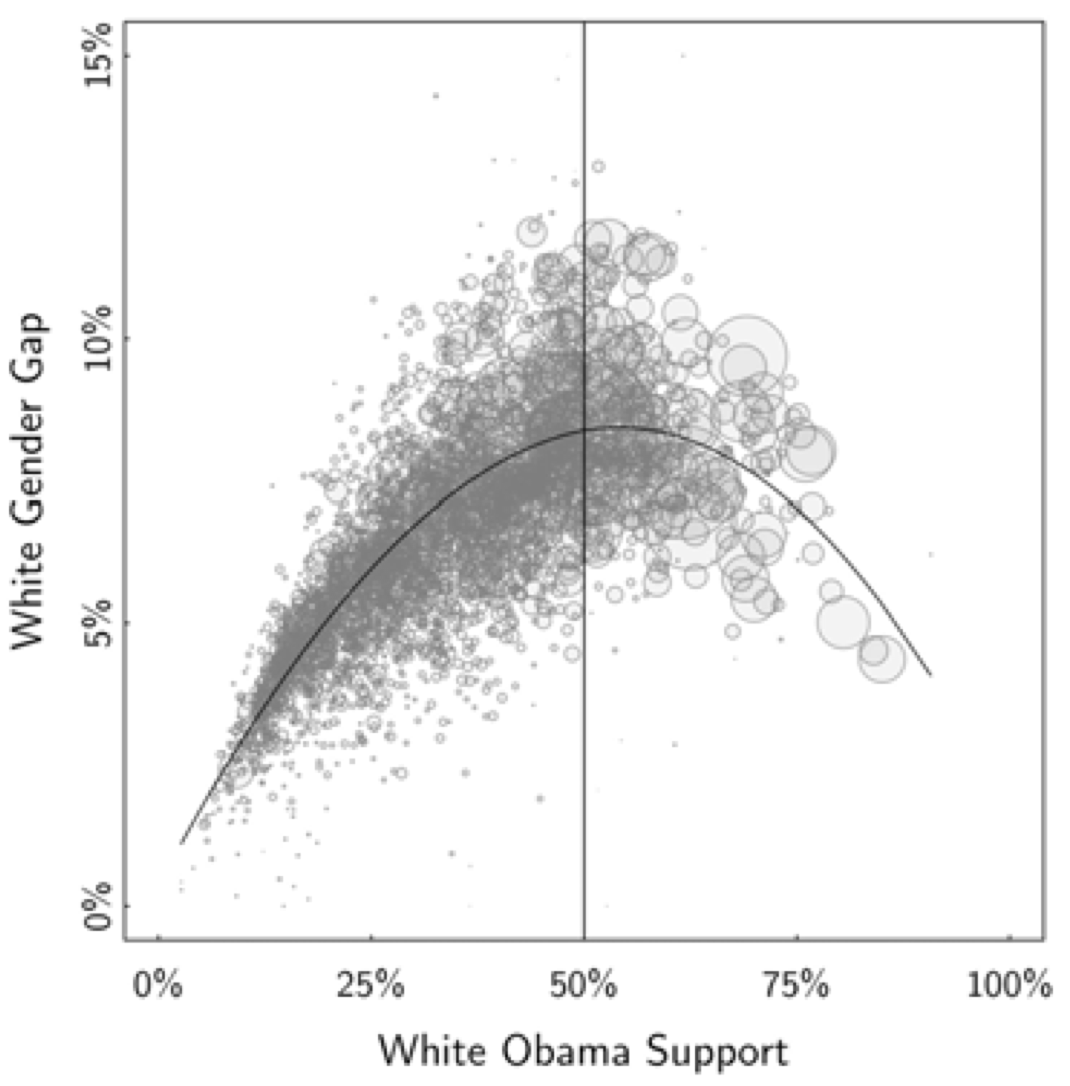}}
\caption{\em From a model fit to survey data during the 2016 U.S. presidential election campaign:  (a) estimate of the gender gap among white voters, (b) estimated gender gap plotted vs.\ Obama's estimated county-level vote share in 2012 among white votes.  The area of each circle is proportional to the number of voters in the county, and the colors on the map represent a range going from dark purple (no difference in voting patterns comparing white men and white women) through light gray (white women supporting Clinton 7.5 percentage points more than white men) to dark green (a white gender gap of 15 percentage points). The striking geographic patterns---a white gender gap that is low in much of the south and high in the west and much of the northeast and midwest---motivates the scatterplot, which reveals that the white gender gap tends to be highest in counties where the white vote is close to evenly split. From Ghitza and Gelman (2020).}
 \label{yair1}
 \end{figure}
 
 \begin{figure}
 \centerline{\includegraphics[width=\textwidth]{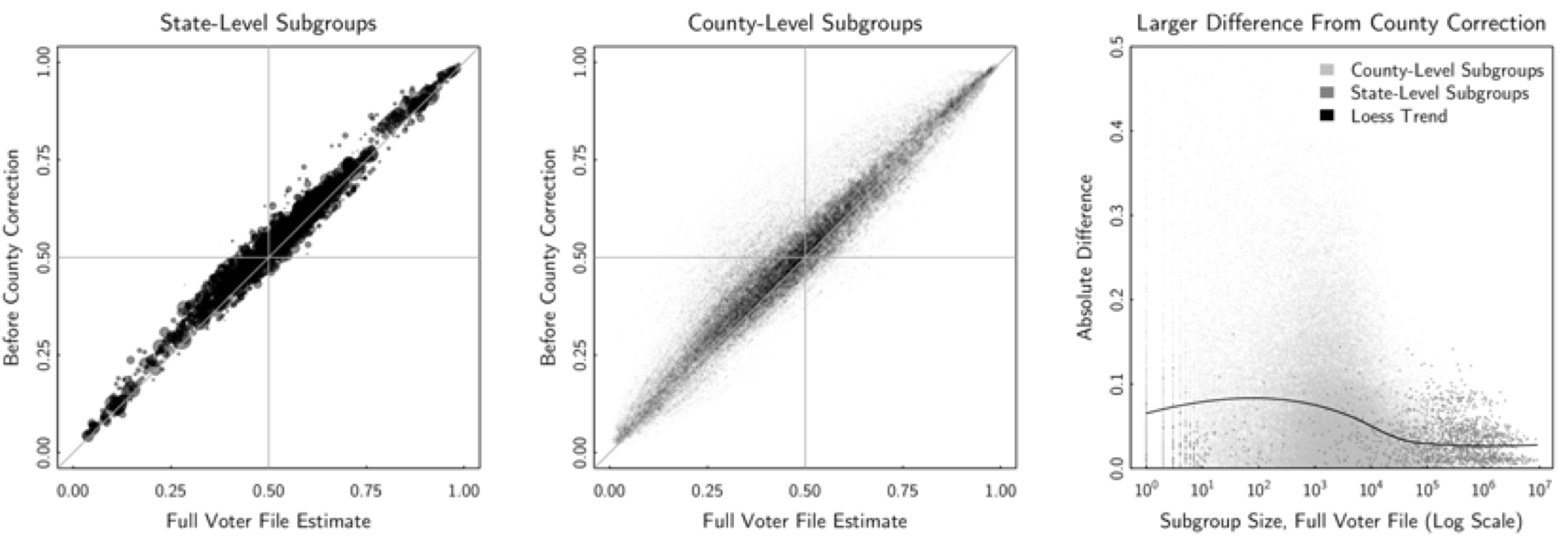}}
\caption{\em Evaluation of one aspect of the model shown in Figure \ref{yair1}, comparing county-level support for Clinton as estimated from two different models.  We include this here to illustrate the way that graphical methods can be used to summarize, understand, evaluate, and compare fitted models.  From Ghitza and Gelman (2020).}
 \label{yair2}
 \end{figure}

We illustrate some uses of graphical model exploration and summary from an analysis by Ghitza and Gelman (2020) of vote preferences in the 2016 U.S. presidential election.  Figure \ref{yair1} shows the estimated gender gap in support for the two candidates, first displayed as a map and then as a scatterplot.  The map, which is the natural first step to visualizing these estimates, reveals some intriguing geographic patterns which are then explored in a more focused way in the scatterplot.  Figure \ref{yair2} evaluates the model by comparing to simpler county-level estimates.
This example demonstrates a general workflow in exploratory graphics, in which the results from inferential summary motivates future exploration.

Gabry et al.\ (2019) present some of our ideas on graphics for Bayesian workflow, and some of this has been implemented in the R package bayesplot (Gabry et al., 2020b, see also Kay, 2020ab, Kumar, 2019).  Probabilistic programming ultimately has the potential to allow random variables to manipulated like any other data objects, with uncertainty implicit in all the plots and calculations (Kerman and Gelman, 2004, 2007), but much more work needs to be done to turn this possibility into reality, going beyond point and interval estimation so that we can make full use of the models we fit.

\section{Modifying a model}\label{building}

Model building is a language-like task in which the modeler puts together existing components (linear, logistic, and exponential functions; additive and multiplicative models; binomial, Poisson, and normal distributions; varying coefficients; and so forth) in order to encompass new data and additional features of existing data, along with links to underlying processes of interest. As mentioned in Section \ref{sec:modular}, most parts of the models can be seen as placeholders that allow replacement or expansion. Alternatively we may find the need to use an approximate model or an approximate algorithm, as discussed in Section \ref{sec:approximate}.

Model expansion can come in response to new data, failures of models fit to existing data, or computational struggles with existing fitting procedures.  
For the election forecast described in Gelman, Hullman, et al.\ (2020), we started with the poll-aggregation model of Linzer (2013) which we had adapted in 2016 but with a notable failure of predictions in certain swing states, which we attributed to poor modeling of correlations of vote swings between states, along with nonsampling errors in the polls (Gelman and Azari, 2017).  In our revision we expanded the model to include both these features.  Sections \ref{golf} and \ref{sec:orbits} give extended examples of iterative model building and evaluation.

\subsection{Constructing a model for the data}
In textbook treatments of statistics, the distribution of data given parameters is typically just given.  In applications, though, we want to set up a data model based on some combination of fit to the data (as found in posterior predictive checks) and domain expertise.  If the model is being chosen from a small menu, we would like to at least be open about that.  Often the most important aspect of the data model is not its distributional form but how the data are linked to underlying parameters of interest.  For example, in election forecasting, our model for polls includes terms for nonsampling error for individual polls and for polling averages, following Shirani-Mehr et al.\ (2018).

A related challenge arises because data are typically preprocessed before they come to us, so that any generative model is necessarily an approximation.  This can arise in meta-analysis or in settings where some large set of predictors have been combined into one or two numerical summaries using a machine learning algorithm or another dimensionality reduction technique.  As always, we need to be concerned about data quality, and the most important aspect of the data model can be bias rather than what would traditionally be considered measurement error.  Understanding this affects Bayesian workflow in that it can make sense to expand a model to allow for systematic error; we give an example in Section \ref{fudge}.

\subsection{Incorporating additional data}
It is sometimes said that the most important aspect of a statistical method is not what it does with the data, but what data are used.  A key part of Bayesian workflow is expanding a model to make use of more data.  This can be as simple as adding regression predictors---but when more parameters are added, it can be necessary to assume that not all of them can have a big effect in the model at the same time.  One way to see this is to consider the addition of a parameter as a relaxation of a prior distribution that was previously concentrated at zero.  For example, we expanded our election model to account for political polarization by adding interaction terms to the regression, allowing the coefficients for the national economic predictor to be lower in recent years.

It sometimes happens that we have two forms of measurement of similar data, thus requiring a generative model for both data sources.  Sometimes this creates technical challenges, as when we are combining direct measurements on a sample with population summary statistics, or integrating measurements of different quality (for example, Lin et al., 1999), or when information is available on partial margins of a table (Deming and Stephan, 1940).  In Weber et al.\ (2018), we fit a pharmacological model with direct data for a set of patients taking one drug but only average data for a set of patients that had received a competitor's product.  In order to avoid the computational cost of modeling as latent data the outcomes of all the averaged patients, we devised an analytic method to approximate the relevant integral so that we could include the averaged data in the likelihood function.

\subsection{Working with prior distributions}\label{priors}

Traditionally in Bayesian statistics we speak of noninformative or fully informative priors, but neither of these generally exist: a uniform prior contains some information, as it depends on the parameterization of the model; a reference prior depends on an assumed asymptotic regime for collecting new, fictional data (Berger et al., 2009);  and even an informative prior rarely includes all available knowledge.  Rather, we prefer to think of a ladder of possibilities:  (improper) flat prior; super-vague but proper prior; very weakly informative prior;
generic weakly informative prior; specific informative prior.  For example, our election model includes random walk terms to allow variation in opinion at the state and national level.  Each of these random walks has a standard deviation parameter corresponding to the unexplained variation (on the logit scale) in one day.  This innovation standard deviation could be given a uniform prior, a super-vague but proper prior (for example, $\mbox{normal}^+(0,1000)$, where we are using the notation $\mbox{normal}^+$ to represent the normal distribution truncated to be positive), a weak prior (for example, $\mbox{normal}^+(0,1)$ on the percentage scale, which would allow unrealistically large day-to-day changes on the order of 0.25 percentage points in the probability of support for a candidate, but would still keep the distribution away from extreme parameter values), or a more informative prior such as $\mbox{normal}^+(0,0.1)$ which does not encapsulate all our prior knowledge but does softly constrain this parameter to within a reasonable range.  Our point here is that in choosing a prior, one is also choosing the amount of subject-matter information to include in the analysis.

Another way to view a prior distribution, or a statistical model more generally, is as a {\em constraint}.  For example, if we fit a linear model plus a spline or Gaussian process, $y=b_0 + b_1 x + g(x) + \mbox{error}$, where the nonlinear function $g$ has bounded variation, then with a strong enough prior on the $g$, we are fitting a curve that is close to linear. The prior distribution in this example could represent prior information, or it could just be considered as part of the model specification, if there is a desire to fit an approximately linear curve.   Simpson et al.\  (2017) provide further discussion on using prior distributions to shrink towards simpler models. This also leads to the more general point that priors are just like any other part of a statistical model in which they can serve different purposes. Any clear distinction between model and prior is largely arbitrary and often depends mostly on the conceptual background of the one making the distinction. 

The amount of prior information needed to get reasonable inference depends strongly on the role of the parameter in the model as well as the depth of the parameter in the hierarchy (Goel and DeGroot, 1981). For instance, parameters that mainly control central quantities (such as the mean or the median) typically tolerate vague priors more than scale parameters, which again are more forgiving of vague priors than parameters that control tail quantities, such as the shape parameter of a generalized extreme value distribution. When a model has a hierarchical structure, parameters that are closer to the data are typically more willing to indulge vague priors than parameters further down the hierarchy.

In Bayesian workflow, priors are needed for a sequence of models. Often as the model becomes more complex, all of the priors need to become tighter. The following simple example of multilevel data (see, for example, Raudenbush and Bryk, 2002) shows why this can be necessary.

Consider a workflow where the data are $y_{ij}$, $i=1,\ldots,n_j, \ j = 1,\ldots, J$. Here $i$ indexes the observation and $j$ indexes the group. Imagine that for the first model in the workflow we elect to ignore the group structure and use a simple normal distribution for the deviation from the mean. In this case some of the information budget will be spent on estimating the overall mean and some of it is spent on the observation standard deviation. If we have a moderate amount of data, the mean will be approximately $\bar{y} = \sum_{i=1}^n y_i/n$ regardless of how weak the prior is. Furthermore, the predictive distribution for a new observation will be approximately $\mbox{normal}(\bar{y}, s)$, where $s$ is the sample standard deviation. Again, this will be true for most sensible priors on the observation standard deviation, regardless of how vague they are. 
 
If the next step in the workflow is to allow the mean to vary by group using a multilevel model, then the information budget still needs to be divided between the  standard deviation and the mean. However, the model now has $J+1$ extra parameters (one for each group plus one for the standard deviation across groups) so the budget for the mean needs to be further divided to model the group means, whereas the budget for the standard deviation needs to be split between the within group variation and the between group variation. But we still have the same amount of data, so we need to be careful to ensure that this model expansion does not destabilize our estimates. This means that as well as putting appropriate priors on our new parameters, we probably need to tighten up the priors on the overall mean and observation standard deviation, lest a lack of information lead to nonsense estimates.

\begin{figure}
 \centerline{\includegraphics[height=2in,angle=0]{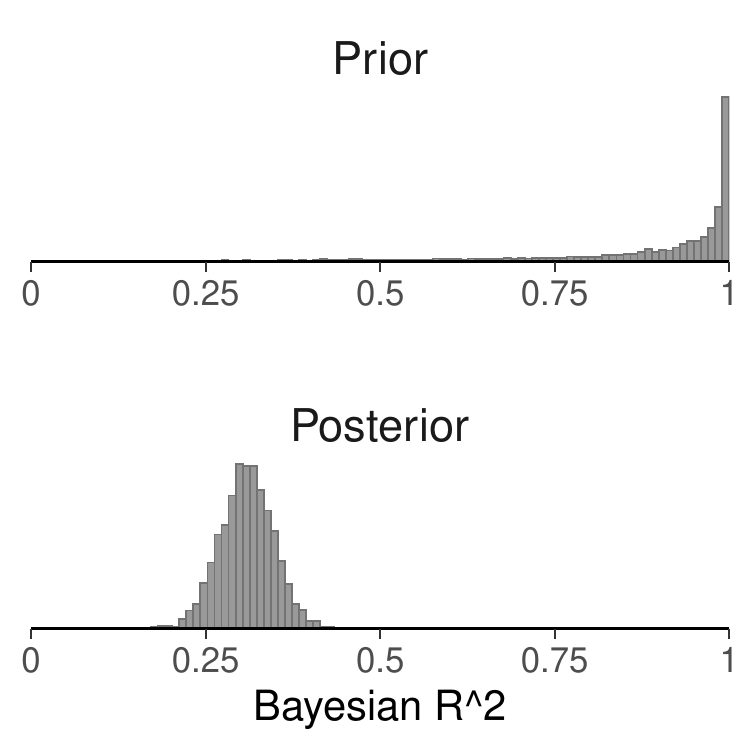}
\hspace{.02in}\includegraphics[height=2in,angle=0]{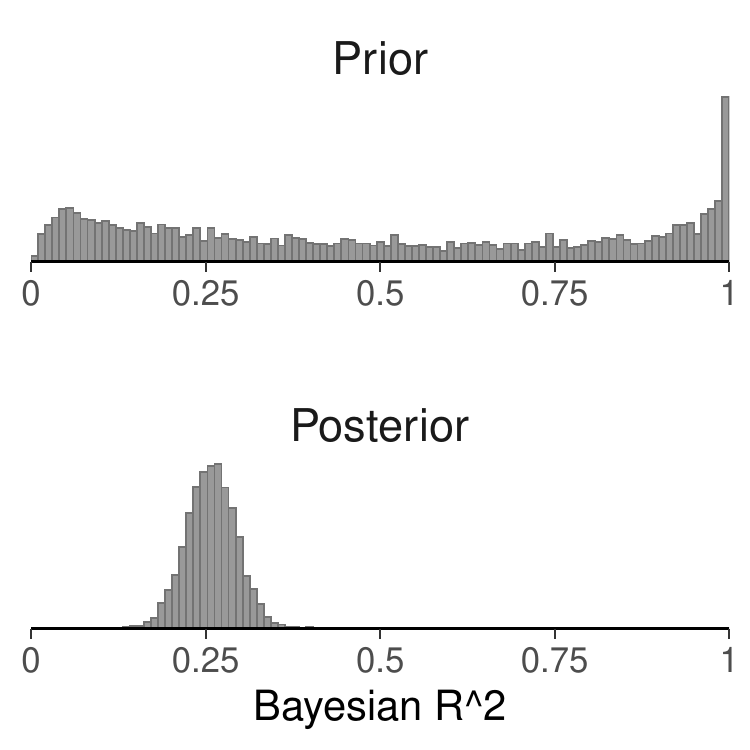}
\hspace{.02in}\includegraphics[height=2in,angle=0]{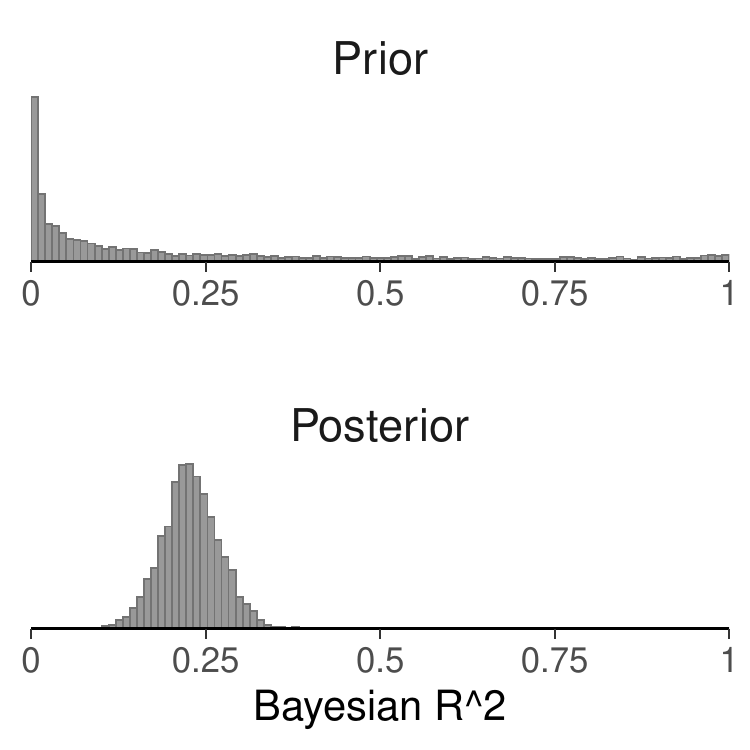}}
\vspace{-.15in}
\caption{\em Prior and posterior distribution of Bayesian $R^2$ for the regression predicting student grades from $26$ predictors, using three different priors for the coefficients: (a) default weak prior, (b) normal prior scaled with the number
  of predictors, and (c) regularized horseshoe prior.  From Section 12.7 of Gelman, Hill, and Vehtari (2020).}
 \label{figure.student.R2.1.2.3}
 \end{figure}

A related issue is the concentration of measure in higher-dimensional space. For example in regression with an increasing number of predictors, the prior on the vector of coefficients needs to have higher density near the mode if we want to keep most of the prior mass near the mode (as the volume further away from the mode increases faster as dimension increases; see, for example, Piironen and Vehtari, 2017).  Consider linear or logistic regression and what happens to the prior on $R^2$ if the marginal prior on weights is fixed. If we want the prior on $R^2$ to stay fixed, the prior on weights needs to get tighter. Figure
\ref{figure.student.R2.1.2.3} uses prior predictive checking (see Section \ref{sec:prior_pred_checks}) to show how the usual weak prior on $26$ weights in linear regression corresponds to a prior strongly favoring higher $R^2$ values, affecting also the posterior. From that perspective, weakly informative but independent priors may jointly be strongly informative if they appear in large numbers.

Priors must be specified for {\em each model} within a workflow.  An expanded model can require additional thought regarding parameterization.  For example, when going from $\mbox{normal}(\mu,\sigma)$ to a $t$-distribution $t_{\nu}(\mu,\sigma)$ with $\nu$ degrees of freedom, we have to be careful with the prior on $\sigma$. The scale parameter $\sigma$ looks the same for the two models, but the variance of the $t$ distribution is actually $\frac{\nu}{\nu-2}\sigma^2$ rather than $\sigma^2$. Accordingly, if $\nu$ is small, $\sigma$ is no longer close to the residual standard deviation.

In general, we need to think in terms of the {\em joint} prior over all the parameters in a model, to be assessed in the context of the generative model for the data, lest unfortunate cancellations or resonances lead to less stabilizing or more informative priors than the modeler actually wants (Gelman et al., 2017, Kennedy et al., 2019).  As discussed in Section \ref{sec:prior_pred_checks}, prior predictive checking is a good general approach to exploring and understanding a prior distribution in the context of a particular data model.

The above examples carry particular messages about priors but also a meta-message about how we think about workflow when constructing statistical models.  Phrases such as ``the information budget still needs to be divided'' represent important but somewhat informal decisions we make about how much effort we put in to include prior information. Such concerns are not always clear in articles or textbooks that present the final model as is, without acknowledging the tradeoffs and choices that have been made.  

\subsection{A topology of models} \label{sec:topology}

Consider a class of models, for simplicity in some particular restricted domain such as autoregressive moving average (ARMA) models, binary classification trees, or linear regressions with some fixed set of input variables.  The models in any of these frameworks can be structured as a partial ordering:  for example, AR(1) is simpler than AR(2) which is simpler than ARMA(2,1), and MA(1) is also simpler than ARMA(2,1), but AR(1) and MA(1) are not themselves ordered.  Similarly, tree splits form a partial ordering, and the $2^k$ possibilities of inclusion or exclusion in linear regression can be structured as the corners of a cube.  As these examples illustrate, each of these model frameworks has its own topology or network structure as determined by the models in the class and their partial ordering.

We speak of this as a topology of models rather than a probability space because we are not necessarily interested in assigning probabilities to the individual models.  Our interest here is not in averaging over models but in navigating among them, and the topology refers to the connections between models and between parameters in neighboring models in the network.

An example implementation of this idea is the Automatic Statistician (Hwang et al., 2016, Gharamani et al., 2019), which searches through models in specified but open-ended classes (for example, time series models and linear regression models), using inference and model criticism to explore the model and data space.  We believe such algorithms can be better understood and, ultimately, improved, through a more formal understanding of the topology of models induced by a statistical modeling language.  From another direction are menu-based packages such as Prophet (Taylor and Lethem, 2018) that allow users to put together models (in this case, for time series forecasting) from some set of building blocks.  It is important in such packages not just to be able to build and fit these models but to understand each model in comparison to simpler or more complicated variants fit to the same data.

However, unlike combining variables, where in many cases a simple and often automated additive model is enough, here each model itself is a high dimensional object. The outputs from different models, as probabilistic random variables, can be added, multiplied, linearly mixed, log-linearly-mixed, pointwisely-mixed, etc, which is within the choice of model topology we need to specify. 

In addition, each model within a framework has its own internal structure involving parameters that can be estimated from data.  And, importantly, the parameters within different models in the network can ``talk with each other'' in the sense of having a shared, observable meaning outside the confines of the model itself.  
In machine learning and applied statistics, two familiar examples with inferential quantities that are shared across models are forecasting and causal inference. In forecasting, an increasingly complicated set of procedures can be used for a particular predictive goal. And in causal inference, a treatment effect can be estimated using a series of regressions, starting from a simple difference and moving to elaborate interaction models adjusting for differences between observed treated and control groups.  Recall that causal inferences are a special case of predictions involving counterfactuals; see, for example, Morgan and Winship (2014).

Thus, the topology of statistical or machine-learning models includes a partial ordering of models, and connections between parameters or functions of parameters and data across different models within the larger framework.  Another twist is that prior distributions add a continuous dimension to the structure, bridging between models.

\section{Understanding and comparing multiple models}\label{comparing}

\subsection{Visualizing models in relation to each other}

The key aspect of Bayesian workflow, which takes it beyond Bayesian data analysis, is that we are fitting many models while working on a single problem. We are not talking here about model selection or model averaging but rather of the use of a series of fitted models to better understand each one.  In the words of Wickham, Cook, and Hofmann (2015), we seek to ``explore the process of model fitting, not just the end result.''   We fit multiple models for several reasons, including:

\begin{itemize}
\item It can be hard to fit and understand a big model, so we will build up to it from simple models.
\item When building our models, we make a lot of mistakes:  typos, coding errors, conceptual errors (for example not realizing that the observations don't contain useful information for some parts of the model), etc.
\item As we get more data, we typically expand our models accordingly.  For example, if we're doing pharmacology and we get data on a new group of patients, we might let certain parameters vary by group.
\item Often we fit a model that is mathematically well specified, but once we fit it to data we realize that there's more we can do, so we expand it.
\item Relatedly, when we first fit a model, we often put it together with various placeholders.  We're often starting with weak priors and making them stronger, or starting with strong priors and relaxing them.
\item We'll check a model, find problems, and then expand or replace it.  This is part of ``Bayesian data analysis''; the extra ``workflow'' part is that we still keep the old model, not for the purpose of averaging but for the purpose of understanding what we are doing.
\item Sometimes we fit simple models as comparisons.  For example, if you're doing a big regression for causal inference, you'll also want to do a simple unadjusted comparison and then see what the adjustments have done.
\item The above ideas are listed as being motivated by statistical considerations, but sometimes we're jolted into action because of computational problems.
\end{itemize}
Given that we are fitting multiple models, we also have to be concerned with researcher degrees of freedom (Simmons et al., 2011), most directly from overfitting if a single best model is picked, or more subtly that if we are not careful, we can consider our inferences from a set of fitted models to bracket some total uncertainty, without recognizing that there are other models we could have fit.  This concern arises in our election forecasting model, where ultimately we only have a handful of past presidential elections with which to calibrate our predictions.

\begin{figure}
  \centerline{\includegraphics[width=.6\textwidth]{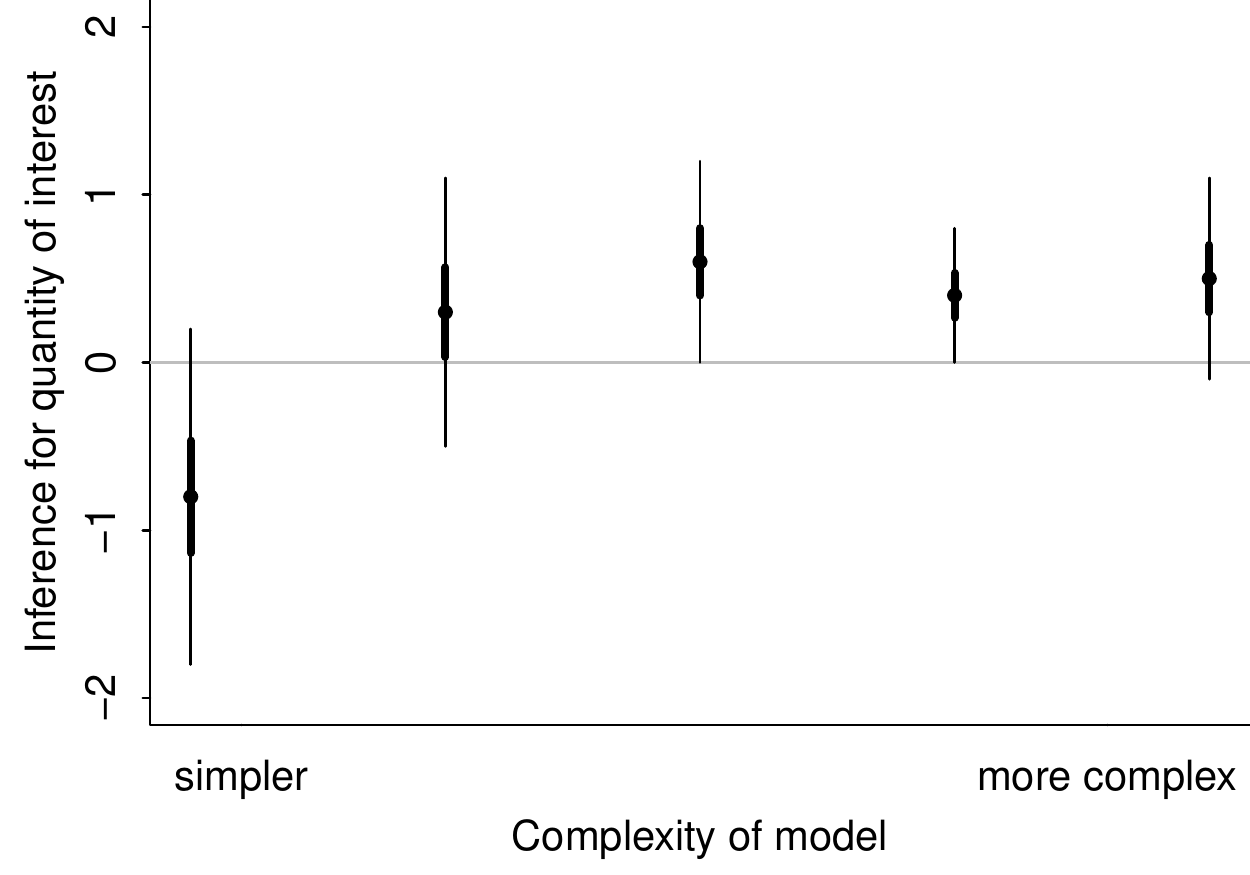}}
  \vspace{-.1in}
\caption{\em Hypothetical graph comparing inferences for a quantity of interest across multiple models.  The goal here is not to perform model selection or model averaging but to understand how inference for a quantity of interest changes as we move from a simple comparison (on the left side of the graph) through the final model (on the right side of the graph), going through various intermediate alternatives.}
\label{sequence}
\end{figure}

Figure \ref{sequence} illustrates the basic idea:  the diagram could represent, for example, a causal effect estimated with a sequence of increasingly complicated models, starting with a simple treatment-control comparison and proceeding through a series of adjustments.  Even if the ultimate interest is only in the final model, it can be useful to understand how the inference changes as adjustments are added.

\begin{figure}[ht]
    \centering
    \includegraphics[width=\textwidth]{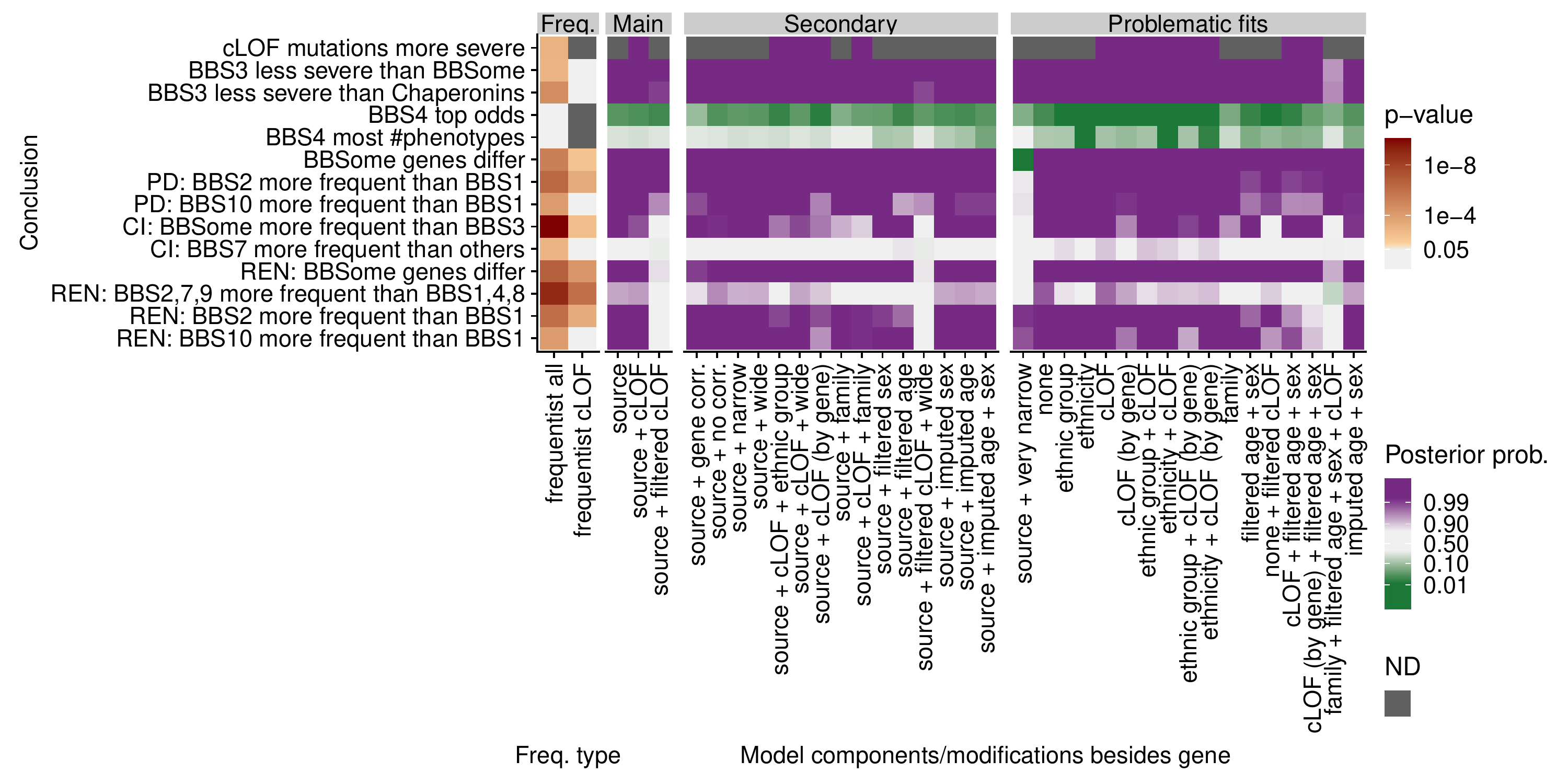}
    \caption{\em The results of a multiverse analysis from the supplementary material of Niederlová et al.\ (2019). The heat map shows statistical evaluation of selected conclusions about the relation between phenotypes (PD, CI, REN, HEART, LIV) and mutations in various genes of the BBSome (BBS01 - BBS8, cLOF - complete loss of function) using a set of Bayesian hierarchical logistic regression models and pairwise frequentist tests. Based on posterior predictive checks the Bayesian models were categorized as ``Main'' (passing all checks), ``Secondary'' (minor problems in some checks), and ``Problematic fits,'' though we see that most conclusions hold across all possible models. The models differ in both predictors that are included and priors (default/wide/narrow/very narrow).}
    \label{fig:multiverse}
\end{figure}

Following the proposed workflow and exploring the topology of models can often lead us to multiple models that pass all the checks. Instead of selecting just one model, we can perform a {\em multiverse analysis}\label{multiverse} and use all of the models and see how the conclusions change across the models (Steegen et al., 2016, Dragicevic et al., 2019, Kale, Kay, and Hullman, 2019). Using multiverse analysis can also relieve some of the effort in validating models and priors:  if the conclusions do not change it is less important to decide which model is ``best.'' Figure~\ref{fig:multiverse} shows an example of a possible output. Other analytical choices (data preprocessing, response distribution, metric to evaluate, and so forth) can also be subject to multiverse analysis.

\subsection{Cross validation and model averaging}

We can use cross validation to compare models fit to the same data (Vehtari et al., 2017). When performing model comparison, if there is non-negligible uncertainty in the comparison (Sivula et al., 2020), we should not simply choose the single model with the best cross validation results, as this would discard all the uncertainty from the cross validation process. Instead, we can maintain this information and use stacking to combine inferences using a weighting that is set up to minimize cross validation error (Yao et al., 2018b).  We think that stacking makes more sense than traditional Bayesian model averaging, as the latter can depend strongly on aspects of the model that have minimal effect on predictions.  For example, for a model that is well informed by the data and whose parameters are on unit scale, changing the prior on parameters from $\mbox{normal}(0,10)$ to $\mbox{normal}(0,100)$ will divide the marginal likelihood by roughly $10^k$ (for a model with $k$ parameters) while keeping all predictions essentially the same.  In addition, stacking takes into account the joint predictions and works well when there are a large number of similar but weak models in the candidate model list. 

In concept, stacking can sometimes be viewed as pointwise model selection. When there are two models and the first model outperforms the second model   20\% of the time, the stacking weights will be close to $(0.2, 0.8)$.  In light of this, stacking fills a gap between independent-error oriented machine learning validation and the grouped structure of modern big data. Model stacking is therefore also an indicator of heterogeneity of model fitting, and this suggests we can further improve the aggregated model with a hierarchical model, so that the stacking is a step toward model improvement rather than an end to itself.  
In the extreme, model averaging can also be done so that different models can apply to different data points (Kamary et al., 2019, Pirš and Štrumbelj, 2019).

In Bayesian workflow, we will fit many models that we will not be interested in including in any average; such ``scaffolds'' include models that are deliberately overly simple (included just for comparison to the models of interest) and models constructed for purely experimental purposes, as well as models that have major flaws or even coding errors.  But even after these mistakes or deliberate oversimplifications have been removed, there might be several models over which to average when making predictions.  In our own applied work we have not generally had many occasions to perform this sort of model averaging, as we prefer continuous model expansion, but there will be settings where users will reasonably want to make predictions averaging over competing Bayesian models, as in Montgomery and Nyhan (2010).

\subsection{Comparing a large number of models}

There are many problems, for example in linear regression with several potentially relevant predictors, where many candidate models are available, all of which can be described as special cases of a single expanded model.  If the number of candidate models is large, we are often interested in finding a comparably smaller model that has the same predictive performance as our expanded model. This leads to the problem of predictor (variable) selection. If we have many models making similar predictions, selecting one of these models based on minimizing cross validation error would lead to overfitting and suboptimal model choices (Piironen and Vehtari, 2017). In contrast, projection predictive variable selection has been shown to be stable and reliable in finding smaller models with good predictive performance (Piironen and Vehtari, 2017, Piironen et al., 2020, Pavone et al., 2020). While searching through a big model space is usually associated with the danger of overfitting, the projection predictive approach avoids it by examining only the projected submodels based on the expanded model's predictions and not fitting each model independently to the data. In addition to variable selection, projection predictive model selection can be used for structure selection in generalized additive multilevel models (Catalina et al., 2020) and for creating simpler explanations for complex nonparametric models (Afrabandpey et al., 2020).


\section{Modeling as software development}\label{sec:modeling_a_software}

Developing a statistical model in a probabilistic programming language means writing code and is thus a form of software development, with several stages:  writing and debugging the model itself; the preprocessing necessary to get the data into suitable form to be modeled; and the later steps of understanding, communicating, and using the resulting inferences.   Developing software is hard. So many things can go wrong because there are so many moving parts that need to be carefully synchronized.

Software development practices are designed to mitigate the problems caused by the inherent complexity of writing computer programs. Unfortunately, many methodologies veer off into dogma, bean counting, or both.  A couple references we can recommend that provide solid, practical advice for developers are Hunt and Thomas (1999) and McConnell (2004).

\subsection{Version control smooths collaborations with others and with your past self}

Version control software, such as Git, should be should be the first
piece of infrastructure put in place for a project. It may seem like a big investment to learn version control, but it's well worth it to be able to type a single command to revert to a previously working version or to get the difference between the current version and an old version.  It's even better when you need to share work with others, even on a paper---work can be done independently and then automatically merged. While version control keeps track of smaller changes in one model, it is useful to keep the clearly different models in different files to allow easy comparison of the models. Version control also helps to keep notes on the findings and decisions in the iterative model building,  increasing transparency of the process. 

Version control is not just for code. It is also for reports, graphs, and data. Version control is a critical part of ensuring that all of these 
components are synchronized and, more importantly, that it is possible to rewind the project
to a previous state. Version control is particularly useful for its ability to
package up and label ``release candidate'' versions of models and data that correspond to milestone reports and publications
and to store them in the same directory without resorting to the dreaded \verb|_final_final_161020.pdf|-style
naming conventions.

When working on models that are used to guide policy decision making, a public version control repository increases transparency about what model, data, inference parameters, and scripts were used for specific reports. An excellent example of this is the Imperial College repository for models and scripts to estimate deaths and cases for COVID-19 (Flaxman et al., 2020). 

\subsection{Testing as you go}

Software design ideally proceeds top down from the goals of the end user back to the technical machinery required to implement it.  For a Bayesian statistical model, top-down design involves at least the data input format, the probabilistic model for the data, and the prior, but may also involve simulators and model testing like simulation-based calibration or posterior predictive checks.  
Software development ideally works bottom up from well-tested foundational functions to larger functional modules.  That way, development proceeds through a series of well-tested steps, at each stage building only on tested pieces.  The advantage to working this way as opposed to building a large program and then debugging it is the same as for incremental model development---it's easier to track where the development went wrong and you have more confidence at each step working with well-tested foundations.  

The key to computational development, either in initial development or modifying code, is modularity.  Big tangled functions are hard to document, harder to read, extraordinarily difficult to debug, and nearly impossible to maintain or modify.  Modularity means building larger pieces out of smaller trusted pieces, such as low-level functions.  Whenever code fragments are repeated, they should be encapsulated as functions.  This results in code that is easier to read and easier to maintain.

As an example of a low-level function, predictors might be rescaled for a generalized linear model by implementing the standardization, $z(v) = (v - \textrm{mean}(v)) / \textrm{sd}(v)$.  Although this function seems simple, subtleties arise, starting with the sd function, which is sometimes defined as $\textrm{sd}(v) = \sqrt{\sum_{i=1}^n(v_i - \textrm{mean}(v))^2/n}$ and sometimes as $\textrm{sd}(v) = \sqrt{\sum_{i=1}^n(v_i - \textrm{mean}(v))^2/(n-1)}$.  If this isn't sorted out at the level of the standardization function, it can produce mysterious biases during inference.  Simple tests that don't rely on the $\textrm{sd}()$ function will sort this out during function development.  If the choice is the estimate that divides by $n-1$, there needs to be decision of what to do when $v$ is a vector of length $1$.  In  cases where there are illegal inputs, it helps to put checks in the input-output routines that let users know when the problem arises rather than allowing errors to percolate through to mysterious divide-by-zero errors later.  

An implementation of cubic splines or an Euler solver for a differential equation is an example of a higher-level function that should be tested before it is used.  As functions get more complicated, they become harder to test because of issues with boundary-condition combinatorics, more general inputs such as functions to integrate, numerical instability or imprecision over regions of their domain which may or may not be acceptable depending on the application, the need for stable derivatives, etc.

\subsection{Making it essentially reproducible}

A lofty goal for any project is to make it fully reproducible in the sense
that another person on another machine could recreate the final report.
This is not the type of reproducibility that is considered in scientific fields,
where the desire is to ensure that an effect is confirmed by new future data (nowadays often called ``replicability'' for a better distinction between the different notions).
Instead this is the more limited (but still vital) goal of ensuring that 
one particular analysis is consistently done. In particular, we would want
to be able to produce analyses and figures that are essentially equivalent to the
original document. Bit-level reproducibility may not be possible, but we would still liken equivalence at a practical
level. In the event that this type of reproduction 
changes the outcome of a paper, we would argue that the original 
results were not particularly robust.

Rather than entering commands on the command line when running models (or
entering commands directly into an interactive programming language like R or
Python), write scripts to run the data through the models and produce
whatever posterior analysis you need. Scripts can be written for the shell, R,
Python, or any other programming language.  The script should be self-contained in the sense that it should run
in a completely clean environment or, ideally, on a different computer. 
This means that the script(s) must
not depend on global variables having been set, other data being read in, or anything
else that is not in the script. 

Scripts are good documentation. It may seem like overkill if running
the project is only a single line of code, but the script provides not
only a way to run the code, but also a form of concrete documentation
for what is being run. For complex projects, we often find that a 
well-constructed series of scripts can be more practical than one large
R Markdown document or Jupyter notebook.

Depending on long-term reproducibility needs, it's important to
choose the right tooling for the job at hand. To guarantee bit-level
reproducibility, and sometimes even just to get a program to run, everything from
hardware, to the operating system, to every piece of software and setting must be
specified with their version number. As time passes between the initial writing of the script and the attempt of reproduction, bit-level
reproducibility can be almost impossible to achieve even if the environment is shipped with the script, as in a Docker container.

\subsection{Making it readable and maintainable}

Treating programs and scripts like other forms of writing for an
audience provides an important perspective on how the code will be
used.  Not only might others want to read and understand a program or model, the
developers themselves will want to read and understand it later.  One of the motivations of
Stan's design was to make models self-documenting in terms of variable
usage (e.g., data or parameters), types (e.g., covariance matrix
or unconstrained matrix) and sizes. This allows us to understand Stan code (or code of other statically typed probabilistic programming languages) to be understandable without the context of the data it is applied on. 

A large part of readability is consistency, particularly in naming
and layout, and not only of programs themselves, but the directories and
files in which they are stored.  Another key principle in coding is to avoid repetition, instead pulling shared code out into functions that can be reused.

Readability of code is not just about
comments---it is also about naming and organization for readability.  Indeed, comments can made code {\em less} readable.  The best
approach is to write readable code, not opaque code with comments.  For example,
we don't want to write this:\\
\vspace{-\baselineskip}
\begin{verbatim}
  real x17;  // oxygen level, should be positive
\end{verbatim}
when we can write this:\\
\vspace{-\baselineskip}
\begin{verbatim}
  real<lower = 0> oxygen_level;
\end{verbatim}
Similarly, we don't want to do this:\\
\vspace{-\baselineskip}
\begin{verbatim}
  target += -0.5 * (y - mu)^2 / sigma^2;  // y distributed normal(mu, sigma)
\end{verbatim}
when we can write,\\
\vspace{-\baselineskip}
\begin{verbatim}
  target += normal_lpdf(y | mu, sigma);
\end{verbatim}
Good practice is to minimize inline code comments and instead write readable code.  As the above examples illustrate, clean code is facilitated by programming languages that gives users the tools they need to use.

User-facing functions should be documented at the function level in terms of their argument types, return types, error conditions, and behavior---that's the application programming interface (API) that users see instead of code internals. The problem with inline code comments aimed at developers is that they quickly go stale during development and wind up doing more harm than good.  Instead, rather than documenting the actual code inline, functions should be reduced to manageable size and names should be chosen so that the code is readable.  Longer variable names are not always better, as they can make the structure of the code harder to scan.  Code documentation should be written assuming the reader understands the programming language well; as such, documentation is only called for when the code strays from idiomatic usage of the language or involves complex algorithms that need external documentation.  When tempted to comment a long expression or block of code, instead consider replacing it with a well-named function.

Related to readability is the maintainability of the workflow code. When fitting a series of similar models, a lot of modules will be shared between them (see Section~\ref{sec:modular}) and so will be the corresponding code. If we had just copied all of the model code each time we wrote a new model, and then discovered an error in one of the shared modules, we would have to fix it in all models manually. This is again an error-prone process. Instead, it can be sensible not only to build models in a modular manner but also keep the corresponding code modular and load it into the models as needed. That way, fixing an error in a module requires changing code in just one rather than many places. Errors and other requirements for later changes will inevitably occur as we move through the workflow and it will save us a lot of time if we prepare our modeling code accordingly.

\section{Example of workflow involving model building and expansion: Golf putting}\label{golf}

We demonstrate the basic workflow of Bayesian modeling using an example of a set of models fit to data on golf putting (Gelman, 2019).

\begin{figure}
    \centering
    \includegraphics[width=.7\textwidth]{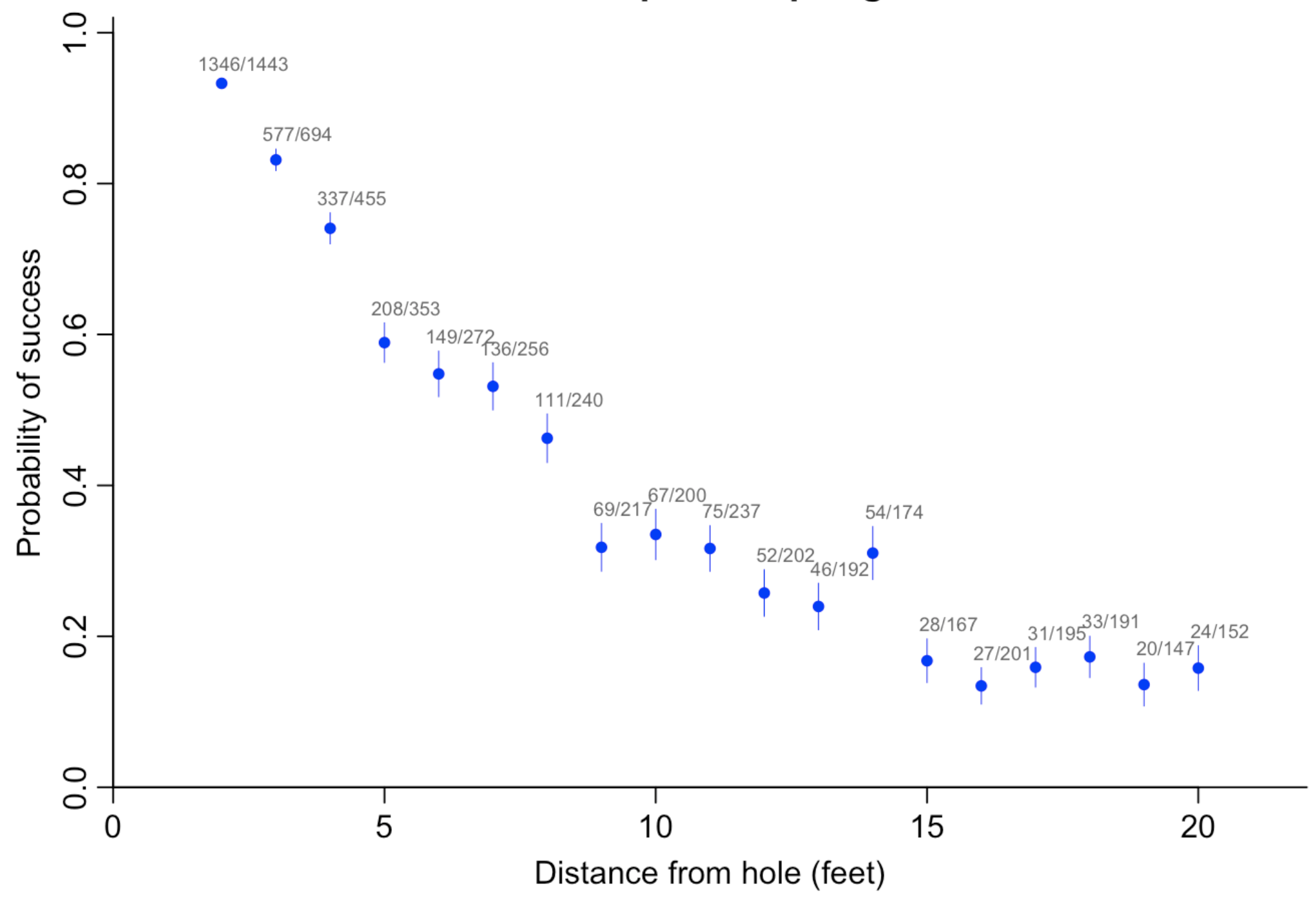}
    \caption{\em Success rate of putts of professional golfers, from a small dataset appearing in Berry (1995). The error bars associated with each point $j$
 in the  graph are simple classical standard deviations, $\sqrt{\hat{p}_j(1-\hat{p}_j)/n_j}$, where $\hat{p}_j=y_j/n_j$ is the success rate for putts taken at distance $x_j$.}
    \label{golf1}
\end{figure}

Figure \ref{golf1} shows data from professional golfers on the proportion of successful putts as a function of (rounded) distance from the hole. Unsurprisingly, the probability of making the shot declines as a function of distance.

\subsection{First model:  logistic regression}

Can we model the probability of success in golf putting as a function of distance from the hole? Given usual statistical practice, the natural starting point would be logistic regression:
$$ y_j \sim \mbox{binomial}\left(n_j,\mbox{logit}^{-1}(a+bx_j)\right), \mbox{ for }j=1, \dots, J.$$

\begin{figure}
    \centering
    \includegraphics[width=.7\textwidth]{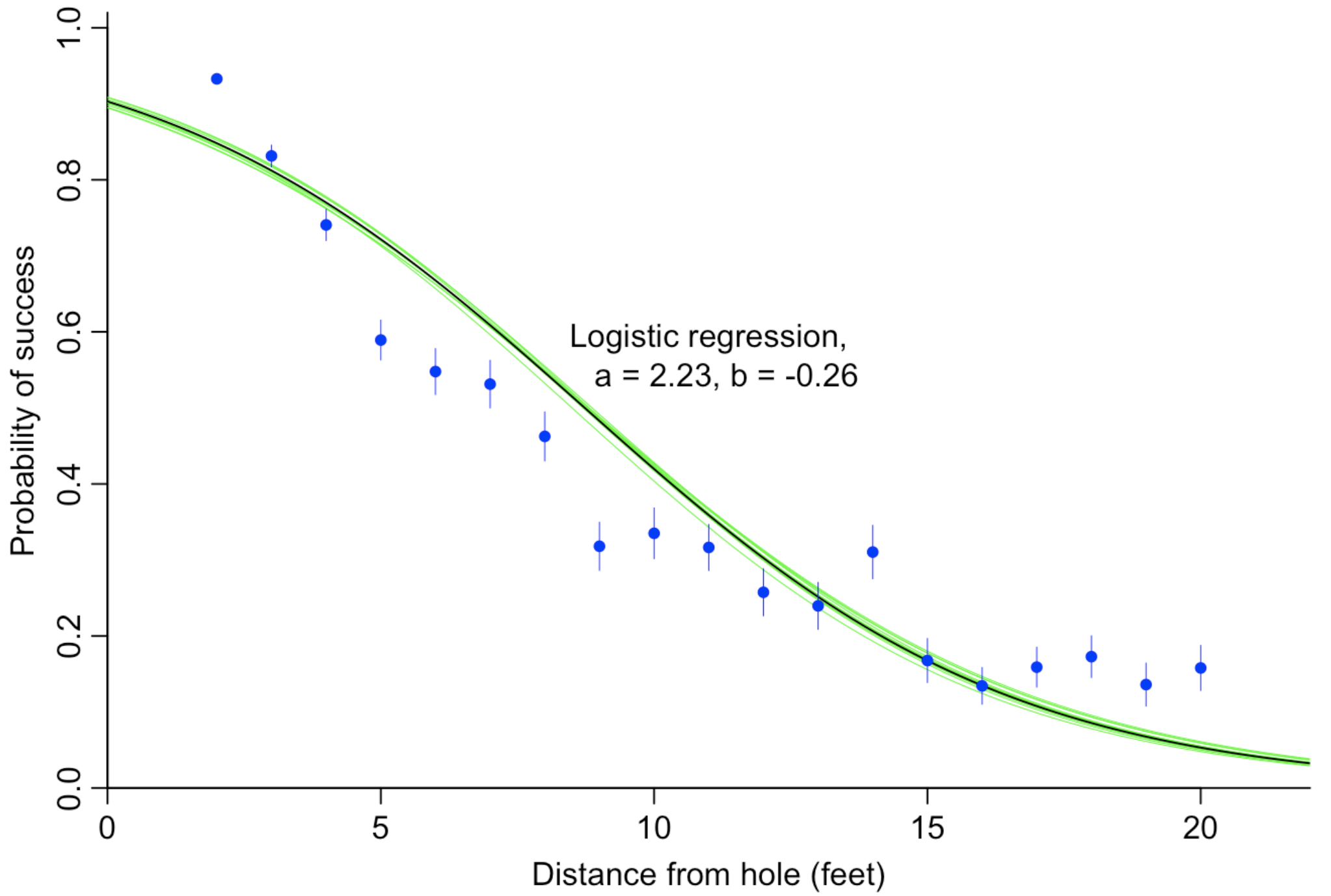}
    \caption{\em Golf data from Figure \ref{golf1} along with fitted logistic regression and draws of the fitted curve, $y =\mbox{logit}^{-1}(a + bx_j)$, given posterior draws of $(a,b)$.}
    \label{golf2}
\end{figure}

Figure \ref{golf2} shows the logistic fitted regression along with draws form the posterior distribution.  Here we fit using a uniform prior on $(a,b)$ which causes no problems given the large sample size.

\subsection{Modeling from first principles}

\begin{figure}
    \centering
    \includegraphics[width=.8\textwidth]{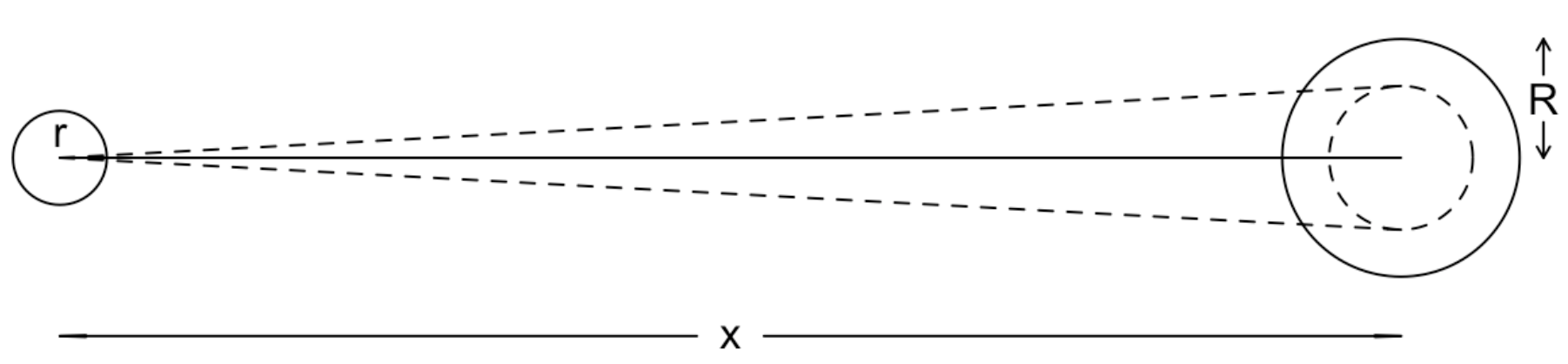}
    \caption{\em Simple geometric model of golf putting, showing the range of angles with which the ball can be hit and still have a trajectory that goes entirely into the hole. Not to scale.}
    \label{golf3}
\end{figure}

We next fit the data using a simple mathematical model of the golf putting process.  Figure \ref{golf3} shows a simplified sketch of a golf shot. The dotted line represents the angle within which the ball of radius $r$ must be hit so that it falls within the hole of radius $R$. This threshold angle is $\sin^{-1}((R-r)/x)$. The graph is intended to illustrate the geometry of the ball needing to go into the hole.

The next step is to model human error. We assume that the golfer is attempting to hit the ball completely straight but that many small factors interfere with this goal, so that the actual angle follows a normal distribution centered at 0 with some standard deviation $\sigma$.

The probability the ball goes in the hole is then the probability that the angle is less than the threshold; that is, $$\mbox{Pr}(|\mbox{angle}|<\sin^{-1}((R-r)/x))=2\Phi\left(\frac{\sin^{-1}((R-r)/x)}{\sigma}\right)-1,$$
where $\Phi$
 is the cumulative normal distribution function. The only unknown parameter in this model is $\sigma$, the standard deviation of the distribution of shot angles.
 
 \begin{figure}
    \centering
    \includegraphics[width=.7\textwidth]{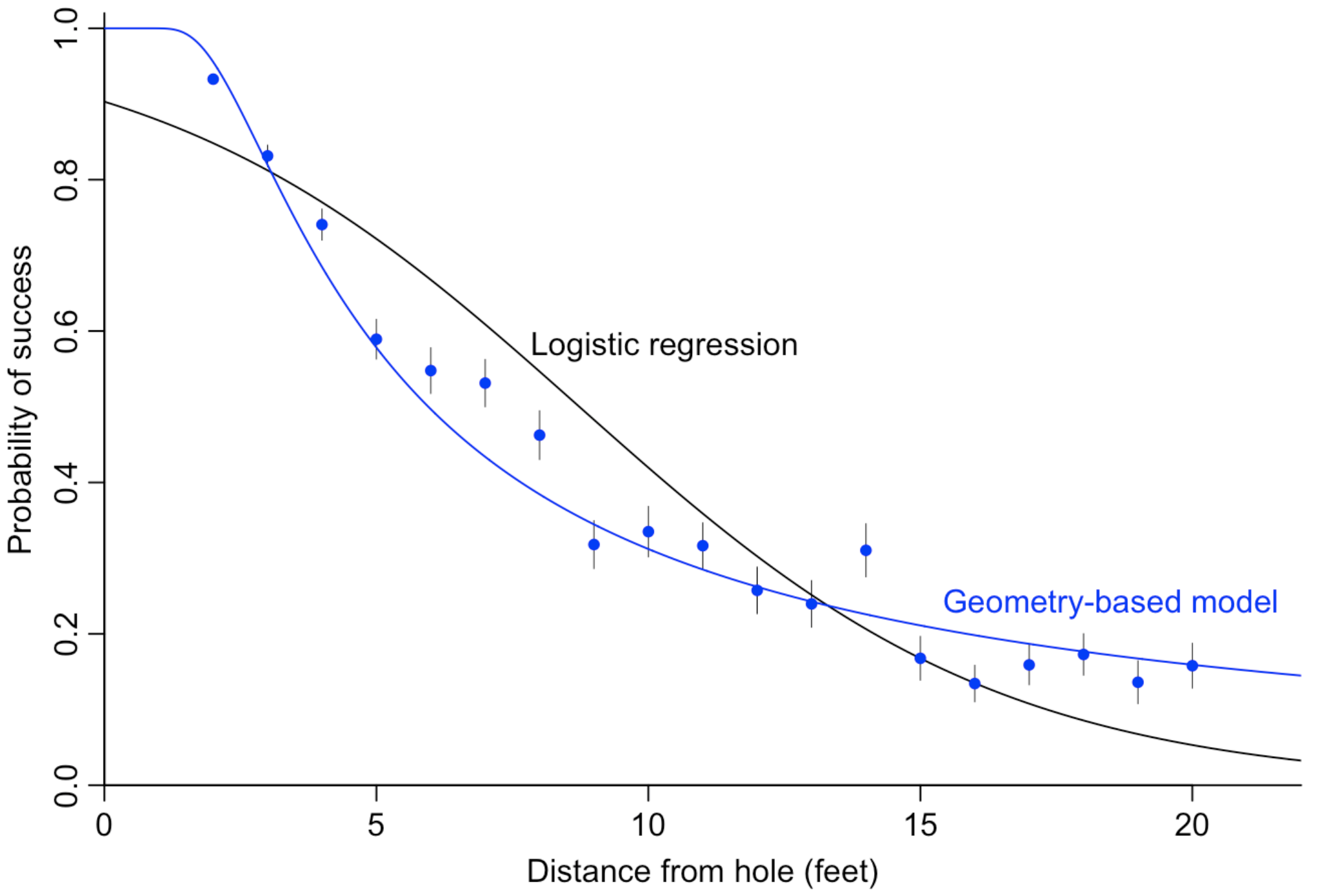}
    \caption{\em Two models fit to the golf data.  The geometry-based model fits much better than the logistic regression even while using one fewer parameter.}
    \label{golf5}
\end{figure}

Fitting this model to the above data with a flat prior on $\sigma$ yields a posterior estimate $\hat{\sigma}=1.53^{\circ}$ with standard error 0.02.  Figure \ref{golf5} shows the fitted model, along with the logistic regression fit earlier. The custom nonlinear model fits the data much better. This is not to say that the model is perfect---any experience of golf will reveal that the angle is not the only factor determining whether the ball goes in the hole---but it seems like a useful start, and it demonstrates the advantages of building up a model directly rather than simply working with a conventional form.

\subsection{Testing the fitted model on new data}

\begin{figure}
    \centering
    \includegraphics[width=.7\textwidth]{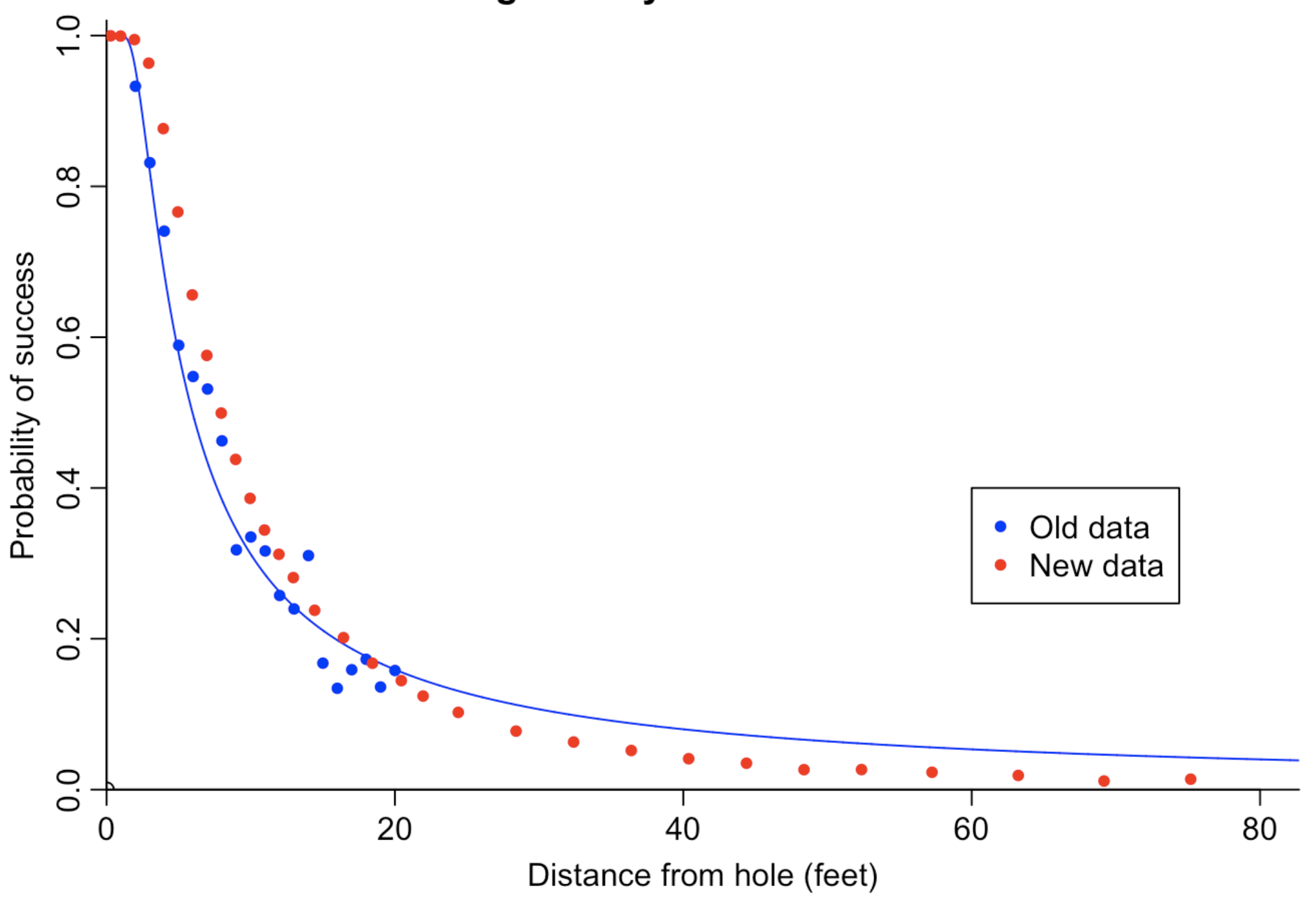}
    \caption{\em Checking the already-fit model to new golf putting data.  At equivalent distances, the success rate is higher for the new data, which may represent improvement over time or just a difference in the datasets.  In addition, the new data show a systematic model failure at higher distances, motivating a model improvement.}
    \label{golf6}
\end{figure}

Several years after fitting the above model, we were presented with a newer and more comprehensive dataset on professional golf putting (Broadie, 2018). For simplicity we just look here at the summary data, probabilities of the ball going into the hole for shots up to 75 feet from the hole. Figure \ref{golf6} these new data, along with our earlier dataset and the already-fit geometry-based model from before, extending to the range of the new data.

Comparing the two datasets in the range 0--20 feet, the success rate is similar for longer putts but is much higher than before for the short putts. This could be a measurement issue, if the distances to the hole are only approximate for the old data, and it could also be that golfers are better than they used to be.

Beyond 20 feet, the empirical success rates become lower than would be predicted by the old model. These are much more difficult attempts, even after accounting for the increased angular precision required as distance goes up.  In addition, the new data look smoother, which perhaps is a reflection of more comprehensive data collection.

\subsection{A new model accounting for how hard the ball is hit}

To get the ball in the hole, the angle is not the only thing you need to control; you also need to hit the ball just hard enough.

Broadie (2018) added this to the geometric model by introducing another parameter corresponding to the golfer’s control over distance. Supposing $u$ is the distance that golfer’s shot would travel if there were no hole, Broadie assumes that the putt will go in if (a) the angle allows the ball to go over the hole, and (b) $u$ is in the range $[x,x+3]$. That is the ball must be hit hard enough to reach the whole but not go too far. Factor (a) is what we have considered earlier; we must now add factor (b).

\begin{figure}
    \centering
    \includegraphics[width=.8\textwidth]{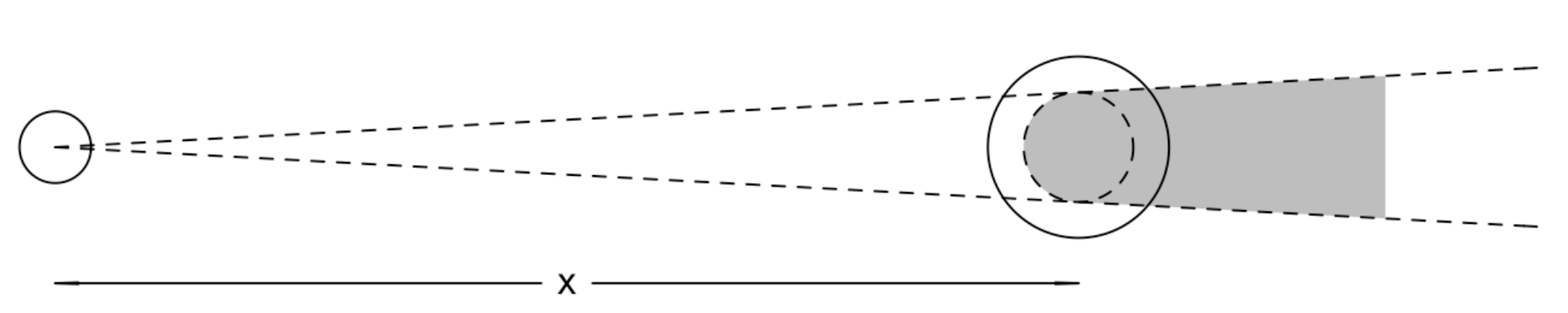}
    \caption{\em Geometric model of golf putting, also including the constraint that the ball must be hit hard enough to reach the hole but not so hard that it hops over. Not to scale.}
    \label{golf7}
\end{figure}

Figure \ref{golf7} illustrates the need for the distance as well as the angle of the shot to be in some range, in this case the gray zone which represents the trajectories for which the ball would reach the hole and stay in it.

Broadie supposes that a golfer will aim to hit the ball one foot past the hole but with a multiplicative error in the shot’s potential distance, so that $u=(x+1)\cdot(1+\epsilon)$, where the error $\epsilon$ has a normal distribution with mean 0 and standard deviation $\sigma_{\rm distance}$. In statistics notation, this model is,
$u\sim\mbox{normal}(x+1,(x+1)\sigma_{\rm distance})$,
and the distance is acceptable if $u\in [x,x+3]$,
an event that has probability $\Phi\left(\frac{2}{(x+1)\sigma_{\rm distance}}\right)-\Phi\left(\frac{-1}{(x+1)\sigma_{\rm distance}}\right).$  
Putting these together, the probability a shot goes in becomes, 
$\left(2\Phi\left(\frac{\sin^{-1}((R-r)/x)}{\sigma_{\rm angle}}\right)-1\right)\left(\Phi\left(\frac{2}{(x+1)\sigma_{\rm distance}}\right)-\Phi\left(\frac{-1}{(x+1)\sigma_{\rm distance}}\right)\right),$ where we have renamed the parameter $\sigma$ from our earlier model to $\sigma_{\rm angle}$ to distinguish it from the new $\sigma_{\rm distance}$ parameter.  

The result is a model with two parameters, $\sigma_{\rm angle}$ and $\sigma_{\rm distance}$. Even this improved geometry-based model is a gross oversimplification of putting, and the average distances in the binned data are not the exact distances for each shot.  But it should be an advance on the earlier one-parameter model; the next step is to see how it fits the data.

We first try to fit this model with flat priors, but the result is computationally unstable, so we assign weakly informative half-normal$(0,1)$ priors. Even after this, we have poor convergence.  Running 4 chains, each with 2000 iterations, yields high values of $\widehat{R}$, indicating poor mixing and making us concerned about the model, following the folk theorem (see Section \ref{folk}).

In this case, rather than examining traceplots and otherwise studying the pathologies of the Markov chain simulation, we just directly to examine the fit of the model, as estimated using the crude estimates of the parameters obtained by averaging the simulations over the poorly-mixing chains.

\begin{figure}[ht]
\centerline{
    \begin{minipage}{.7\textwidth}
    \includegraphics[width=\textwidth]{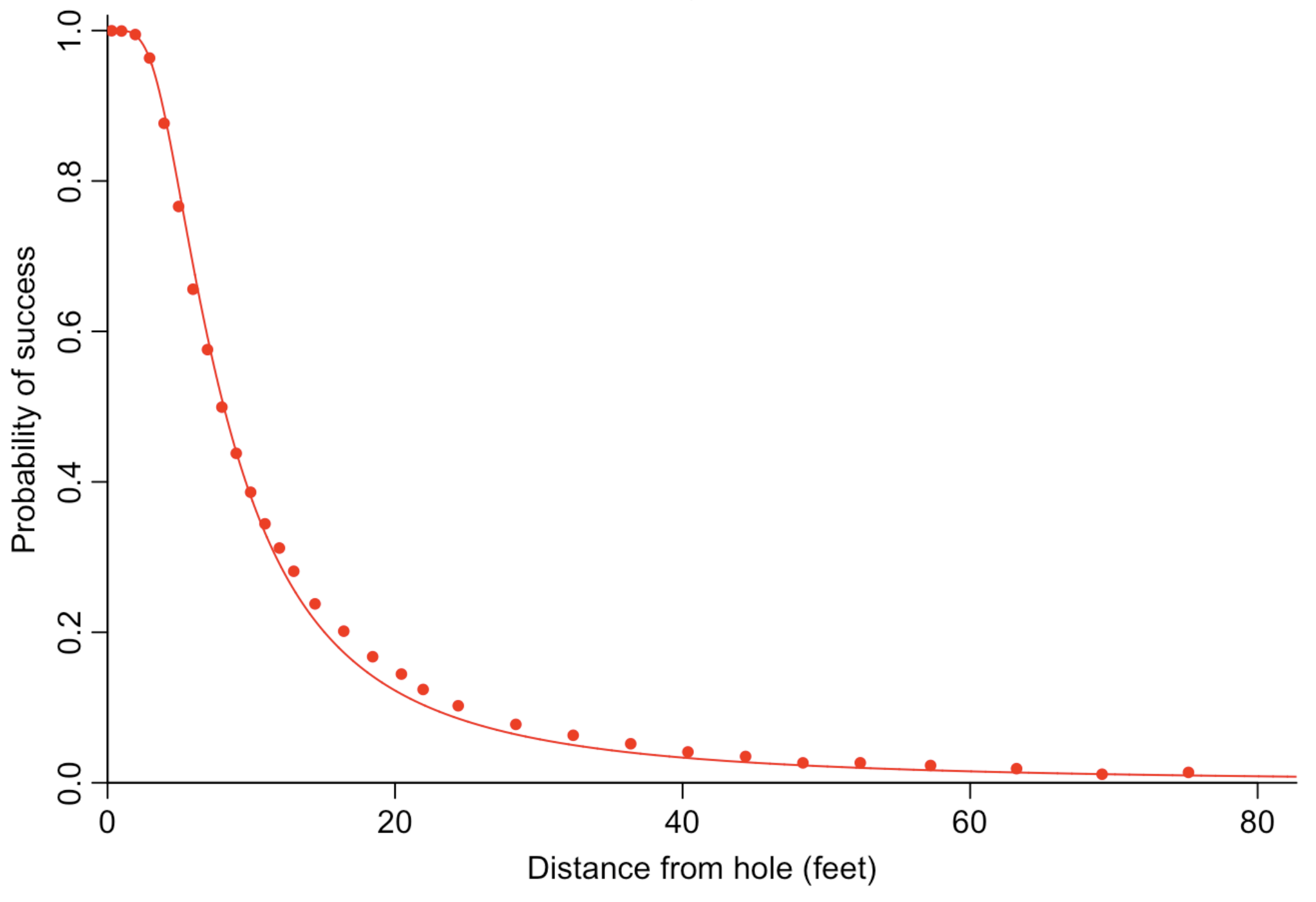}
    \end{minipage}
    \begin{minipage}{.3\textwidth}
    \begin{small}
    \begin{tabular}{rrr}
    $x$ & $n$ & $y$ \\\hline
0.28 & 45198  & 45183\\
0.97 & 183020 & 182899\\
1.93 & 169503 & 168594\\
2.92 & 113094 & 108953\\
3.93 & 73855 & 64740 \\
\dots & \dots & \dots
    \end{tabular}
    \end{small}
    \end{minipage}
    }
    \caption{\em Working to understand the poor convergence of the expanded golf putting model that had showed poor convergence.  (a) A graph of data and fitted model reveals problems with the fit near the middle of the curve, and we realized that the poor behavior of the Markov simulation arose from the model trying too hard to fit the data at the upper left of the curve. (b) Data for putts from the shortest distances, with $x$ being the average distance for the putts in each bin (presumably 0--0.5 feet, 0.5--1.5 feet, 1.5--2.5, etc.).  Sample sizes are very large in the initial bins, hence the binomial model tries to fit these points nearly exactly.}
    \label{golf8}
\end{figure}

Figure \ref{golf8}a shows the result.  The overall fit is not terrible, but there are problems in the middle of the curve, and after some thought we realized that the model is struggling because the binomial likelihood is constraining it too strongly at the upper left part of the curve where the counts are higher.  Look at how closely the fitted curve hugs the data at the very lowest values of $x$.

Figure \ref{golf8}b displays the data, which was given to us in binned form, for putts from the shortest distances.  Because of the vary large sample sizes, the binomial model tries very hard to fit these probabilities as exactly as possible.  The likelihood function gives by far its biggest weight to these first few data points.  If we were sure the model was correct, this would be the right thing to do, but given inevitable model error, the result is a problematic fit to the entire curve.  In addition, the poor MCMC convergence is understandable: there are no parameter values that fit all the data, and it is difficult for the chains to move smoothly between values that fit the bulk of the data and those that fit the first few data points.

\subsection{Expanding the model by including a fudge factor}\label{fudge}

As the data are binned, the individual putt distances have been rounded to the bin center values, which has the biggest effect in very short distances. We could incorporate a rounding error model for the putting distances, but we opt for a simpler error model.
To allow for a model that can fit reasonably well to all the data without being required to have a hyper-precise fit to the data at the shortest distances, we took the data model, $y_j\sim\mbox{binomial}(n_j,p_j)$, and added an independent error term to each observation. There is no easy way to add error directly to the binomial distribution---we could replace it with its overdispersed generalization, the beta-binomial, but this would not be appropriate here because the variance for each data point $j$ would still be roughly proportional to the sample size $n_j$, and our whole point here is to get away from that assumption and allow for model misspecification---so instead we first approximate the binomial data distribution by a normal and then add independent variance; thus:
$$y_j/n_j\sim\mbox{normal}\left(p_j,\sqrt{p_j(1-p_j)/n_j+\sigma_y^2}\right).$$  This model has its own problems and would fall apart if the counts in any cell were small enough, but it is transparent and easy to set up and code, and so we try it out, with the understanding that we can clean it up later on if necessary.

After assigning independent half-normal$(0,1)$ priors to all three parameters of this new model, it fits with no problem in Stan, yielding the posterior mean estimates $\sigma_{\rm angle}=1.02^{\circ}$, $\sigma_{\rm distance}=0.08$ (implying that shots can be hit to an uncertainty of about 8\% in distance), and $\sigma_y=0.003$ (implying that the geometric model sketched in figure \ref{golf7} fits the aggregate success rate as a function of distance to an accuracy of 0.3 percentage points).

\begin{figure}
 \centerline{
    \includegraphics[width=.5\textwidth]{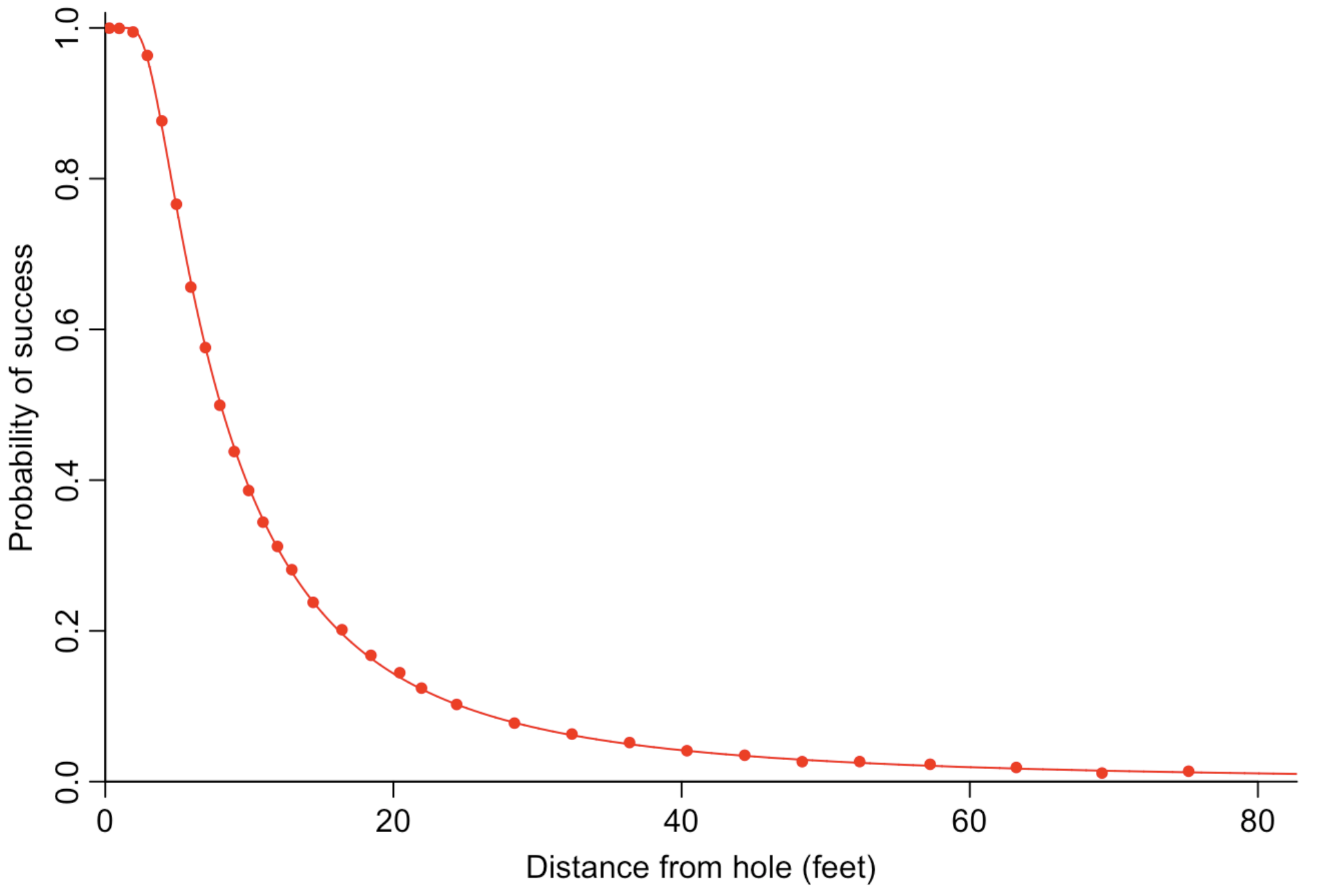}\includegraphics[width=.5\textwidth]{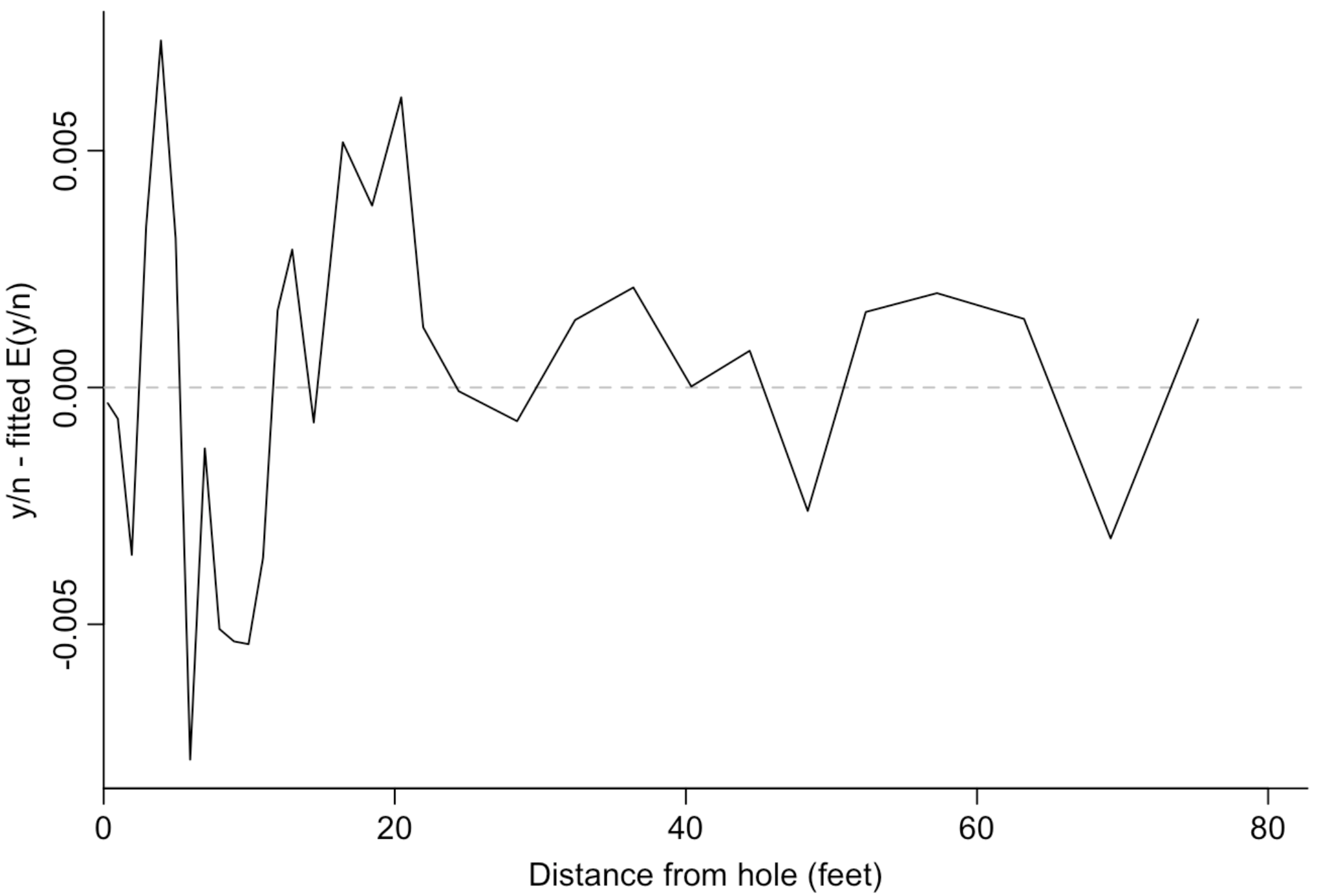}}
    \caption{\em (a) With the additional error term added, the model sketched in Figure \ref{golf7} provides an excellent fit to the expanded golf putting data. (b) The residuals from the fitted model are small and show no pattern, thus we see no obvious directions of improvement at this point.}
    \label{golf9}
\end{figure}

Figure \ref{golf9} shows the fitted model and the residuals, $y_j/n_j-\hat{p}_j$, as a function of distance. The fit is good, and the residuals show no strong pattern, also they are low in absolute value---the model predicts the success rate to within half a percentage point at most distances, suggesting not that the model is perfect but that there are no clear ways to develop it further just given the current data.

There are various ways the model could be improved, most obviously by breaking down the data and allowing the two parameters to vary by golfer, hole, and weather conditions.  As discussed earlier, a key motivation for model expansion is to allow the inclusion of more data, in this case classifying who was taking the shot and where.

\subsection{General lessons from the golf example}

This was an appealing example in that a simple one-parameter model fit the initial dataset, and then the new data were fit by adding just one more parameter to capture uncertainty in distance as well as angle of the shot.  A notable feature of the model expansion was that the binomial likelihood was too strong and made it difficult for the new model to fit all the data at once.  Such problems of stickiness---which appear in the computation as well as with the inference---are implicit in any Bayesian model, but they can become more prominent when as sample size increases.

This is an example of the general principle that bigger data need bigger models.  In this case, we expanded our second model by adding an error term which had no underlying golf interpretation, but allowed the model to flexibly fit the data.  This is similar to how, in a multi-center trial, we might allow the treatment effect to vary by area, even if we are not particularly interested in this variation, just because this can capture otherwise unexplained aspects of the data, and it is also similar to the idea in classical analysis of variance of including a fully saturated interaction term to represent residual error.

The golf example also illustrates the way that inferences from a sequence of models can be compared, both by graphing predictions along with data and by studying the way that parameter estimates change as the model is expanded.  For example, when we add uncertainty in the distance of the shot, our estimate of the angular uncertainty decreases.  Finally, we recognize that even the final fitted model is a work in progress, and so we want to work in a probabilistic programming environment where we can expand it by allowing the parameters to vary by player and condition of the course.

\section{Example of workflow for a model with unexpected multimodality: Planetary motion}\label{sec:orbits}

The previous example was relatively straightforward in that we built a model and gradually improved it.  Next we consider a case study in which we start with a complicated model, encounter problems with our inference, and have to figure out what is going on.

Section~\ref{sec:fast} alludes to measurements of a planet's motion. 
Let us now investigate this example from a slightly different perspective.
While simple in appearance, this problem illustrates many of the concepts we have discussed and highlights that the workflow draws on both statistical and field expertise.
It also cautions us that the workflow is not an automated process; each step requires careful reasoning.
For many problems we have encountered finding the right visualization tools is often the key to understand our model, its limitations, and how to improve it.
This example is no exception.
We monitor various intermediate quantities as prescribed in Section~\ref{sec:intermediate_quant}, and make extensive use of predictive checks (Sections~\ref{sec:prior_pred_checks} and \ref{sec:post_pred_checks}).

\subsection{Mechanistic model of motion}

Instead of fitting an ellipse, we use a mechanistic model based on elementary notions of classical mechanics.
This allows us to estimate quantities of physical interest, such as stellar mass, and more readily apply domain knowledge, as well as  track the planet's trajectory in space and time. 

We can describe the planet's motion using Newton's laws, which is a second-order differential equation or equivalently a system of two first-order differential equations, which yields Hamilton's formulation:
\begin{eqnarray*}
  \frac{\mathrm d q}{\mathrm d t} & = & \frac{p}{m} \\
  \frac{\mathrm d p}{\mathrm d t} & = & -\, \frac{k}{r^3} (q - q_*),
\end{eqnarray*}
where
\begin{itemize}
    \item $q(t)$ is the planet's position vector over time,
    \item $p(t)$ is the planet's momentum vector over time,
    \item $m$ is the planet's mass (assumed to be 1 in some units),
    \item $k = GmM$, with $G = 10^{-3}$, the gravitational constant in the appropriate units, and $M$ the stellar mass; hence $k =10^{-3} M$,
    \item and $r = \sqrt{(q - q_*)^T(q - q_*)}$ is the distance between the planet and the star, with $q_*$ denoting the star's position (assumed to be fixed).
\end{itemize}
The planet moves on a plane, hence $p$ and $q$ are each vectors of length 2.
The differential equations tell us that the change in position is determined by the planet's momentum and that the change in momentum is itself driven by gravity.

We would like to infer the gravitational force between the star and the planet, and in particular the latent variable $k$.
Other latent variables include the initial position and momentum of the planet, respectively $q_0$ and $p_0$, the subsequent positions of the planet, $q(t)$, and the position of the star, $q_*$.
Realistically, an astronomer would use cylindrical coordinates but for simplicity we stick to Cartesian coordinates.
We record the planet's position over regular time intervals, and assume measurements $q_{\mathrm{obs}, 1},\dots,q_{\mathrm{obs}, n}$ at times $t_1,\dots,t_n$, where each observation $q_{\mathrm{obs}, i}$ is two-dimensional with independent normally distributed errors,
 $$q_{\mathrm{obs}, i} \sim \mathrm{N}_2(q(t_i), \sigma^2 I).$$

We follow our general workflow and fit the model using fake data to see if we can recover the assumed parameter values.  We run an MCMC sampler for this model with Stan, which supports a numerical ordinary differential equation (ODE) solver.
A first attempt fails dramatically: the chains do not converge and take a long time to run.
This is an invitation to start with a simpler model, again working in the controlled setting offered by simulated data, where we know the true value of each parameter.

\subsection{Fitting a simplified model}

Ideally, we would find a simplification which is more manageable but still demonstrates the issues our algorithm encounters.
Our first simplified model only estimates $k$, with prior $k \sim \mathrm{normal}^+(0, 1)$, and an assumed true value of $k = 1$.  We set the other parameters of the model at $m=1$, $q_*=(0,0)$, $q_0=(1,0)$, and $p_0=(0,1)$.
Since the parameter space is 1-dimensional, we can compute the posterior distribution using quadrature; nevertheless we use MCMC, because our goal is to understand the challenges that frustrate our sampling algorithm.

\begin{figure}
    \centering
    \includegraphics{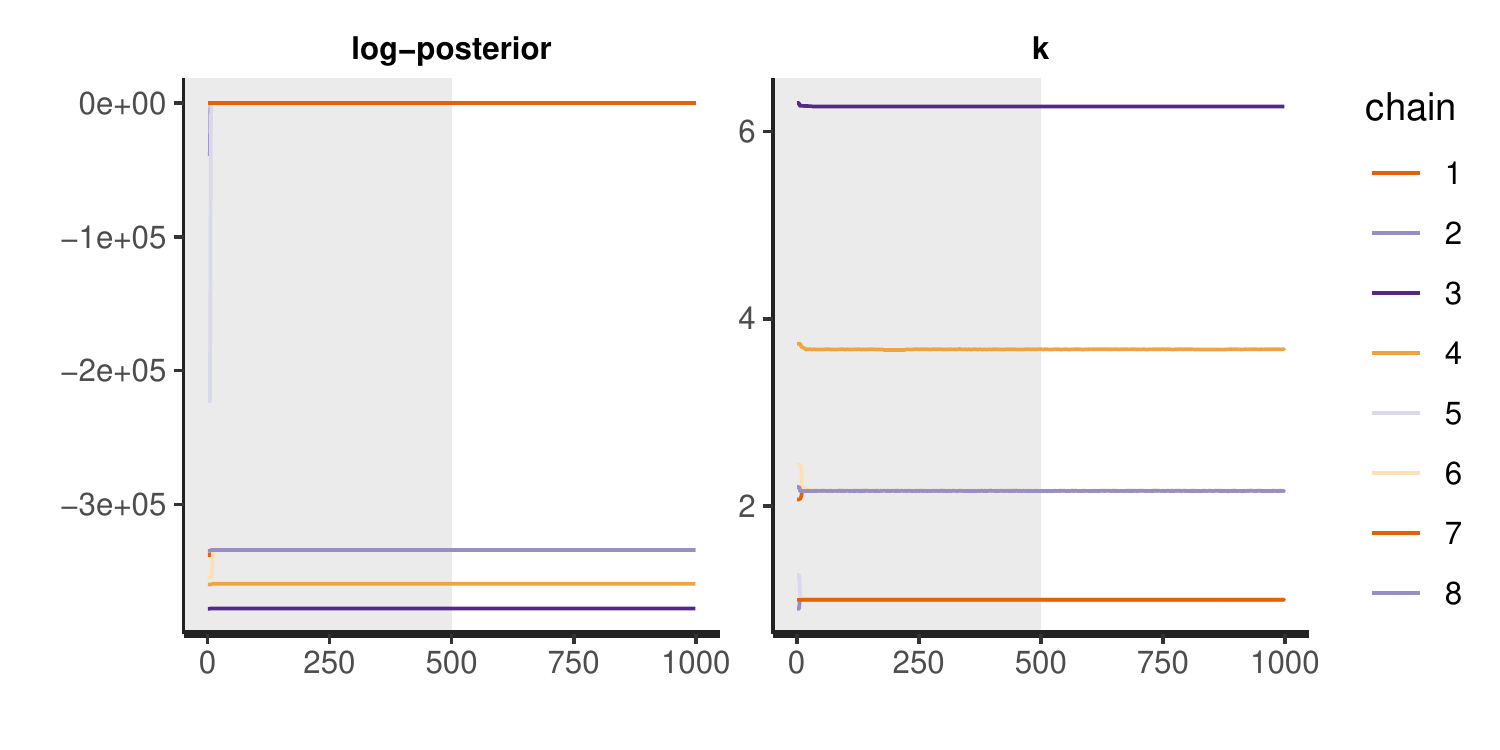}
    \vspace{-.2in}
    \caption{\em Traceplots for our simplified model of planetary motion. The chains fail to mix, and they converge to several different local modes, depending on their initial values. The varying log posterior suggests some modes are more consistent with the data than others. The shaded area represents samples during the warmup phase.}
    \label{fig:PMtrace}
\end{figure}

We run 8 chains, each with 500 iterations for the warmup phase and 500 more for the sampling phase, and we see:
\begin{itemize}
    \item The run time varies widely between chains, ranging from $\sim\,2$ seconds to $\sim\,2000$ seconds. While not necessarily a concern in itself, this indicates the chains are behaving in substantially different ways.
    \item $\widehat{R}$ for some parameters are large, implying that the chains have not mixed. Typically, we are comfortable with $\widehat{R} < 1.01$.  When $\widehat{R}>2$, this is an indication that the chains are not mixing well.
\end{itemize}
Faced with these issues, we examine the traceplots (Figure~\ref{fig:PMtrace}).
The chains seem to be stuck at local modes and do not cohesively explore the posterior space.
Some chains have much lower log posterior densities than others.
When doing posterior predictive checks for these chains specifically, we find that the simulated data does not agree with the observations.
For reasons we will uncover, the chains with the lowest log posterior and highest $k$ also turn out to be the ones with the longest run times.
Departing from Stan's defaults, we also plot iterations during the warmup phase.
The plot now clearly indicates that which mode each chain converges to is determined by its initial value, suggesting these modes are strongly attractive for the Markov chain.
This is an important practical point: the right plot can help us diagnose the problem almost instantaneously, but unfortunately, and despite our best efforts, the default plot need not be the right plot.

\begin{figure}
    \centering
    \includegraphics[width=3.5in]{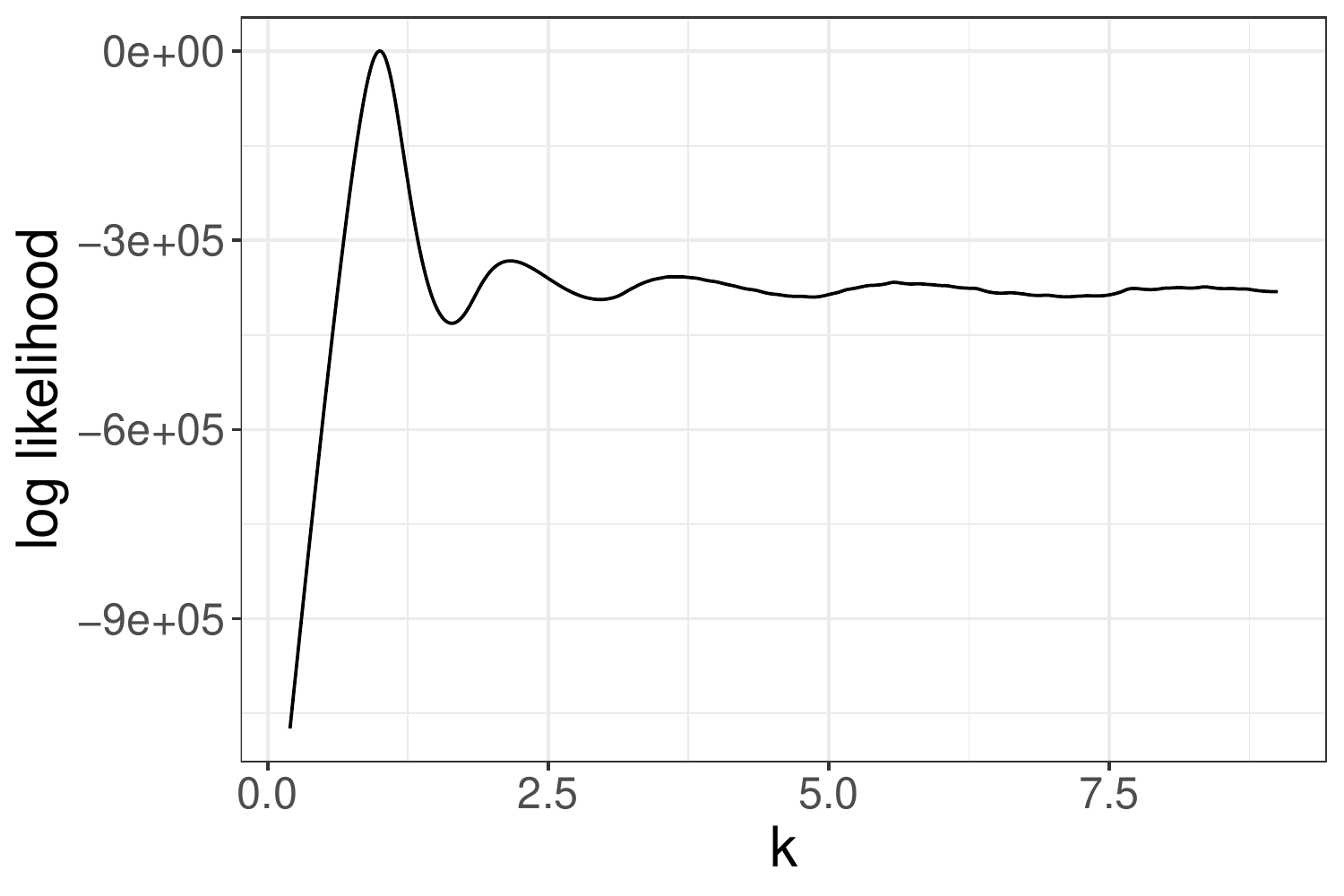}
    \vspace{-.1in}
    \caption{\em Log likelihood across various values of $k$ for our simplified planetary motion model. There is a dominating mode near $k = 1$, followed by local modes as $k$ increases. These modes are due to the cyclical trajectory, which allows the possibility of approximate aliasing.}
    \label{fig:PMloglk}
\end{figure}

\begin{figure}
    \centering
    \includegraphics[width=4in]{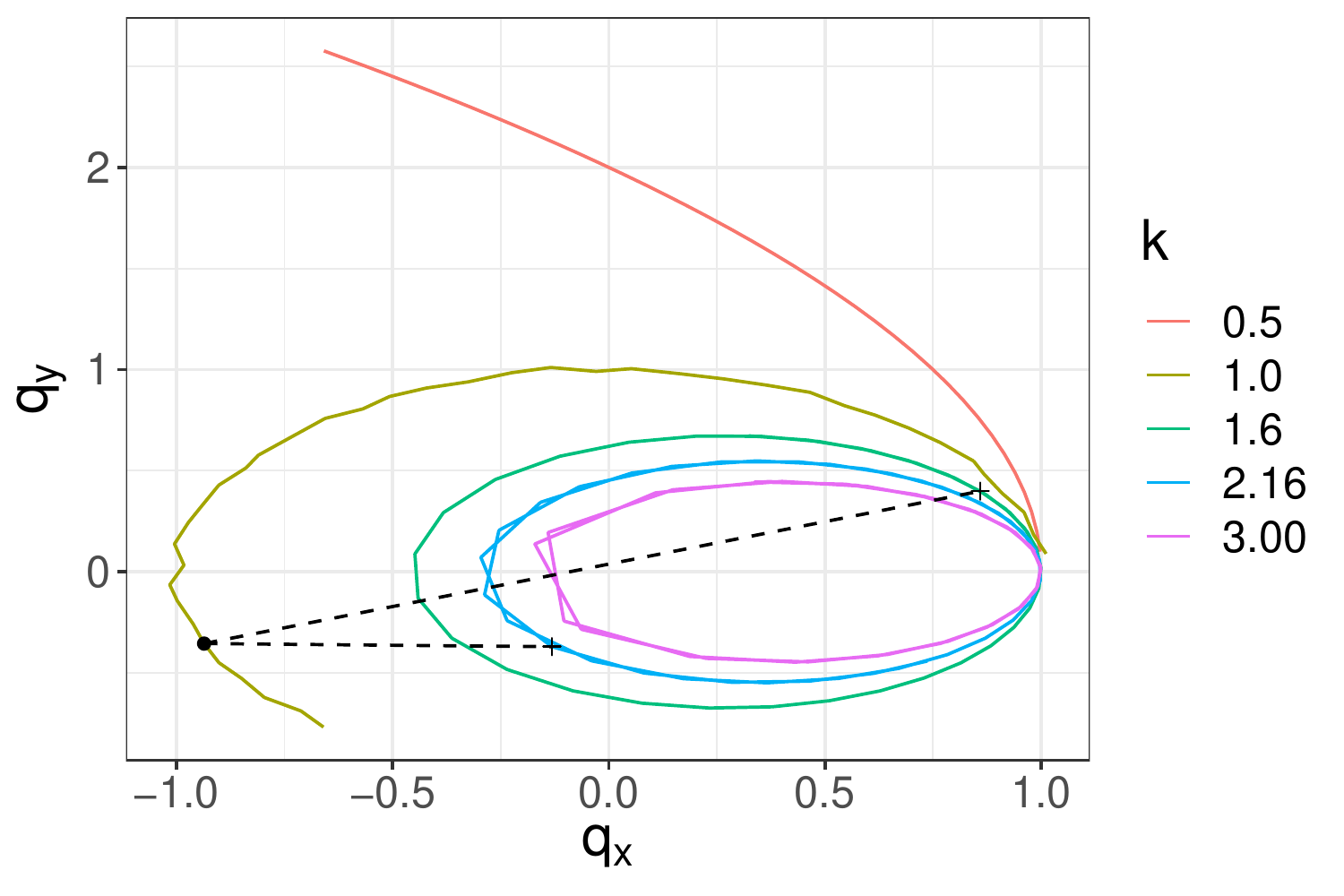}
    \vspace{-.1in}
    \caption{\em Orbits simulated for various values of $k$ for the planetary motion model. There is no strong degeneracy in $k$, but as this parameter changes, some of the simulated points by chance get closer to their observed counterparts, inducing wiggles in the tails of the log likelihood and creating local modes.
    For example, the $35^\mathrm{th}$ observation (solid dot, on the $k = 1$ orbit) is closer to the corresponding position simulated with $k = 2.2$ (cross on the blue line) than the one obtained with $k = 1.6$ (cross on the green line).}
    \label{fig:PMorbits}
\end{figure}

It is important to figure out whether these modes describe a latent phenomenon of interest which we must account for in our analysis or whether they are caused by a mathematical artifact.
Because we have the luxury of fitting a simplified model, we can exactly work out what is going on and use the gained insight for more elaborate models.
Figure~\ref{fig:PMloglk} plots the likelihood computed via a quadrature scheme and confirms the existence of local modes.
To understand how these modes arise, we may reason about the log likelihood as a function that penalizes distance between $q_\mathrm{obs}$ and $q(k)$, the positions of the planet simulated for a certain value of $k$. Indeed,
$$
\log p(q_\mathrm{obs} | k) = C - \frac{1}{2 \sigma} ||q_\mathrm{obs} - q(k)||^2_2,
$$
where $C$ is a constant that does not depend on $k$.
Figure~\ref{fig:PMorbits} displays the planet's simulated motion given different values of $k$.
Recall that $k$ controls the strength of the gravitational interaction: a higher value implies a closer and shorter orbit.
The assumed value is $k = 1$.
The other values of $k$ fail to generate data consistent with the observed data.
For $k < 1$, the trajectory can drift arbitrarily far away from the observed ellipse.
But for $k > 1$, the simulated ellipse must be contained inside the observed ellipse, which bounds the distance between $q_\mathrm{obs}$ and $q$.
Finally, as we change $k$ and rotate the ellipse, some of the observed and simulated positions happen to become relatively close, which induces local modes that appear as wiggles in the tail of the likelihood.
The parameter values at the mode do not induce simulations that are in close agreement with the data; but they do better than neighboring parameter values, which is enough to create a bump in the likelihood.

The tail modes are a mathematical artifact of the periodic structure of the data and do not characterize a latent phenomenon of interest.
Moreover they only contribute negligible probability mass.
Hence, any chain that does not focus on the dominating mode wastes our precious computational resources.
So, what can we do? \\

\noindent
\textit{Building stronger priors.} One option is to build a more informative prior to reflect our belief that a high value of $k$ is implausible; or that any data generating process that suggests the planet undergoes several orbits over the observation time is unlikely. When such information is available, stronger priors can indeed improve computation. This is unfortunately not the case here. A stronger prior would reduce the density at the mode, but the wiggles in the tail of the joint would persist. Paradoxically, with more data, these wiggles become stronger: the model is fundamentally multi-modal.
Note also that our current prior, $k \sim \mathrm{normal}^+(0, 1)$, is already inconsistent with the values $k$ takes at the minor modes. In principle we could go a step further and add a hard constraint on orbital time or velocity to remove the modes.\\

\noindent
\textit{Reweighting draws from each chain.} One issue is that the Markov chains fail to transition from one mode to the other, meaning some chains sample over a region with a low probability mass. We can correct our Monte Carlo estimate using a re-weighting scheme, such as stacking. This strategy likely gives us reasonable Monte Carlo estimates, but: (i) we will not comprehensively explore all the modes with 8 chains, so stacking should really be treated as discarding the chains stuck at local modes and (ii) we still pay a heavy computational price, as the chains in minor modes take up to $\sim$1000 times longer to run. \\

\noindent
\textit{Tuning the starting points.} We did not specify the Markov chain's starting points and instead relied on Stan's defaults, which sample the initial point from a $\mathrm{uniform}(-2, 2)$ over the unconstrained space, that is, the starting points are drawn from
$  \log k^{(0)} \sim \mathrm{uniform}(-2, 2)$.
This default, designed for unconstrained parameters on the unit scale, indulges values of $k$ which are widely inconsistent with our prior and our domain expertise. In a non-asymptotic regime, the Markov chain does not always ``forget'' its starting point, and it is unlikely to do so here even if we run the chain for many \textit{many} more iterations. We can therefore not ignore this tuning parameter of our algorithm. An alternative to the default is to sample $k^{(0)}$ from our prior, thereby imposing that the chains start in a range of values deemed reasonable. In this setup, the chains converge quickly and our computation focuses on the relevant region. \\

Whether with stacking, tuned starting points, or perhaps another mean, we need to give MCMC a helping hand to avoid local modes. In general, ignoring non-converging chains is poor practice, so we want to highlight how the here presented process differs from that. First, we examine all the chains, using posterior predictive checks, and work out exactly how the local modes arise. We decisively demonstrate that they do not describe data generating processes of interest, nor do they contribute more than a negligible amount of probability mass. Only then do we redirect our computational resources to focus on the mode around which all the probability mass concentrates.

\subsection{Bad Markov chain, slow Markov chain?}

Recall that the chains that yielded the lowest log posteriors were also the ones that were the slowest---an instance of the folk theorem of statistical computing (see Section \ref{folk}).
We can in fact show that Hamilton's equations become more difficult to solve as $k$ increases.
The intuition is the following: if the gravitational interaction is strong, then the planet moves at a much faster rate.
From a numerical perspective, this means each time step, $\mathrm d t$, incurs a greater change in $q(t)$ and the integrator's step size must accordingly be adjusted.

There is wisdom is this anecdote: an easy deterministic problem can become difficult in a Bayesian analysis.
Indeed Bayesian inference requires us to solve the problem across a range of parameter values, which means we must sometimes confront unsuspected versions of the said problem.
In our experience, notably with differential equation based models in pharmacology and epidemiology, we sometime require a more computationally expensive stiff solver to tackle difficult ODEs generated during the warmup phase.

Other times slow computation can alert us that our inference is allowing for absurd parameter values and that we need either better priors or more reasonable initial points.

Unfortunately this goes against the ``fail fast'' principle outlined in Section~\ref{sec:fast}.
Our current tools tend to be much slower when the fit is bad, hence an important research topic in Bayesian computational workflow is to flag potential problems quickly to avoid wasting too much time on dead ends.

\subsection{Building up the model}

Starting from the simplified model, we now gradually build our way back to the original model.
This turns out to be not quite straightforward, but we can put what we have learned from the simplified model to good use.
Most inference problems we encounter across the models we fit can be traced back to the interaction between the likelihood and the cyclical observations---an elementary notion, once grasped, but which would have been difficult to discover in a less simple setting than the one we used.

Here is an example.
In the complete model, we estimate the position of the star, $q_*$, and find that the chains converge to many different values, yielding simulations which, depending on the chain, agree or disagree with the observations.
There is however, based on the traceplots, no obvious connection between the starting points and the neighborhoods of convergence.
It is difficult to examine this type of connections because the model has now 7 parameters, some with strong posterior correlations.
Fortunately, we can reason about the physics of the problem and realize that tweaking the star's position, $q_*$, and implicitly, $r$, the star-planet distance, is not unlike modifying $k$.
Recall that
$$
  \frac{\mathrm d p}{\mathrm d t} = -\, \frac{k}{r^3} (q - q_*),
$$
whence both $k$ and $r$ control the gravitational interaction.

\begin{figure}
    \centerline{
    \includegraphics[width = .5\textwidth]{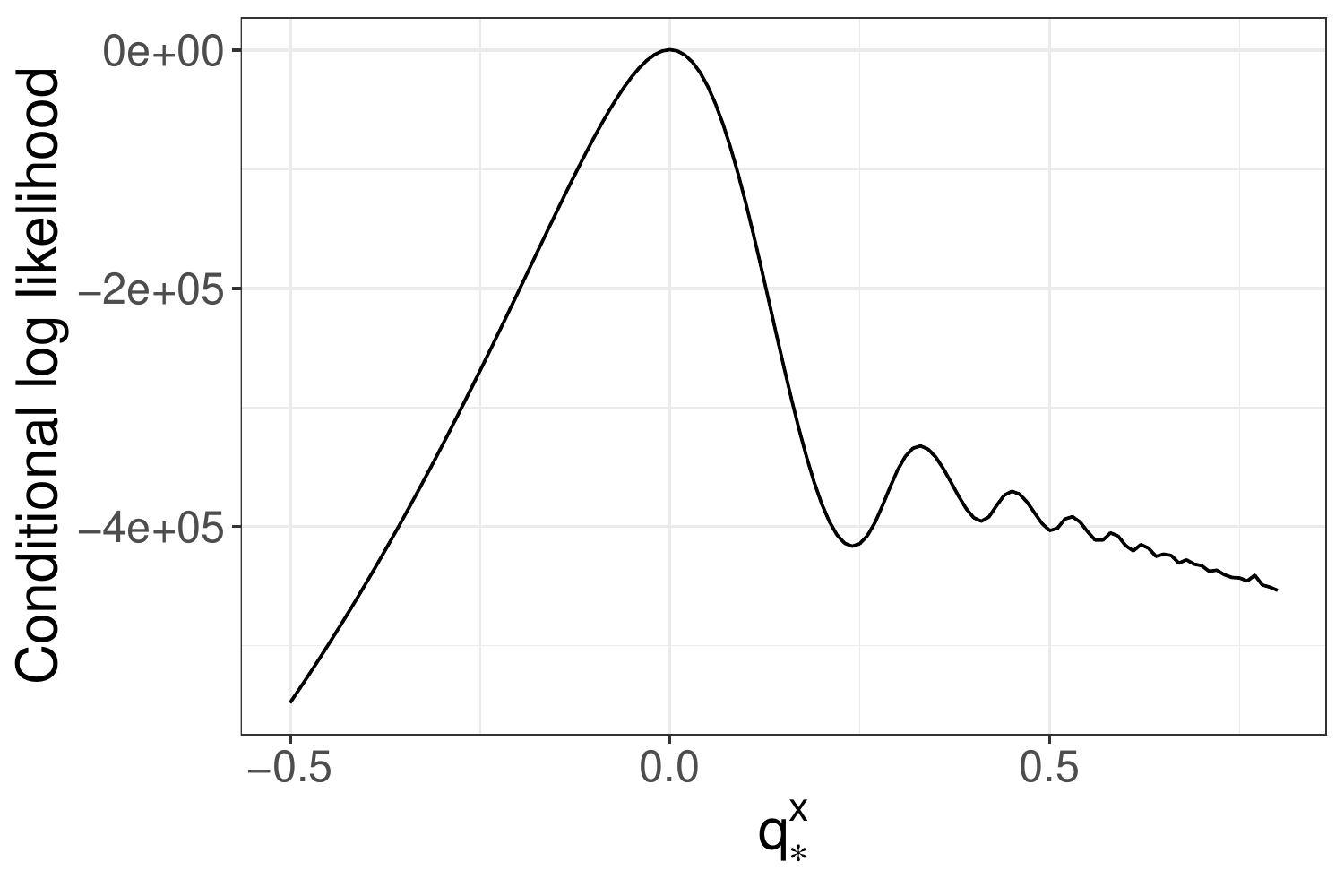}
    \includegraphics[width = .5\textwidth]{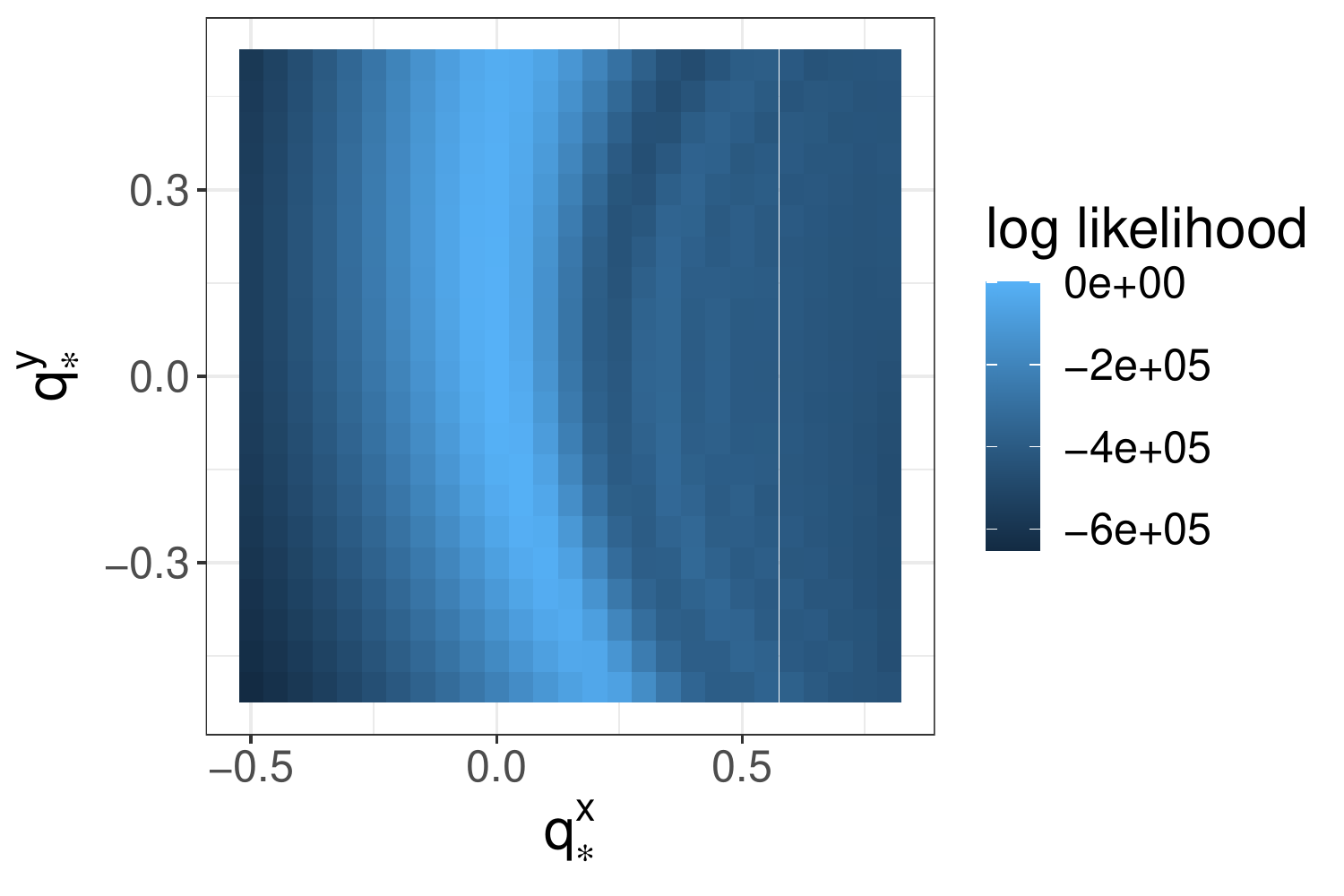}}
    \vspace{-.1in}
    \caption{\em For the planetary motion example, log conditional likelihood when varying the star's position, $q_*$. {\em Left:} All parameters are kept fixed, except for the $x$-coordinate of $q_*$. {\em Right:} This time both coordinates of $q_*$ are allowed to vary. The quadrature computation allows us to expose the multimodality of the problem.}
    \label{fig:PMloglk2}
\end{figure}

We cannot do a quadrature across all 7 parameters of our model.
Instead, we look at the \textit{conditional} likelihood, wherein we keep all parameters ($k$, $q_0$, and $p_0$) fixed, except for $q_*$.
In a sense, this amounts to investigating another simplification of our model.
Figure~\ref{fig:PMloglk2} shows the suspected modes, thereby supporting our conjecture.
At this point, with some level of confidence, we construct starting points to guide our Markov chains towards the dominating mode and obtain a good fit of the complete model.

\subsection{General lessons from the planetary motion example}
When we fail to fit a model, examining a simplified model can help us understand the challenges that frustrate our inference algorithm.
In practice it is difficult to find a simplification which is manageable and still exhibits the pathology we wish to understand.
Reasoning about the topology surrounding our model, as alluded to in section~\ref{sec:topology}, can help us in this process.
A straightforward way of simplifying is to fix some of the model parameters.

In the planetary motion example, we are confronted with a multi-modal posterior distribution.
This geometry prevents our chains from cohesively exploring the parameter space and leads to biased Monte Carlo estimates.
It is important to understand how these local modes arise and what their contribution to the posterior probability mass might be. We do so using posterior predictive checks.
It is not uncommon for minor modes, with negligible probability mass, to ``trap'' a Markov chain.
The possibility of such ill-fitting modes implies we should always run multiple chains, perhaps more than our current default of $4$.

This case study also raises the question of what role starting points may play.
Ideally a Markov chain forgets its initial value but in a non-asymptotic regime this may not be the case.
This is not a widely discussed topic but nevertheless one of central importance for practitioners and thence for Bayesian workflow.
Just as there is no universal default prior, there is no universal default initial point.
Modelers often need to depart from defaults to insure a numerically stable evaluation of the joint density and improve MCMC computation.
At the same time we want dispersed initial points in order to have reliable convergence diagnostics and to potentially explore all the relevant modes.
Like for other tuning parameters of an inference algorithm, picking starting points can be an iterative process, with adjustments made after a first attempt at fitting the model.


We do not advocate mindlessly discarding misbehaving chains. It is important to analyze where this poor behavior comes from, and whether it hints at serious flaws in our model and in our inference. Our choice to adjust the initial estimates is based on: (a) the realization that the defaults are widely inconsistent with our expertise and (b) the understanding that the local modes do not describe a latent phenomenon of interest, as shown by our detailed analysis of how cyclical data interacts with a normal likelihood.

\section{Discussion}\label{discussion}

\subsection{Different perspectives on statistical modeling and prediction}

Consider three different ways to think about modeling and prediction:
\begin{itemize}
\item {\em Traditional statistical perspective.}  
In textbooks, statistical inference is typically set up as a problem in which a model has been chosen ahead of time, and in the Bayesian context the goal is to accurately summarize the posterior distribution.  Computation is supposed to be done as long as necessary to reach approximate convergence.
\item {\em Machine learning perspective.}
In machine learning, the usual goal is prediction, not parameter estimation, and computation can stop when cross validation prediction accuracy has plateaued.
\item {\em Model exploration perspective.}
In applied statistical work, much of our modeling effort is spent in exploration, trying out a series of models, many of which will have terrible fit to data, poor predictive performance, and slow convergence (see also Section \ref{folk}).
\end{itemize}

These three scenarios imply different inferential goals.  In a traditional statistical modeling problem, it can make sense to run computation for a long time, using approximations only when absolutely necessary.  Another way of saying this is that in traditional statistics, the approximation might be in the choice of model rather than in the computation. In machine learning, we want to pick an algorithm that trades off predictive accuracy, generalizability, and scalability, so as to make use of as much data as possible within a fixed computational budget and predictive goal.  In model exploration, we want to cycle through many models, which makes approximations attractive. But there is a caveat here: if we are to efficiently and accurately explore the {\em model space} rather than the algorithm space, we require any approximation to be sufficiently faithful as to reproduce the salient features of the posterior.

The distinction here is not about inference vs.\ prediction, or exploratory vs.\ confirmatory analysis. Indeed all parameters in inference can be viewed as some quantities to predict, and all of our modeling can be viewed as having exploratory goals (Gelman, 2003). Rather, the distinction is how much we trust a given model and allow the computation to approximate. 

As the examples in Sections~\ref{golf} and \ref{sec:orbits} illustrate, problems with a statistical model in hindsight often seem obvious, but we needed the workflow to identify them and understand the obviousness.  Another important feature of these examples, which often happens with applied problems, is that particular challenges in modeling arise in the context of the data at hand: had the data been different, we might never have encountered these particular issues, but others might well have arisen.  This is one reason that subfields in applied statistics advance from application to application, as new wrinkles become apparent in existing models.  A central motivation of this paper is to make more transparent the steps by which we can uncover and then resolve these modeling problems.

\subsection{Justification of iterative model building}\label{sec:justification_of_iterative}

We view the process of {\em model navigation} as the next transformative step in data science. The first big step of data science, up to 1900 or so, was {\em data summarization}, centering on the gathering of relevant data and summarizing through averages, correlations, least-squares fits, and the like.  The next big step, beginning with Gauss and Laplace and continuing to the present day, was {\em modeling}:  the realization that a probabilistic model with scientific content could greatly enhance the value from any given dataset, and also make it more feasible to combine different sources of data.  We are currently in the midst of another big step, {\em computation}:  with modern Bayesian and machine learning methods, increases in algorithmic and computational efficiency have effected a qualitative improvement in our ability to make predictions and causal inferences.  We would like to move beyond good practice and workflow in particular case studies, to a formulation of the process of model navigation, ``facilitating the exploration of model space'' (Devezer et al., 2019).

In the ideal world we would build one perfect model and solve the math. In the real world we need to take into account the limitations of humans and computers, and this should be included in models of science and models of statistics (Navarro, 2019, Devezer et al., 2020).

From a human perspective, our limited cognitive capabilities make it easier to learn gradually. Iterative model building starting from a simple model is gradual learning and helps us better understand the modeled phenomenon. Furthermore, building a rich model takes effort, and it can be efficient in human time to start from a simpler model and stop when the model seems to be sufficiently good.  We gave an example in Section \ref{golf}.
One goal of workflow is to make the process easier for humans even in the idealized setting where exact computation can be performed automatically.

Iterating over a set of models is also useful from a computational standpoint. 
Given a proper posterior, computation in Bayesian inference is theoretically solved. 
In practice we must contend with finite computational resources.
An algorithm for which asymptotic guarantees exist can fail when run for a finite time.
There is no fully automated computation that yields perfect results, at least not across the vast range of models practitioners care about. 
Another goal of workflow is to avoid some computational problems and be able to efficiently diagnose the remaining ones.
Here too deconstructing the model into simpler versions can be helpful: it is easier to understand computational challenges when there are fewer moving parts.
Hence, even if a mathematical description of a model is given, correctly implementing the model tends to require iteration.

Not only is iterative model building beneficial from a cognitive and computational standpoint, but the complexity of complicated computational models makes it difficult for the human user to disentangle computational concerns, modeling issues, data quality problems, and bugs in the code. By building models iteratively, we can employ software engineering techniques in our modeling procedure. Simple model components can be checked to make sure they act in an expected way before more complicated components are added. 

\subsection{Model selection and overfitting}\label{overfitting}

A potential issue with the proposed iterative workflow is that model improvement is conditioned on discrepancy between the currently considered model and the data, and thus at least some aspects of the data are used more than once. This ``double dipping'' can in principle threaten the frequency properties of our inferences, and it is important to be aware of the possibility of overfitting arising from model selection, as considered for example by Fithian et al.\ (2015) and Loftus (2015). A related issue is the garden of forking paths, the idea that different models would have been fit had the data come out differently (Gelman and Loken, 2013). We do not advocate selecting the best fit among some such set of models. Instead, we describe a process of building to a more complex model taking the time to understand and justify each decision.  


To put it in general terms: suppose we fit model $M_1$, then a posterior predictive check reveals problems with its fit to data, so we move to an improved $M_2$ that, we hope, includes more prior information and makes more sense with respect to the data and the applied problem under study.  But had the data been different, we would have been satisfied with $M_1$.  The steps of model checking and improvement, while absolutely necessary, represent an aspect of fitting to data that is not captured in the likelihood or the prior.

This is an example of the problem of post-selection inference (Berk et al., 2013, Efron, 2013).  Much of the research in this area has been on how to adjust $p$-values and confidence intervals to have appropriate frequency properties conditional on the entire fitting procedure, but Bayesian inferences are subject to this concern as well.  For example, here is the story of one of the adaptations we made in election forecasting (Gelman, Hullman, et al., 2020):

\begin{quotation}
\noindent
A few weeks after we released our first model of the election cycle for {\em The Economist}, we were disturbed at the narrowness of some of its national predictions. In particular, at one point the model gave Biden 99\% chance of winning the national vote. Biden was clearly in the lead, but 99\% seemed like too high a probability given the information available at that time. Seeing this implausible predictive interval motivated us to refactor our model, and we found some bugs in our code and some other places where the model could be improved—including an increase in between-state correlations, which increased uncertainty of national aggregates. The changes in our model did not have huge effects---not surprisingly given that we had tested our earlier model on 2008, 2012, and 2016---but the revision did lower Biden’s estimated probability of winning the popular vote to 98\%. This was still a high value, but it was consistent with the polling and what we’d seen of variation in the polls during the campaign.
\end{quotation}

\noindent
The errors we caught were real, but if we had not been aware of these particular problematic predictions, we might have never gone back to check.  This data-dependence of our analysis implies a problem with a fully Bayesian interpretation of the probability statements based on the final model we settled on.  And, in this case, model averaging would not resolve this problem:  we would not want to average our final model with its buggy predecessor.  We might want to average its predictions with those of some improved future model, but we can't do that either, as this future model does not yet exist!

That said, we think that Bayesian workflow as described here will avoid the worst problems of overfitting.  Taylor and Tibshirani (2015) warn of the problem of inference conditional on having ``searched for the strongest associations.''  But our workflow does not involve searching for optimally-fitting models or making hard model selection under uncertainty.  Rather, we use problems with fitted models to reassess our modeling choices and, where possible, include additional information.  For our purposes, the main message we take from concerns about post-selection inference is that our final model should account for as much information as possible, and when we might be selecting among a large set of possible models, we instead embed these in a larger model, perform predictive model averaging, or use all of the models simultaneously (see Section~\ref{comparing}).  As discussed by Gelman, Hill, and Yajima (2012), we expect that would work better than trying to formally model the process of model checking and expansion.

 We also believe that our workflow enables practitioners to perform severe tests of many of the assumptions that underlie the models being examined (Mayo, 2018). Our claim is that often a model whose assumptions withstood such severe tests is, despite being the result of data-dependent iterative workflow, more trustworthy than a preregistered model that has not been tested at all.

On a slightly different tack, iterative model building is fully justified as a way to understand a fixed, complex model. This is an important part of workflow as it is well known that components in complex models can interact in complex ways. For example, Hodges and Reich (2010) describe how structured model components such as spatial effects can have complex interactions with linear covariate effects. 

\subsection{Bigger datasets demand bigger models}

Great progress has been made in recent decades on learning from data using methods developed in statistics, machine learning, and applied fields ranging from psychometrics to pharmacology.  Hierarchical Bayesian modeling, deep learning, and other regularization-based approaches are allowing researchers to fit larger, more complex models to real-world data, enabling information aggregation and partial pooling of inferences from different sources of data.

While the proposed workflow offers advantages regardless of the size of the dataset, the case of big data deserves a special mention.  ``Big data'' is sometimes defined as being too big to fit in memory on your machine, but here we use the term more generally to also include datasets that are so large that our usual algorithms do not run in reasonable time.  In either case, the definition is relative to your current computing capacity and inferential goals.

It is frequently assumed that big data can alleviate the need for careful modeling. We do not believe this is the case.  Quantity does not always substitute for quality.  Big data is messy data. Big data prioritizes availability over randomization, which means Big data is almost always observational rather than from designed experiments. Big data frequently uses available proxies rather than direct measurements of underlying constructs of interest. To make relevant inferences from big data, we need to extrapolate from sample to population, from control to treatment group, and from measurements to latent variables. All these steps require statistical assumptions and adjustment of some sort, which in the Bayesian framework is done using probability modeling and mathematical connections to inferential goals.  For example, we might fit a multilevel model for data given respondents' demographic and geographic characteristics and then poststratify to connect the predictions from this model to the goal of inference about the general population.

Each of these steps of statistical extrapolation should be more effective if we adjust for more factors---that is, include more information---but we quickly reach a technical barrier. Models that adjust for many of factors can become hard to estimate, and effective modeling requires (a) regularization to get more stable estimates (and in turn to allow us to adjust for more factors), and (b) modeling of latent variables (for example parameters that vary by person when modeling longitudinal data), missingness, and measurement error.

A key part of Bayesian workflow is adapting the model to the data at hand and the questions of interest.  The model does not exist in isolation and is not specified from the outside; it emerges from engagement with the application and the available data.

\subsection{Prediction, generalization, and poststratification}\label{generalization}

The three core tasks of statistics are generalizing from sample to population, generalizing from control to treatment group, and generalizing from observed data to underlying constructs of interest.   In machine learning and causal inference, the terms ``domain adaptation'' and ``transportability'' have been used to represent the challenges of taking inference from a particular dataset and applying it in a new problem (Blitzer, Dredze, and Pereira, 2007, Pearl and Bareinboim, 2011).  Many statistical tools have been developed over the years to attack problems of generalization, for example weighting and poststratification in sample surveys, matching and regression in causal inference, and latent variable modeling in areas such as psychometrics and econometrics where there is concern about indirect or biased observations.

Bayesian methods enter in various ways, including the idea of hierarchical modeling or partial pooling to appropriately generalize across similar but not identical settings, which has been rediscovered in many fields (e.g., Henderson, 1950, Novick et al., 1972, Gelman and Hill, 2007, Finkel and Manning, 2009, Daumé, 2009), regularization to facilitate the use of large nonparametric models (Hill, 2011), and multilevel modeling for latent variables (Skrondal and Rabe-Hesketh, 2004), and there are connections between transportability and Bayesian graph models (Pearl and Bareinboim, 2014).

Bayesian workflow does not stop with inference for the fitted model.  We are also interested in inferences for new real-world situations, which implies that the usual prior and data model is embedded in a larger framework including predictions and inferences for new settings, including potentially different modes of measurement and treatment assignments.  Statistical models can also go into production, which provides future opportunities for feedback and improvement.

Just as the prior can often only be understood in the context of the likelihood (Gelman et al., 2017, Kennedy et al., 2019), so should the model be understood in light of how it will be used.  For example, Singer et al.\ (1999) and Gelman, Stevens, and Chan (2003) fit a series of models to estimate the effects of financial incentives on response rates in surveys.  The aspects of these models that will be relevant for predicting effects for small incentives in mail surveys are different from what is relevant for predictions for large incentives in telephone surveys.  This is related to the discussion of sensitivity analysis in Sections \ref{influence} and \ref{multiverse}.  For another illustration of this point, Rubin (1983) gives an example where the choice of transformation has minor effects on inference for the median of a distribution while having large effects on inference for the mean.

\subsection{Going forward}

All workflows have holes, and we cannot hope to exhaust all potential flaws of applied data analysis. In the prior and posterior predictive check, a wrong model will pass the check silently for overfitting all output at the observed values. In simulation based calibration, an incorrect computation program can satisfy the diagnostics if the posterior stays in the prior. In cross validation, the consistency relies on conditional independence and stationarity of the covariates. In causal inference, there are always untestable causal assumptions, no matter how many models we have fit.  More generally, statistics relies on some extrapolation, for which some assumption is always needed.  To ultimately check the model and push the workflow forward, we often need to collect more data, along the way expanding the model, and an appropriate  experiment design will be part of this larger workflow.

This article has focused on data analysis: the steps leading from data and assumptions to scientific inferences and predictions.  Other important aspects of Bayesian statistics not discussed here include design, measurement, and data collection (coming before the data analysis) and decision making and communication (coming after data analysis).  We also have not gone into the details of computing environments or the social and economic aspects of collaboration, sharing of data and code, and so forth.

The list of workflow steps we have provided is too long to be a useful guide to practice.  What can be done?  Rather than giving users a 25-item checklist, we hope that we can clarify these processes so they can be applied in a structured or even automated framework.  Our rough plan is as follows:
\begin{itemize}
\item Abstract these principles from our current understanding of best practice, thus producing the present article.
\item Apply this workflow to some applied problems and write these up as case studies.
\item Implement as much of the workflow as possible in software tools for general application.
\end{itemize}
Automating what can be automated should enable the statistician or applied researcher to go beyond button pushing and integrate data with domain expertise.
The ultimate goal of this project is to enable ourselves and other data analysts to use statistical modeling more effectively, and to allow us to build confidence in the inferences and decisions that we come to.

This article is a review, a survey of the territory, a reminder of methods we have used, procedures we have followed, and ideas we would like to pursue.  To be useful to practitioners, we need worked examples with code.  We would also like to provide more structure:  if not a checklist, at least a some paths to follow in a Bayesian analysis. Some guidance is given in the Stan User's Guide (Stan Development Team, 2020), and we are working on a book on Bayesian workflow using Stan to provide such a resource to novice and experienced statisticians alike.  That said, we believe this paper has value as a first step toward putting the many different activities of Bayesian workflow under a single roof.

\section*{References}

\noindent

\bibitem Afrabandpey, H., Peltola, T., Piironen, J., Vehtari, A., and Kaski, S. (2020). Making Bayesian predictive models interpretable: A decision theoretic approach. {\em Machine Learning} {\bf 109}, 1855--1876.

\bibitem Akaike, H. (1973).  Information theory and an extension of the maximum likelihood principle.  In {\em Proceedings of the Second International Symposium on Information Theory}, ed.\ B. N. Petrov and F. Csaki, 267--281.  Budapest:  Akademiai Kiado.  Reprinted in {\em Breakthroughs in Statistics}, ed.\ S. Kotz, 610--624.  New York: Springer (1992).

\bibitem Berger, J. O., Bernardo, J. M., and Sun, D. (2009). The formal definition of reference priors. {\em Annals of Statistics} {\bf 37}, 905--938.

\bibitem Berk, R., Brown, L., Buja, A., Zhang, K., and Zhao, L. (2013).  Valid post-selection inference.  {\em Annals of Statistics} {\bf 41}, 802--837.

\bibitem  Berry, D. (1995). {\em Statistics: A Bayesian Perspective}. Duxbury Press.

\bibitem Betancourt, M. (2017a). A conceptual introduction to Hamiltonian Monte Carlo. \url{arxiv.org/abs/1701.02434}

\bibitem Betancourt, M. (2017b).  Identifying Bayesian mixture models. {\em Stan Case Studies} {\bf 4}. \url{mc-stan.org/users/documentation/case-studies/identifying_mixture_models.html}

\bibitem Betancourt, M. (2018).  Underdetermined linear regression. \url{betanalpha.github.io/assets/case_studies/underdetermined_linear_regression.html}

\bibitem Betancourt, M. (2020a).  Towards a principled Bayesian workflow.  \url{betanalpha.github.io/assets/case_studies/principled_bayesian_workflow.html}

\bibitem Betancourt, M. (2020b).  Robust Gaussian Process Modeling. \url{github.com/betanalpha/knitr_case_studies/tree/master/gaussian_processes}

\bibitem Betancourt, M., and Girolami, M. (2015). Hamiltonian Monte Carlo for hierarchical models. In {\em Current Trends in Bayesian Methodology with Applications}, ed.\ S. K. Upadhyay, U. Singh, D. K. Dey, and A. Loganathan, 79--102.

\bibitem Blei, D. M., Kucukelbir, A., and McAuliffe, J. D. (2017). Variational inference: A review for statisticians. {\em Journal of the American Statistical Association} {\bf 112}, 859--877.

\bibitem Blitzer, J., Dredze, M., and Pereira, F. (2007). Biographies, Bollywood, boom-boxes and blenders: Domain adaptation for sentiment classification.  In {\em Proceedings of the 45th Annual Meeting of the Association of Computational Linguistics}, 440--447.

\bibitem Box, G. E. P. (1980).  Sampling and Bayes inference in scientific modelling and robustness.  {\em Journal of the Royal Statistical Society A} {\bf 143}, 383--430.

\bibitem Broadie, M. (2018).  Two simple putting models in golf.  \url{statmodeling.stat.columbia.edu/wp-content/uploads/2019/03/putt_models_20181017.pdf}

\bibitem Bryan, J. (2017).  Project-oriented workflow.  \url{www.tidyverse.org/blog/2017/12/workflow-vs-script}

\bibitem Bürkner, P.-C. (2017). brms: An R Package for Bayesian multilevel models using Stan. {\em Journal of Statistical Software} {\bf 80}, 1--28. 

\bibitem Carpenter, B. (2017).  Typical sets and the curse of dimensionality. {\em Stan Case Studies} {\bf 4}. \url{mc-stan.org/users/documentation/case-studies/curse-dims.html}

\bibitem Carpenter, B. (2018).  Predator-prey population dynamics:
The Lotka-Volterra model in Stan. {\em Stan Case Studies} {\bf 5}. \url{mc-stan.org/users/documentation/case-studies/lotka-volterra-predator-prey.html}
  
\bibitem Carpenter, B., Gelman, A., Hoffman, M., Lee, D., Goodrich, B., Betancourt, M., Brubaker, M., Guo, J., Li, P., and Riddell, A. (2017).  Stan: A probabilistic programming language. {\em Journal of Statistical Software} {\bf 76} (1).

\bibitem Chen, C.,  Li, O., Barnett, A., Su, J., and Rudin, C. (2019).  This looks like that: Deep learning for interpretable image recognition.  {\em 33rd Conference on Neural Information Processing Systems}.  \url{papers.nips.cc/paper/9095-this-looks-like-that-deep-} \url{learning-for-interpretable-image-recognition.pdf}

\bibitem Chiu, W. A., Wright, F. A., and Rusyn, I. (2017).  A tiered, Bayesian approach to estimating of population variability for regulatory decision-making.  {\em ALTEX} {\bf 34}, 377--388.

\bibitem Chung, Y., Rabe-Hesketh, S., Gelman, A., Liu, J. C., and Dorie, A. (2013).  A non-degenerate penalized likelihood estimator for hierarchical variance parameters in multilevel models.  {\em Psychometrika} {\bf 78}, 685--709.

\bibitem Chung, Y., Rabe-Hesketh, S., Gelman, A., Liu, J. C., and Dorie, A. (2014).  Nonsingular covariance estimation in linear mixed models through weakly informative priors.  {\em Journal of Educational and Behavioral Statistics} {\bf 40}, 136--157. 
  
\bibitem Clayton, D. G. (1992). Models for the analysis of cohort and case-control studies with inaccurately measured exposures. In {\em Statistical Models for Longitudinal Studies of Exposure and Health}, ed.\ J. H. Dwyer, M. Feinleib, P. Lippert, and H. Hoffmeister, 301--331. Oxford University Press.

\bibitem Cook, S., Gelman, A., and Rubin, D. B. (2006).  Validation of software for Bayesian models using posterior quantiles. {\em Journal of Computational and Graphical Statistics} {\bf 15}, 675--692. 

\bibitem Daumé, H. (2009).  Frustratingly easy domain adaptation. \url{arxiv.org/abs/0907.1815}
 
\bibitem Deming, W. E., and Stephan, F. F. (1940).  On a least squares adjustment of a sampled frequency table when the expected marginal totals are known.  {\em Annals of Mathematical Statistics} {\bf 11}, 427--444.

\bibitem  Devezer, B., Nardin, L. G., Baumgaertner, B., and Buzbas, E. O. (2019).  Scientific discovery in a model-centric framework: Reproducibility, innovation, and epistemic diversity.  {\em PLoS One} {\bf 14}, e0216125. 

\bibitem Devezer, B., Navarro, D. J., Vanderkerckhove, J., and  Buzbas, E. O. (2020). The case for formal methodology in scientific reform.  \url{doi.org/10.1101/2020.04.26.048306}

\bibitem Dragicevic, P., Jansen, Y., Sarma, A., Kay, M., and Chevalier, F. (2019). Increasing the transparency of research papers with explorable multiverse analyses. {\em Proceedings of the 2019 CHI Conference on Human Factors in Computing Systems},  paper no.\ 65.

\bibitem Efron, B. (2013).  Estimation and accuracy after model selection.  {\em Journal of the American Statistical Association} {\bf 109}, 991--1007.

\bibitem Finkel, J. R., and Manning, C. D. (2009). Hierarchical Bayesian domain adaptation.  In {\em Proceedings of Human Language Technologies: The 2009
Annual Conference of the North American Chapter of the Association for
Computational Linguistics}, 602--610.

\bibitem Fithian, W., Taylor, J., Tibshirani, R., and Tibshirani, R. J. (2015).  Selective sequential model selection.  \url{arxiv.org/pdf/1512.02565.pdf}

\bibitem Flaxman, S., Mishra, S., Gandy, A., et al.\ (2020). Estimating the effects of non-pharmaceutical interventions on COVID-19 in Europe. {\em Nature} {\bf 584}, 257--261. Data and code at \url{github.com/ImperialCollegeLondon/covid19model}

\bibitem Fuglstad, G. A., Simpson, D., Lindgren, F., and Rue, H. (2019). Constructing priors that penalize the complexity of Gaussian random fields.  {\em Journal of the American Statistical Association} {\bf 114}, 445--452.

\bibitem Gabry, J., et al.\ (2020a).  rstanarm: Bayesian applied regression modeling via Stan, version 2.19.3.  \url{cran.r-project.org/package=rstanarm}

\bibitem Gabry, J., et al.\ (2020b).  bayesplot: Plotting for Bayesian models, version 1.7.2.
\url{cran.r-project.org/package=bayesplot}
  
  
\bibitem Gabry, J., Simpson, D., Vehtari, A., Betancourt, M., and Gelman, A. (2019).  Visualization in Bayesian workflow (with discussion and rejoinder). {\em Journal of the Royal Statistical Society A} {\bf 182}, 389--441.

\bibitem Gelman, A., Bois, F. Y., and Jiang, J. (1996). Physiological pharmacokinetic analysis using population modeling and informative prior distributions. {\em Journal of the American Statistical Association} {\bf 91}, 1400--1412.
  
\bibitem Gelman, A. (2003).  A Bayesian formulation of exploratory data analysis and goodness-of-fit testing.  {\em International Statistical Review} {\bf 71}, 369--382.

\bibitem Gelman, A. (2004).  Parameterization and Bayesian modeling. {\em Journal of the American Statistical Association} {\bf 99}, 537--545.

\bibitem Gelman, A. (2011).   Expanded graphical models: Inference, model comparison, model checking, fake-data debugging, and model understanding.  \url{www.stat.columbia.edu/~gelman/presentations/ggr2handout.pdf}

\bibitem Gelman, A. (2014).  How do we choose our default methods?  In {\em Past, Present, and Future of Statistical Science}, ed.\ X. Lin, C. Genest, D. L. Banks, G. Molenberghs, D. W. Scott, and J. L. Wang.  London:  CRC Press.

\bibitem Gelman, A. (2019).  Model building and expansion for golf putting.  {\em Stan Case Studies} {\bf 6}.  \url{mc-stan.org/users/documentation/case-studies/golf.html}
 
\bibitem Gelman, A., et al.\ (2020). Prior choice recommendations. \url{github.com/stan-dev/stan/wiki/Prior-Choice-Recommendations}

\bibitem Gelman, A., and Azari, J. (2017).  19 things we learned from the 2016 election (with discussion). {\em Statistics and Public Policy} {\bf 4}, 1--10.

\bibitem Gelman, A., Carlin, J. B., Stern, H. S., Dunson, D. B., Vehtari, A., and  Rubin, D. B. (2013). {\em Bayesian Data Analysis}, third edition.  London:  CRC Press.
 
\bibitem Gelman, A., and Hill, J. (2007).  {\em Data Analysis Using Regression and Multilevel/Hierarchical Models}.  Cambridge University Press.

\bibitem Gelman, A., Hill, J., and Vehtari, A. (2020).  {\em Regression and Other Stories}.  Cambridge University Press.

\bibitem Gelman, A., Hill, J., and Yajima, M. (2012). Why we (usually) don't have to worry about multiple comparisons. {\em Journal of Research on Educational Effectiveness} {\bf 5}, 189--211. 

\bibitem Gelman, A., Hullman, J., Wlezien, C., and Morris, G. E. (2020).  Information, incentives, and goals in election forecasts.  {\em Judgment and Decision Making} {\bf 15}, 863--880.

\bibitem Gelman, A., and Loken, E. (2013). The garden of forking paths: Why multiple comparisons can be a problem, even when there is no ``fishing expedition'' or ``p-hacking'' and the research hypothesis was posited ahead of time. \url{www.stat.columbia.edu/~gelman/research/unpublished/forking.pdf}

\bibitem Gelman, A., Meng, X. L., and Stern, H. S. (1996).  Posterior predictive assessment of model fitness via realized discrepancies
(with discussion).  {\em Statistica Sinica} {\bf 6}, 733--807.

\bibitem Gelman, A., Simpson, D., and Betancourt, M. (2017). The prior can often only be understood in the context of the likelihood. {\em Entropy} {\bf 19}, 555.

\bibitem Gelman, A., Stevens, M., and Chan, V. (2003). Regression modeling and meta-analysis for decision making: A cost-benefit analysis of a incentives in telephone surveys.  {\em Journal of Business and Economic Statistics} {\bf 21}, 213--225.

\bibitem Gharamani, Z., Steinruecken, C., Smith, E., Janz, E., and Peharz, R. (2019). The Automatic Statistician: An artificial intelligence for data science. \url{www. automaticstatistician.com/index}

\bibitem Ghitza, Y., and Gelman, A. (2020).  Voter registration databases and MRP: Toward the use of large scale databases in public opinion research. {\em Political Analysis} {\bf 28},  507--531.

\bibitem Giordano, R. (2018).  StanSensitivity.  \url{github.com/rgiordan/StanSensitivity}
  
\bibitem Giordano, R., Broderick, T., and Jordan, M. I. (2018).  Covariances, robustness, and variational Bayes. {\em Journal of Machine Learning Research} {\bf 19}, 1981--2029.

\bibitem Goel, P. K., and DeGroot, M. H. (1981). Information about hyperparameters in hierarchical models. {\em Journal of the American Statistical Association} {\bf 76}, 140--147.

\bibitem Grinsztajn, L., Semenova, E., Margossian, C. C., and Riou, J. (2020).  Bayesian workflow for disease transmission modeling in Stan.  \url{mc-stan.org/users/documentation/case-studies/boarding_school_case_study.html}

\bibitem Grolemund, G., and Wickham, H. (2017). {\em R for Data Science}.  Sebastopol, Calif.:  O'Reilly Media.

\bibitem Gunning, D. (2017).  Explainable artificial intelligence (xai). U.S. Defense Advanced Research Projects Agency (DARPA) Program.

\bibitem Henderson, C. R. (1950).  Estimation of genetic parameters (abstract).  {\em Annals of Mathematical Statistics} {\bf 21}, 309--310.
  
\bibitem Hill, J. L. (2011).  Bayesian nonparametric modeling for causal inference. {\em Journal of Computational and Graphical Statistics} {\bf 20}, 217--240.

\bibitem Hodges, J. S., and Reich, B. J. (2010). Adding spatially-correlated errors can mess up the fixed effect you love. {\em American Statistician} {\bf 64}, 325--334.

\bibitem Hoffman, M., and Ma, Y. (2020). Black-box variational inference as a parametric approximation to Langevin dynamics. {\em Proceedings of Machine Learning and Systems}, in press. 

\bibitem Hunt, A., and Thomas, D. (1999).  {\em The Pragmatic Programmer}.  Addison-Wesley.

\bibitem Hwang, Y., Tong, A. and Choi, J. (2016). The Automatic Statistician: A relational perspective. ICML 2016: Proceedings of the 33rd International Conference on Machine Learning.

\bibitem Jacquez, J. A. (1972).  {\em Compartmental Analysis in Biology and Medicine}.  Elsevier.

\bibitem Kale, A., Kay, M., and Hullman, J. (2019). Decision-making under uncertainty in research synthesis: Designing for the garden of forking paths. {\em Proceedings of the 2019 CHI Conference on Human Factors in Computing Systems}, paper no.\ 202.

\bibitem Kamary, K., Mengersen, K., Robert, C. P., and Rousseau, J. (2019). Testing hypotheses via a mixture estimation model. \url{arxiv.org/abs/1412.2044}
  
\bibitem Katz, J. (2016).  Who will be president?  \url{www.nytimes.com/interactive/2016/upshot/presidential-polls-forecast.html}

\bibitem Kay, M. (2020a). ggdist: Visualizations of distributions and uncertainty. R package version 2.2.0. \url{mjskay.github.io/ggdist}. doi:10.5281/zenodo.3879620.

\bibitem Kay, M. (2020b). tidybayes: Tidy data and geoms for Bayesian models. R package version 2.1.1. \url{mjskay.github.io/tidybayes}. doi:10.5281/zenodo.1308151.

\bibitem Kennedy, L., Simpson, D., and Gelman, A. (2019). The experiment is just as important as the likelihood in understanding the prior: A cautionary note on robust cognitive modeling. {\em Computational Brain and Behavior} {\bf 2}, 210--217.

\bibitem Kerman, J., and Gelman, A. (2004).  Fully Bayesian computing. \url{www.stat.columbia.edu/~gelman/research/unpublished/fullybayesiancomputing-nonblinded.pdf}
  
\bibitem Kerman, J., and Gelman, A. (2007).  Manipulating and summarizing posterior simulations using random variable objects. {\em Statistics and Computing} {\bf 17}, 235--244.

\bibitem Kucukelbir, A., Tran, D., Ranganath, R., Gelman, A., and Blei, D. M. (2017). Automatic differentiation variational inference. {\em Journal of Machine Learning Research} {\bf 18}, 1--45.

\bibitem
Kumar, R., Carroll, C., Hartikainen, A., and Martin, O. A. (2019).
ArviZ a unified library for exploratory analysis of Bayesian models in Python. {\em Journal of Open Source Software}, doi:10.21105/joss.01143.

\bibitem Lambert, B., and Vehtari, A. (2020). $R^*$: A robust MCMC convergence diagnostic with uncertainty using gradient-boosted machines.  \url{arxiv.org/abs/2003.07900}

\bibitem Lee, M. D., Criss, A. H., Devezer, B., Donkin, C., Etz, A., Leite, F. P., Matzke, D., Rouder, J. N., Trueblood, J. S., White, C. N., and Vandekerckhove, J. (2019).  Robust modeling in cognitive science.  {\em Computational Brain and Behavior} {\bf 2}, 141--153.

\bibitem Lin, C. Y., Gelman, A., Price, P. N., and Krantz, D. H. (1999).  Analysis of local decisions using hierarchical modeling, applied to home radon measurement and remediation (with discussion).  {\em Statistical Science} {\bf 14}, 305--337.

\bibitem Lindley, D. V. (1956). On a measure of the information provided by an experiment. {\em Annals of Mathematical Statistics} {\bf 27}, 986--1005. 

\bibitem Lins, L., Koop, D., Anderson, E. W., Callahan, S. P., Santos, E., Scheidegger, C. E., Freire, J., and Silva, C. T. (2008).  Examining statistics of workflow evolution provenance: A first study. In {\em Scientific and Statistical Database Management, SSDBM 2008}, ed.\ B. Ludäscher and N. Mamoulis, 573--579.  Berlin: Springer.
  
\bibitem Linzer, D. A. (2013). Dynamic Bayesian forecasting of presidential elections in the states. {\em Journal of the American Statistical Association} {\bf 108}, 124--134.

\bibitem Liu, Y., Harding, A., Gilbert, R., and Journel, A. G. (2005).  A workflow for multiple-point geostatistical simulation. In {\em Geostatistics Banff 2004}, ed.\ O. Leuangthong and C. V. Deutsch.  Dordrecht:  Springer.

\bibitem Loftus, J. (2015).  Selective inference after cross-validation.  \url{arxiv.org/pdf/1511.08866.pdf}
 
\bibitem Long, J. S. (2009).  {\em The Workflow of Data Analysis Using Stata}.  London:  CRC Press.

\bibitem Mallows, C. L. (1973).  Some comments on $C_p$.  {\em Technometrics} {\bf 15}, 661--675.

\bibitem Margossian, C. C, and Gelman, A (2020). Bayesian Model of Planetary Motion: exploring ideas for a modeling workflow when dealing with ordinary differential equations and multimodality. \textit{Technical Report}, \url{https://github.com/stan-dev/example-models/tree/case-study/planet/knitr/planetary_motion}

\bibitem Margossian, C. C., Vehtari, A., Simpson, D., and Agrawal, R. (2020a). Hamiltonian Monte Carlo using an adjoint-differentiated Laplace approximation: Bayesian inference for latent Gaussian models and beyond. \textit{Advances in Neural Information Processing Systems 34}, \textit{page to appear}, \url{arXiv:2004.12550}

\bibitem Margossian, C. C., Vehtari, A., Simpson, D., and Agrawal, R. (2020b). Approximate Bayesian inference for latent Gaussian models in Stan. \textit{Stan Con 2020}, \url{researchgate.net/publication/343690329_Approximate_Bayesian_inference_for_latent_Gaussian_models_in_Stan}

\bibitem Mayo, D. (2018).  {\em Statistical Inference as Severe Testing: How to Get Beyond the Statistics Wars}.  Cambridge University Press.

\bibitem McConnell, S. (2004).  {\em Code Complete}, second edition.  Microsoft Press.

\bibitem Meng, X. L., and van Dyk, D. A. (2001).  The art of data augmentation.  {\em Journal of Computational and Graphical Statistics} {\bf 10}, 1--50.

\bibitem Merkle, E. C., Furr, D., and Rabe-Hesketh, S. (2019). Bayesian comparison of latent variable models: Conditional versus marginal likelihoods. {\em Psychometrika} {\bf 84}, 802--829.

\bibitem Millar, R. B. (2018). Conditional vs marginal estimation of the predictive loss of hierarchical models using WAIC and cross-validation. {\em Statistics and Computing} {\bf 28}, 375--385.

\bibitem Modrák, M. (2018). Reparameterizing the sigmoid model of gene regulation for Bayesian inference. {\em Computational Methods in Systems Biology. CMSB 2018. Lecture Notes in Computer Science, vol.\ 11095}, 309--312. 

\bibitem Montgomery, J. M., and Nyhan, B. (2010). Bayesian model averaging: Theoretical developments and practical applications. {\em Political Analysis} {\bf 18}, 245--270. 

\bibitem Morgan, S. L., and Winship, C. (2014). {\em Counterfactuals and Causal Inference: Methods and Principles for Social Research}, second edition. Cambridge University Press.
  
\bibitem Morris, G. E., Gelman, A., and Heidemanns, M. (2020).  How the Economist presidential forecast works.  \url{projects.economist.com/us-2020-forecast/president/how-this-works}

\bibitem Navarro, D. J. (2019).  Between the devil and the deep blue sea: Tensions between scientific judgement and statistical model selection. {\em Computational Brain and Behavior} {\bf 2}, 28--34.

\bibitem Navarro, D. J. (2020). If mathematical psychology did not exist we might need to invent it: A comment on theory building in psychology. {\em Perspectives on Psychological Science}.  \url{psyarxiv.com/ygbjp}

\bibitem  Neal, R. M. (1993). Probabilistic inference using Markov chain Monte Carlo methods. Technical Report CRG-TR-93-1, Department of Computer Science, University of Toronto.

\bibitem Neal, R. M. (2011).  MCMC using Hamiltonian dynamics.  In {\em Handbook of Markov Chain Monte Carlo}, ed.\ S. Brooks, A. Gelman, G. L. Jones, and X. L. Meng, 113--162.  London:  CRC Press.

\bibitem Niederlová, V., Modrák, M., Tsyklauri, O., Huranová, M., and Štěpánek, O. (2019). Meta-analysis of genotype-phenotype associations in Bardet-Biedl Syndrome uncovers differences among causative genes. {\em Human Mutation} {\bf 40}, 2068--2087.

\bibitem Nott, D. J., Wang, X., Evans, M., and Englert, B. G. (2020). Checking for prior-data conflict using prior-to-posterior divergences. {\em Statistical Science} {\bf 35}, 234--253.

\bibitem Novick, M. R., Jackson, P. H., Thayer, D. T., and Cole, N. S. (1972).
Estimating multiple regressions in $m$-groups:  a cross validation study.
{\em British Journal of Mathematical and Statistical Psychology} {\bf 25},
33--50.

\bibitem O'Hagan, A., Buck, C. E., Daneshkhah, A., Eiser, J. R., Garthwaite, P. H., Jenkinson, D. J., Oakely, J. E., and Rakow, T. (2006). {\em Uncertain Judgements: Eliciting Experts' Probabilities}. Wiley.

\bibitem Paananen, T., Piironen, J., Bürkner, P.-C., and Vehtari, A. (2020). Implicitly adaptive importance sampling. {\em Statistics and Computing}, in press.

\bibitem Pearl, J., and Bareinboim, E. (2011).  Transportability of causal and statistical relations: A formal approach. In {\em Data Mining Workshops (ICDMW), 2011 IEEE 11th International Conference}, 540--547.
  
\bibitem Pearl, J., and Bareinboim, E. (2014).  External validity: From do-calculus to transportability across populations. {\em Statistical Science} {\bf 29}, 579--595.

\bibitem Piironen, J., and Vehtari, A. (2017).  Sparsity information and regularization in the horseshoe and other shrinkage priors. {\em Electronic Journal of Statistics} {\bf 11}, 5018--5051.

\bibitem Pirš, G., and Štrumbelj, E. (2009). Bayesian combination of probabilistic classifiers using multivariate normal mixtures. {\em Journal of Machine Learning Research} {\bf 20}, 1--18.
  
\bibitem Price, P. N., Nero, A. V., and Gelman, A. (1996). Bayesian prediction of mean indoor radon concentrations for Minnesota counties. {\em Health Physics} {\bf 71}, 922--936.

\bibitem Rasmussen, C. E., and Williams, C. K. I. (2006).  {\em Gaussian Processes for Machine Learning}.  MIT Press.

\bibitem Raudenbush, S. W., and Bryk, A. S. (2002).
{\em Hierarchical Linear Models}, second edition.  Sage Publications.

\bibitem Richardson, S., and Gilks, W. R. (1993). A Bayesian approach to measurement error problems in epidemiology using conditional independence models. {\em American Journal of Epidemiology} {\bf 138}, 430--442.

\bibitem Riebler, A., Sørbye, S. H., Simpson, D., and Rue, H. (2018).  An intuitive Bayesian spatial model for disease mapping that accounts for scaling.  {\em Statistical Methods in Medical Research} {\bf 25}, 1145--1165.

\bibitem Robert, C., and Casella, G. (2011). A short history of Markov chain Monte Carlo: Subjective recollections from incomplete data. {\em Statistical Science} {\bf 26}, 102--115.

\bibitem Rubin, D. B. (1984).  Bayesianly justifiable and relevant frequency calculations for the applied statistician.  {\em Annals of Statistics} {\bf 12}, 1151--1172.

\bibitem Rudin, C. (2018). Please stop explaining black box models for high stakes decisions. NeurIPS 2018 Workshop on Critiquing and Correcting Trends in Machine Learning.  \url{arxiv.org/abs/1811.10154}

\bibitem Rue, H., Martino, S., and Chopin, N. (2009). Approximate Bayesian inference for latent Gaussian models by using integrated nested Laplace approximations. {\em Journal of the Royal Statistical Society B} {\bf 71}, 319--392.

\bibitem Sarma, A., and Kay, M. (2020). Prior setting in practice: Strategies and rationales used in choosing prior distributions for Bayesian analysis. {\em Proceedings of the 2020 CHI Conference on Human Factors in Computing Systems}.

\bibitem Savage, J. (2016).  What is modern statistical workflow?  \url{khakieconomics.github.io/2016/08/29/What-is-a-modern-statistical-workflow.html}

\bibitem Shi, X., and Stevens, R. (2008).  SWARM: a scientific workflow for supporting bayesian approaches to improve metabolic models.  {\em CLADE '08: Proceedings of the 6th International Workshop on Challenges of Large Applications in Distributed Environments}, 25--34.

\bibitem Shirani-Mehr, H., Rothschild, D., Goel, S., and Gelman, A. (2018). Disentangling bias and variance in election polls. {\em Journal of the American Statistical Association} {\bf 118}, 607--614.

\bibitem Simmons, J., Nelson, L., and Simonsohn, U. (2011).  False-positive psychology:  Undisclosed flexibility in data collection and analysis allow presenting anything as significant.  {\em Psychological Science} {\bf 22}, 1359--1366.

\bibitem Simpson, D., Rue, H., Riebler, A., Martins, T. G., and Sørbye, S. H. (2017).  Penalising model component complexity: A principled, practical approach to constructing priors.  {\em Statistical Science} {\bf 32}, 1--28.

\bibitem Singer, E., Van Hoewyk, J., Gebler, N., Raghunathan, T., and McGonagle, K. (1999).  The effects of incentives on response rates in interviewer-mediated surveys.  {\em Journal of Official Statistics} {\bf 15}, 217--230.

\bibitem Sivula, T., Magnusson, M, and Vehtari, A. (2020). Uncertainty in Bayesian leave-one-out cross-validation based model comparison.  \url{arxiv.org./abs/2008.10296}

\bibitem Skrondal, A. and Rabe-Hesketh, S. (2004). {\em Generalized Latent Variable Modeling: Multilevel, Longitudinal and Structural Equation Models}. London:  CRC Press.

\bibitem Smith, A. (2013). {\em Sequential Monte Carlo Methods in Practice}. New York: Springer.
  
\bibitem Stan Development Team (2020). {\em Stan User's Guide}.  \url{mc-stan.org}

\bibitem Steegen, S., Tuerlinckx, F., Gelman, A., and Vanpaemel, W. (2016). Increasing transparency through a multiverse analysis. {\em Perspectives on Psychological Science} {\bf 11}, 702--712.

\bibitem Stone, M. (1974).  Cross-validatory choice and assessment of statistical predictions (with discussion). {\em Journal of the Royal Statistical Society B} {\bf 36}, 111--147.

\bibitem Stone, M. (1977). An asymptotic equivalence of choice of model cross-validation and Akaike's criterion. {\em Journal of the Royal Statistical Society B} {\bf 36}, 44--47.

\bibitem Talts, S., Betancourt, M., Simpson, D., Vehtari, A., and Gelman, A. (2020).  Validating Bayesian inference algorithms with simulation-based calibration. \url{www.stat.columbia.edu/~gelman/research/unpublished/sbc.pdf}

\bibitem Taylor, J., and Tibshirani, R. J. (2015).  Statistical learning and selective inference.  {\em Proceedings of the National Academy of Sciences} {\bf 112}, 7629--7634.
  
\bibitem Taylor, S. J., and Lethem, B. (2018).  Forecasting at scale. {\em American Statistician} {\bf 72}, 37--45.

\bibitem Tierney, L., and Kardane, J.B. (1986).
Accurate approximations for posterior moments and marginal densities.
{\em Journal of the American Statistical Association},
81(393):82–86.

\bibitem Turner, K. J., and Lambert, P. S. (2015).  Workflows for quantitative data analysis in the social sciences.  {\em International Journal on Software Tools for Technology Transfer} {\bf 17}, 321--338.

\bibitem Unwin, A., Volinsky, C., and Winkler, S. (2003). Parallel coordinates for exploratory modelling analysis. {\em Computational Statistics and Data Analysis} {\bf 43}, 553--564.

\bibitem Vehtari, A. (2019).  Cross-validation for hierarchical models.  \url{avehtari.github.io/modelselection/rats_kcv.html}

\bibitem Vehtari A., Gabry J., Magnusson M., Yao Y., Bürkner P., Paananen T., Gelman A. (2020). loo:
Efficient leave-one-out cross-validation and WAIC for Bayesian models. R package version 2.3.1, \url{mc-stan.org/loo}.

\bibitem Vehtari, A., and Gabry, J. (2020).  Bayesian stacking and pseudo-BMA weights using the loo package.  \url{mc-stan.org/loo/articles/loo2-weights.html}
  
\bibitem Vehtari, A., Gelman, A., and Gabry, J. (2017). Practical Bayesian model evaluation using leave-one-out cross-validation and WAIC. {\em Statistics and Computing} {\bf 27}, 1413--1432.

\bibitem Vehtari, A., Gelman, A., Simpson, D., Carpenter, D., and Bürkner, P.-C. (2020).  Rank-normalization, folding, and localization: An improved R-hat for assessing convergence of MCMC. {\em Bayesian Analysis}. 

\bibitem Vehtari, A., Gelman, A., Sivula, T., Jylanki, P., Tran, D., Sahai, S., Blomstedt, P., Cunningham, J. P., Schiminovich, D., and Robert, C. P. (2020). Expectation propagation as a way of life: A framework for Bayesian inference on partitioned data. {\em Journal of Machine Learning Research} {\bf 21}, 1--53.

\bibitem Vehtari, A., Simpson, D., Gelman, A., Yao, Y., and Gabry, J. (2015). Pareto smoothed importance sampling.  \url{arxiv.org/abs/1507.02646}

\bibitem Wang, W., and Gelman, A. (2015).   Difficulty of selecting among multilevel models using predictive accuracy. {\em Statistics and Its Interface} {\bf 8} (2), 153--160.
  
\bibitem Weber, S., Gelman, A., Lee, D., Betancourt, M., Vehtari, A., and Racine-Poon, A. (2018).  Bayesian aggregation of average data: An application in drug development. {\em Annals of Applied Statistics} {\bf 12}, 1583--1604. 

\bibitem Wickham, H. (2006). Exploratory model analysis with R and GGobi. \url{had.co.nz/model-vis/2007-jsm.pdf}

\bibitem Wickham, H., Cook, D., and Hofmann, H. (2015).  Visualizing statistical models: Removing the blindfold.  {\em Statistical Analysis and Data Mining: The ASA Data Science Journal} {\bf 8}, 203--225.

\bibitem Wickham, H., and Groelmund, G. (2017).  {\em R for Data Science}.  Sebastopol, Calif.:  O'Reilly.
  
\bibitem Wilson, G.,  Aruliah, D. A., Brown, C. T., Hong, N. P. C.,  Davis, M., Guy, R. T., Haddock, S. H. D., Huff, K. D., Mitchell, I. M., Plumbley, M. D., Waugh, B., White, E. P., and Wilson, P. (2014).  Best practices for scientific computing.  {\em PLoS Biology} {\bf 12}, e1001745.

\bibitem Wilson, G., Bryan, J., Cranston, K., Kitzes, J. Nederbragt, L., and Teal, T. K. (2017). Good enough practices in scientific computing. {\em PLoS Computational Biololgy} {\bf 13}, e1005510. 

\bibitem Yao, Y., Cademartori, C.,  Vehtari, A., and Gelman, A. (2020).   Adaptive path sampling in metastable posterior distributions.
\url{arxiv.org/abs/2009.00471}

\bibitem Yao, Y., Vehtari, A., and Gelman, A. (2020).  Stacking for non-mixing Bayesian computations: The curse and blessing of multimodal posteriors.  \url{arxiv.org/abs/2006.12335}

\bibitem Yao, Y., Vehtari, A., Simpson, D., and Gelman, A. (2018a). Yes, but did it work?: Evaluating variational inference.  In {\em Proceedings of  International Conference on Machine Learning}, 5581--5590. 

\bibitem Yao, Y., Vehtari, A., Simpson, D., and Gelman, A. (2018b).  Using stacking to average Bayesian predictive distributions (with discussion). {\em Bayesian Analysis} {\bf 13}, 917--1003.

\bibitem Yu, B., and Kumbier, K. (2020).  Veridical data science.  {\em Proceedings of the National Academy of Sciences} {\bf 117}, 3920--3929.

\bibitem Zhang, Y. D., Naughton, B. P., Bondell, H. D., and Reich, B. J. (2020). Bayesian regression using a prior on the model fit: The R2-D2 shrinkage prior. {\em Journal of the American Statistical Association}, doi:10.1080/01621459.2020.1825449

\end{document}